
%
%
%
%
\input psfig
\magnification=\magstep1
\vsize=46pc
\overfullrule=0pt
\def\lb{\lbrack}
\def\rb{\rbrack}
\def\min{{\rm min}}
\def\q#1{[#1]}
\def\bibitem#1{\parindent=9.5mm\item{\hbox to 7.5 mm{$\q{#1}$\hfill}}}
\def\mn{\medskip\smallskip\noindent}
\def\sn{\smallskip\noindent}
\def\bn{\bigskip\noindent}
\def\figindents{\leftskip=3 true pc \rightskip=2 true pc}
\font\helv=cmss10
\font\helvit=cmssi10
\font\helvits=cmssi8
\font\iteight=cmti8
\font\amsmath=msbm10
\font\amsmaths=msbm8
\font\fraktur=eufm10
\def\Parity{\hbox{\fraktur P}}
\def\HH{\hbox{\extra H}}
\def\Real{\hbox{\amsmath R}}
\def\Complex{\hbox{\amsmath C}}
\def\ComplexBar{\overline{\Complex}}
\def\Zed{\hbox{\amsmath Z}}
\def\zed{\hbox{\amsmaths Z}}
\def\Rational{\hbox{\amsmath Q}}
\font\extra=cmss10 scaled \magstep0

\setbox22 = \hbox{{{\extra 1}}}
\def\One{{{\extra 1}}\hskip-\wd22\hskip 2.5 true pt{{\extra 1}}}
\def\id{\hbox{{\extra\One}}}
\def\mod{\phantom{l} {\rm mod} \phantom{l}}
\def\vac{\mid \!v \rangle\, }
\def\avac{\langle v \! \mid\, }
\def\state#1{\mid \! #1 \rangle\, }
\def\rrangle{\rangle \hskip -1pt \rangle}
\def\llangle{\langle \hskip -1pt \langle}
\def\astate#1{\langle #1 \! \mid\, }
\def\pstate#1{\Vert #1 \rrangle\, }
\def\apstate#1{\llangle #1 \Vert\, }
\def\norm#1{\Vert #1 \Vert}
\def\abs#1{\vert #1 \vert}
\font\bigboldfont=cmbx10 scaled \magstep2
\def\chaptitle#1{{\bigboldfont \leftline{#1}}
\vskip-10pt
\line{\hrulefill}}
\def\chapsubtitle#1{\leftline{\bf #1}
\vskip-10pt
\line{\hrulefill}}
\countdef\contcnt=111
\def\contents#1#2{\advance\contcnt by #2\line{\bf #1 \dotfill
         \hbox to 5mm{\hfil \number\contcnt}}}
\def\contentssub#1#2{\advance\contcnt by #2\line{\qquad #1 \dotfill
         \hbox to 5mm{\hfil \number\contcnt}}}
\def\vl{\hskip 1pt \vrule}
\def\normstate#1{\langle #1 \! \mid \!#1 \rangle\, }
\def\normvac{\normstate{v}}
\def\ab{\bar{\alpha}}
\def\a{\alpha}
\def\si{\sigma}
\def\Ga{\Gamma}
\def\om{\omega}
\def\la{\lambda}
\def\vphi{\varphi}
\def\rt{\tilde{r}}
\def\D{{\cal D}}
\def\E{{\cal E}}
\def\EE{\hbox{{\extra E}}}
\def\H{{\cal H}}
\def\HN{\H_N}

\def\O{{\cal O}}
\def\GS#1{\state{{\rm GS}}^{#1}}
\def\bsin#1{\sin{\left(#1\right)}}
\def\bcos#1{\cos{\left(#1\right)}}
\def\cPh{\bcos{\textstyle{P \over 2}}}
\def\phit{{\phi \over 3}}
\def\Ph{{P \over 2}}
\def\PMQ#1{P_{{\rm min},#1}}
\def\re{{\rm Re}}

\def\Ddots{\mathinner{\mkern1mu\raise1pt\hbox{.}\mkern2mu
       \raise4pt\hbox{.}\mkern2mu\raise7pt\vbox{\kern7pt\hbox{.}}\mkern1mu}}
\def\mskp{\mkern-6mu}
\def\hr{\hbox{\raise2.5pt\hbox to 1.0cm{\hrulefill}}}
\def\Hr#1{\mathop{\hr}\limits_{#1}}
\def\hrr{\hbox{\raise1pt\hbox to 1.0cm{\hrulefill
 }\hskip-1.0cm\raise4pt\hbox to 1.0cm{\hrulefill}}}
\def\Hrr#1{\mathop{\hrr}\limits_{#1}}
\def\hrf{\hbox{\raise1pt\hbox to 1.0cm{\hrulefill
 }\hskip-1.0cm\raise2pt\hbox to 1.0cm{\hrulefill
 }\hskip-1.0cm\raise3pt\hbox to 1.0cm{\hrulefill
 }\hskip-1.0cm\raise4pt\hbox to 1.0cm{\hrulefill}}}
\def\Hrf#1{\mathop{\hrf}\limits_{#1}}
\def\bull{\bullet}
\def\bu#1{\mathop\bull\limits^{#1}}
\def\longTO{\mathop{\longrightarrow}}
\def\cab{{\cal C}}
\def\cnab#1{{\bar{\cal C}_{#1}}}
\def\chab{{\widehat{\cal C}}}
\def\Rab{{\cal R}}
\def\Uab{{\cal U}}
\def\cabt{\tilde{\cab}}
\def\Rabt{\tilde{\Rab}}
%
\def\verline#1#2#3{\rlap{\kern#1mm\raise#2mm
                   \hbox{\vrule height #3mm depth 0 pt}}}
\def\horline#1#2#3{\rlap{\kern#1mm\raise#2mm
                   \vbox{\hrule width #3mm depth 0 pt}}}
\def\Verline#1#2#3{\rlap{\kern#1mm\raise#2mm
                   \hbox{\vrule height #3mm width 0.7pt depth 0 pt}}}
\def\Horline#1#2#3{\rlap{\kern#1mm\raise#2mm
                   \vbox{\hrule height 0.7pt width #3mm depth 0 pt}}}
\def\putbox#1#2#3{\setbox117=\hbox{#3}
                  \dimen121=#1mm
                  \dimen122=#2mm
                  \dimen123=\wd117
                  \dimen124=\ht117
                  \divide\dimen123 by -2
                  \divide\dimen124 by -2
                  \advance\dimen121 by \dimen123
                  \advance\dimen122 by \dimen124
                  \rlap{\kern\dimen121\raise\dimen122\hbox{#3}}}
\def\leftputbox#1#2#3{\setbox117=\hbox{#3}
                  \dimen122=#2mm
                  \dimen124=\ht117
                  \divide\dimen124 by -2
                  \advance\dimen122 by \dimen124
                  \rlap{\kern#1mm\raise\dimen122\hbox{#3}}}
\def\topputbox#1#2#3{\setbox117=\hbox{#3}
                  \dimen121=#1mm
                  \dimen123=\wd117
                  \divide\dimen123 by -2
                  \advance\dimen121 by \dimen123
                  \rlap{\kern\dimen121\raise#2mm\hbox{#3}}}

\def\cross#1#2{
  \putbox{#1}{#2}{$\times$}}
\def\vt{\tilde{v}}
\def\vact{\mid \!\vt \rangle\, }
\def\avact{\langle \vt \! \mid\, }
\def\normvact{\normstate{\vt}}
\def\Ct{\tilde{C}}
\def\siz{\si^z}
\def\six{\si^x}
\def\siy{\si^y}
\def\Om{\Omega}
\def\xv{{\vec{x}}}
\def\ei{{\vec{e}_i}}
\def\e{{\vec{e}}}
\def\La{\Lambda}
\def\F{{\cal F}}
\font\Cal=cmbsy10
\def\w{{\cal W}}
\def\bw{\hbox{\Cal W}}
\def\A{{\cal A}}
\def\AC{{\cal C}}
\def\G{{\cal G}}
\def\Lie{{\cal L}}
\def\orb#1{\hbox{Orb}\left(#1\right)}
\def\pl{\hbox{$+$}}
\def\mi{\hbox{$-$}}
\def\Cdot{\hbox{$\cdot$}}
\def\OPE#1#2{#1 \circ #2}
\def\clOPE#1#2{\{#1,#2\}}
\def\oh{{\textstyle{1 \over 2}}}
\def\tr{{\rm tr}}
\def\I{{\cal I}}
\def\K{{\cal K}}
\def\Qp{{\cal Q}}
\def\secIntro{1}
\def\secSsup{2}
\def\secSsa{\secSsup.1}
\def\secSsb{\secSsup.2}
\def\secSsc{\secSsup.3}
\def\secSsd{\secSsup.4}
\def\secW{3}
\def\secB{\secW.1}
\def\secIOPD{\secW.2}
\def\secC{\secW.3}
\def\secX{4}
\def\secF{\secX.1}
\def\secD{\secX.2}
\def\secIOPC{\secX.3}
\def\secE{\secX.4}
\def\secI{\secX.5}
\def\secY{5}
\def\secG{\secY.1}
\def\secGG{\secY.2}
\def\secIOPE{\secY.3}
\def\secFORM{\secY.4}
\def\secZ{6}
\def\secwA{\secZ.1}
\def\secwB{\secZ.2}
\def\secwC{\secZ.3}
\def\secwCl{\secZ.4}
\def\secwCO{\secZ.5}
\def\secwSQ{\secZ.6}
\def\secJ{7}
\def\appA{A}
\def\appB{B}
\def\ahn{1}
\def\albCONF{2}
\def\albcoy{3}
\def\mccoyadv{4}
\def\albertiniA{5}
\def\albertiniB{6}
\def\tang{7}
\def\alcaraz{8}
\def\altschuler{9}
\def\baxterA{10}
\def\yang{11}
\def\perkadv{12}
\def\baakeA{13}
\def\baakeB{14}
\def\bbss{15}
\def\bai{16}
\def\bakinf{17}
\def\bakri{18}
\def\baf{19}
\def\bal{20}
\def\baxterB{21}
\def\baxterC{22}
\def\baym{23}
\def\bpz{24}
\def\bersh{25}
\def\blg{26}
\def\unilet{27}
\def\unicos{28}
\def\supwir{29}
\def\blm{30}
\def\ajl{31}
\def\dBT{32}
\def\bouwschou{33}
\def\bogo{34}
\def\camp{35}
\def\cardyA{36}
\def\cardyB{37}
\def\cardy{38}
\def\rietalB{39}
\def\chrihen{40}
\def\dkcoy{41}
\def\daviesA{42}
\def\daviesB{43}
\def\dijkgraaf{44}
\def\dogra{45}
\def\dunne{46}
\def\wowoPhD{47}
\def\wirrep{48}
\def\proceedings{49}
\def\supwirrep{50}
\def\howcl{51}
\def\elitzur{52}
\def\enriquez{53}
\def\lykyanov{54}
\def\flrep{55}
\def\fatzamA{56}
\def\fateev{57}
\def\fatzamB{58}
\def\FORT{59}
\def\laszlorep{60}
\def\fehort{61}
\def\flohrdipl{62}
\def\frenkel{63}
\def\fkacr{64}
\def\gehlenunpub{65}
\def\gehlenph{66}
\def\hoeger{67}
\def\weA{68}
\def\lett{69}
\def\gehri{70}
\def\gehlen{71}
\def\gehlenA{72}
\def\gehritell{73}
\def\schuetz{74}
\def\stringA{75}
\def\stringB{76}
\def\gepner{77}
\def\goddard{78}
\def\lacaze{79}
\def\jones{80}
\def\grabo{81}
\def\han{82}
\def\hela{83}
\def\henkel{84}
\def\hokimA{85}
\def\hokimB{86}
\def\hokimC{87}
\def\diplom{88}
\def\commute{89}
\def\automos{90}
\def\perturb{91}
\def\horn{92}
\def\klausWAN{93}
\def\hornfeck{94}
\def\klausREP{95}
\def\kneu{96}
\def\hkn{97}
\def\kohmoto{98}
\def\kato{99}
\def\katobook{100}
\def\hgkpriv{101}
\def\kau{102}
\def\kedem{103}
\def\kogut{104}
\def\krallm{105}
\def\nakan{106}
\def\lehmann{107}
\def\rietalA{108}
\def\scm{109}
\def\mcroan{110}
\def\mussardo{111}
\def\narganes{112}
\def\onsager{113}
\def\ostlund{114}
\def\perk{115}
\def\poly{116}
\def\reedsimon{117}
\def\rellich{118}
\def\roanA{119}
\def\roanB{120}
\def\tarasov{121}
\def\toppan{122}
\def\rva{123}
\def\weyl{124}
\def\wilczynski{125}
\def\yildirim{126}
\def\yuwu{127}
\def\don{128}
\def\zam{129}
\def\zamPA{130}
\def\zamPB{131}
\def\zamPC{132}
\def\zamzam{133}
\def\fullbox{\vrule height 1.3ex width 1.2ex depth -.1ex} 
\def\emptybox{\hbox{\vrule\vbox{\hrule\phantom{\fullbox}\hrule}\vrule}}
\def\begindef#1{{\bf Def.\ #1:} \helv}
\def\beginddef#1#2{{\bf Def.\ #1:} \helv (#2)}
\def\enddef{\rm}
\def\begintheorem#1{{\bf Theorem #1:} \it}
\def\beginttheorem#1#2{{\bf Theorem #1:} \it(#2)}
\def\endTheorem{\rm \vskip-10pt \rightline{\fullbox} \noindent}
\def\endtheorem{\rm \quad \fullbox}
\def\beginlemma#1{{\bf Lemma #1:} \it}
\def\beginllemma#1#2{{\bf Lemma #1:} \it(#2)}

\def\endlemma{\rm \quad \fullbox}
\def\beginprop#1{{\bf Proposition #1:} \it}
\def\beginpprop#1#2{{\bf Proposition #1:} \it(#2)}
\def\endprop{\rm \quad \fullbox}
\def\beginclaim#1{{\bf Claim #1:} \it}
\def\endclaim{\rm \quad \fullbox}
\def\beginremark{{\it Remark:}}
\def\beginremarks{{\it Remarks:}}
\def\endremark{}
\def\beginproof{{\it Proof:}}
\def\endproof{\quad \emptybox}
\def\begincorollary{{\bf Corollary:} \helv}
\def\endcorollary{\rm}
\def\ThSUPa{I}
\def\ThSUPb{II}
\def\ThSUPc{III}
\def\ThSUPX{IV}
\def\ThA{V}
\def\ThB{VI}
\def\LemA{I}
\def\LemCla{II}
\def\LemClA{III}
\def\PropA{I}
\def\PropSpec{II}
\def\PropSQ{III}
\def\DefSUPa{I}
\def\DefSUPb{II}
\def\DefSUPc{III}
\def\DefA{IV}
\def\DefSpec{V}
\def\DefB{VI}
\def\DefClA{VII}
\def\ClaimA{I}
\def\ClaimB{II}
%
%
\font\LARGE=cmr12 scaled \magstep4
\font\Large=cmr12 scaled \magstep2
\font\large=cmbx10 scaled \magstep3
\font\bigF=cmr10 scaled \magstep2
\font\bigf=cmr10 scaled \magstep1
\pageno=-3
\def\folio{
\ifnum\pageno<1 \footline{\hfil} \else\number\pageno \fi}
\phantom{not-so-FUNNY}
\rightline{ BONN--IR--95--12\break}
\rightline{ hep-th/9503104\break}
\rightline{ Ph.D.\ thesis Bonn\break}
\rightline{ March 1995\break}
\vskip 2.0truecm
\centerline{\large Quantum Spin Models}
\vskip 6pt
\centerline{\large \phantom{g} and \phantom{g}}
\vskip 6pt
\centerline{\large Extended Conformal Algebras}
\vskip 1.0truecm
\centerline{\bigF A.\ Honecker}
\bigskip\medskip
\centerline{\it Physikalisches Institut der Universit\"at Bonn}
\centerline{\it Nu{\ss}allee 12, 53115 Bonn, Germany}
\vskip 1.1truecm
\centerline{\bf Abstract}
\vskip 0.2truecm
\noindent
First, an algebraic criterion for integrability is discussed --the so-called
`superintegrability'-- and some results on the classification of
superintegrable quantum spin Hamiltonians based on $sl(2)$ are obtained.
\par\noindent
Next, the massive phases of the $\Zed_n$-chiral Potts quantum spin chain (a
model that violates parity) are studied in detail. It is shown that the
excitation spectrum of the massive high-temperature phase can be explained in
terms of $n-1$ fundamental quasiparticles. We compute correlation functions
from
a perturbative and numerical evaluation of the groundstate for the
$\Zed_3$-chain. In addition to an exponential decay we observe an oscillating
contribution. The oscillation length seems to be related to the asymmetry of
the
dispersion relations. We show that this relation is exact at special values of
the parameters for general $\Zed_n$ using a form factor expansion.
\par\noindent
Finally, we discuss several aspects of extended conformal algebras
($\w$-algebras). We observe an analogy between boundary conditions for
$\Zed_n$-spin chains and $\w$-algebras and then turn to statements about
the structure of $\w$-algebras. In particular, we briefly summarize results
on unifying structures present in the space of all quantum $\w$-algebras.
\vfill
\eject
\pageno=-2
\def\Sk{\hskip 0.5pt}
\setbox114=\hbox{\LARGE.\hskip1.8mm.}
\setbox115=\hbox{\LARGE A}
\centerline{\LARGE
UNIVERSIT\raise\ht115\hbox{\lower\ht114
   \hbox{\LARGE.\hskip1.8mm.}}\hskip-\wd115A{\hskip0.1mm}T BONN}
\vskip 8pt
\centerline{\LARGE
P{\Sk}h{\Sk}y{\Sk}s{\Sk}i{\Sk}k{\Sk}a{\Sk}l{\Sk}i{\Sk}s{\Sk}c\Sk
h{\Sk}e{\Sk}s{\Sk}
I{\Sk}n{\Sk}s{\Sk}t{\Sk}i{\Sk}t{\Sk}u{\Sk}t}
\vskip 2cm
\centerline{\Large Quantum Spin Models}
\vskip 6pt
\centerline{\Large \phantom{g} and \phantom{g}}
\vskip 6pt
\centerline{\Large Extended Conformal Algebras}
\vskip 2.5cm
\centerline{\bigf von}
\mn
\centerline{\bigf Andreas Honecker}
\vfill
\noindent
Dieser Forschungsbericht wurde als Dissertation von der
Mathematisch-Natur\-wissen\-schaft\-lichen Fakult\"at der Universit\"at
Bonn angenommen.
\bn\mn
\settabs \+& \phantom{XXXXXXXXXXXXXX}  & \phantom{XXXXXXXXXXX} & \cr
\+ & Angenommen am:       & 13.02.95                           & \cr
\+ & Referent:            & Prof.\ Dr.\ G\"unter von Gehlen    & \cr
\+ & Korreferent:         & Prof.\ Dr.\ Werner Nahm            & \cr
\eject
\pageno=-1
\font\Greek=cmmi10
\chardef\Gra='013
\chardef\Grb='014
\chardef\Grc='015
\chardef\Grd='016
\chardef\Gre="22
\chardef\Grzeta="10
\chardef\Greta="11
\chardef\Grtheta="23
\chardef\Gri="13
\chardef\Grk="14
\chardef\Grl="15
\chardef\Grm="16
\chardef\Grn="17
\chardef\Grp="19
\chardef\Grr="1A
\chardef\Grs="1B
\chardef\Grss="26
\chardef\Grt="1C
\chardef\Gru="1D
\chardef\Grphi="27
\chardef\Grchi="1F
\chardef\Gro="21
\chardef\GrA="41
\chardef\GrB="42
\chardef\GrC='000
\chardef\GrD='001
\chardef\GrE="45
\chardef\GrL="03
\chardef\GrS="06
\chardef\GrY="07
\chardef\GrPhi="08
\chardef\GrO="0A
\chardef\GrAccIcH="2D
\chardef\GrAcccH="2C
\def\GrAccIc{\rlap{\raise3pt\hbox{\GrAccIcH}}\hskip1pt}
\def\GrAccc#1{\rlap{\kern1pt\raise2.6pt\hbox{\GrAcccH}}#1}
\def\GrAccic#1{\rlap{\kern1pt\raise2.6pt\hbox{\GrAccIcH}}#1}
\def\GrAcciic#1{\rlap{\raise2.6pt\hbox{\GrAccIcH}}#1}
\def\GrGrave#1{{\it\`{\Greek #1}}}
\def\GrAccut#1{{\it\'{\Greek #1}}}
\def\GrTilde#1{{\it\~{\Greek #1}}}
\def\GrTildec#1{\rlap{\kern2pt\raise2pt\hbox{\GrAcccH}
               }\rlap{\kern1pt\raise1.5pt\hbox{\rm\~{}}}#1}
\phantom{x}
\vskip9cm
{\it
\leftskip=2.15cm
\rightskip=2.15cm
\par\noindent
{\helvit F\"ur meine Eltern}
\vskip1cm
\par\noindent
{\Greek \GrAccIc{}E\Grp\Gre\Gri\Grd\GrGrave\Greta}
{\Greek \Grt\GrGrave{o}}
{\Greek \Gre\GrAccic\Gri\Grd\GrAccut\Gre\Grn\Gra\Gri}
{\Greek \Grk\Gra\GrGrave\Gri}
{\Greek \Grt\GrGrave{o}}
{\Greek \GrAccic\Gre\Grp\GrAccut\Gri\Grs\Grt\Gra\Grs\Grtheta\Gra\Gri}
{\Greek \Grs\Gru\Grm\Grb\Gra\GrAccut\Gri\Grn\Gre\Gri}
{\Greek \Grp\Gre\Grr\GrGrave\Gri}
{\Greek \Grp\GrAccut\Gra\Grs\Gra\Grss}
{\Greek \Grt\GrGrave\Gra\Grss}
{\Greek \Grm\Gre\Grtheta\GrAccut{o}\Grd{o}\Gru\Grss},
{\Greek \GrTildec\Gro\Grn}
{\Greek \Gre\GrAccic\Gri\Grs\GrGrave\Gri\Grn}
{\Greek \GrAccic\Gra\Grr\Grchi\Gra\GrGrave\Gri}
{\Greek \GrAccic{\GrGrave\Greta}}
{\Greek \Gra\GrAcciic{\GrAccut\Gri}\Grt\Gri\Gra}
{\Greek \GrAccic{\GrGrave\Greta}}
{\Greek \Grs\Grt{o}\Gri\Grchi\Gre\GrTilde\Gri\Gra},
{\Greek \GrAccic\Gre\Grk}
{\Greek \Grt{o}\GrTilde\Gru}
{\Greek \Grt\Gra\GrTilde\Gru\Grt\Gra}
{\Greek \Grc\Grn\Gro\Grr\GrAccut\Gri\Grzeta\Gre\Gri\Grn}
({\Greek \Grt\GrAccut{o}\Grt\Gre}
{\Greek \Grc\GrGrave\Gra\Grr}
{\Greek o\GrAccic\Gri\GrAccut{o}\Grm\Gre\Grtheta\Gra}
{\Greek \Grc\Gri\Grn\GrAccut\Gro\Grs\Grk\Gre\Gri\Grn}
{\Greek \GrAccc{\GrAccut\Gre}\Grk\Gra\Grs\Grt{o}\Grn},
{\Greek \GrAccc{\GrAccut{o}}\Grt\Gra\Grn}
{\Greek \Grt\GrGrave\Gra}
{\Greek \Gra\GrAcciic{\GrAccut\Gri}\Grt\Gri\Gra}
{\Greek \Grc\Grn\Gro\Grr\GrAccut\Gri\Grs\Gro\Grm\Gre\Grn}
{\Greek \Grt\GrGrave\Gra}
{\Greek \Grp\Grr\GrTilde\Gro\Grt\Gra}
{\Greek \Grk\Gra\GrGrave\Gri}
{\Greek \Grt\GrGrave\Gra\Grss}
{\Greek \GrAccic\Gra\Grr\Grchi\GrGrave\Gra\Grss}
{\Greek \Grt\GrGrave\Gra\Grss}
{\Greek \Grp\Grr\GrAccut\Gro\Grt\Gra\Grss}
{\Greek \Grk\Gra\GrGrave\Gri}
{\Greek \Grm\GrAccut\Gre\Grchi\Grr\Gri}
{\Greek \Grt\GrTilde\Gro\Grn}
{\Greek \Grs\Grt{o}\Gri\Grchi\Gre\GrAccut\Gri\Gro\Grn}),
{\Greek \Grd\GrTilde\Greta\Grl{o}\Grn}
{\Greek \GrAccc{\GrAccut{o}}\Grt\Gri}
{\Greek \Grk\Gra\GrGrave\Gri}
{\Greek \Grt\GrTilde\Greta\Grss}
{\Greek \Grp\Gre\Grr\GrGrave\Gri}
{\Greek \Grphi\GrAccut\Gru\Grs\Gre\Gro\Grss}
{\Greek \GrAccic\Gre\Grp\Gri\Grs\Grt\GrAccut\Greta\Grm\Greta\Grss}
{\Greek \Grp\Gre\Gri\Grr\Gra\Grt\GrAccut\Gre{o}\Grn}
{\Greek \Grd\Gri{o}\Grr\GrAccut\Gri\Grs\Gra\Grs\Grtheta\Gra\Gri}
{\Greek \Grp\Grr\GrTilde\Gro\Grt{o}\Grn}
{\Greek \Grt\GrGrave\Gra}
{\Greek \Grp\Gre\Grr\GrGrave\Gri}
{\Greek \Grt\GrGrave\Gra\Grss}
{\Greek \GrAccic\Gra\Grr\Grchi\GrAccut\Gra\Grss}.
\vskip1cm
\noindent
{\hfill \it {\Greek API{\GrS}TOTE{\GrL}O{\GrY}{\GrS}}
``{\Greek {\GrPhi\GrY\GrS}IKH\GrS} {\Greek AKPOA{\GrS}E\GrO\GrS}.''
  {\Greek A}.\ {1.}}\par\noindent
{\hfill \helvit (Aristoteles' Acht B\"ucher Physik, 1.\ Buch, 1.)}
\par\noindent}
\vfill
\eject
\pageno=0
\contcnt=1
\chaptitle{Contents}
\sn
\mn
\contents{\secIntro.\ Introduction}{0}\sn
\mn
\contents{\secSsup.\ Superintegrable quantum spin models}{4}\sn
\contentssub{\secSsa.\ The general setting}{0}\sn
\contentssub{\secSsb.\ First considerations}{1}\sn
\contentssub{\secSsc.\ The quasi superintegrable case}{1}\sn
\contentssub{\secSsd.\ Results}{1}\sn
\mn
\contents{\secW.\ The chiral Potts quantum chain}{3}\sn
\contentssub{\secB.\ Preliminaries}{0}\sn
\contentssub{\secIOPD.\ Duality of spectra}{4}\sn
\contentssub{\secC.\ Generalities about perturbation theory}{1}\sn
\mn
\contents{\secX.\ Excitation spectrum}{2}\sn
\contentssub{\secF.\ Analytic results for the superintegrable $\Zed_3$-chain}
            {0}\sn
\contentssub{\secD.\ High-temperature expansions}{1}\sn
\contentssub{\secIOPC.\ Spectrum in the low-temperature phase}{5}\sn
\contentssub{\secE.\ Evidence for quasiparticle spectrum}{2}\sn
\contentssub{\secI.\ Convergence of single-particle excitations}{3}\sn
\mn
\contents{\secY.\ Correlation functions}{5}\sn
\contentssub{\secG.\ Correlation functions of the high-temperature phase}
            {0}\sn
\contentssub{\secGG.\ Numerical computation of the correlation functions}
            {5}\sn
\contentssub{\secIOPE.\ Correlation functions in the low-temperature regime}
            {2}\sn
\contentssub{\secFORM.\ Form factors and correlation functions}{3}\sn
\mn
\contents{\secZ.\ Extended conformal algebras}{5}\sn
\contentssub{\secwA.\ Quantum $\w$-algebras}{0}\sn
\contentssub{\secwB.\ Automorphisms of Casimir $\w$-algebras and
                boundary conditions}{4}\sn
\contentssub{\secwC.\ Consequences for spin quantum chains}{3}\sn
\contentssub{\secwCl.\ Classical $\w$-algebras and Virasoro structure}{3}\sn
\contentssub{\secwCO.\ Classical cosets and orbifolds}{4}\sn
\contentssub{\secwSQ.\ The structure of quantum $\w$-algebras}{5}\sn
\mn
\contents{\secJ.\ Conclusion}{7}
\mn
\contentssub{Acknowledgements}{2}\sn
\contentssub{Appendix {\appA}: Perturbation expansions for two-particle states}
                    {1}\sn
\contentssub{Appendix {\appB}: Proof of the symmetries of the
Hamiltonian}{4}\sn
\contentssub{References}{2}\sn
\contentssub{Translations of citations}{7}
\mn
\line{\hrulefill}
\vfill
\eject
{\helvits
{\par\noindent
\leftskip=1cm
\baselineskip=11pt
{\iteight Mathematik} und {\iteight Physik} sind die beiden theoretischen
Erkenntnisse der Vernunft, welche ihre {\iteight Objecte} a priori bestimmen
sollen, die erstere ganz rein, die zweite wenigstens zum Theil rein, dann aber
auch nach Massgabe anderer Erkenntnisquellen als der der Vernunft.
\par\noindent}
\sn
\rightline{Immanuel Kant, Vorrede zur zweiten Auflage der
                                  ``Kritik der reinen Vernunft''}
\par\noindent}
\bn
\chaptitle{\secIntro.\ Introduction}
\mn
Theoretical physicists have paid much attention to low dimensions in the
last years. On the one hand, models in low dimensions can more easily be
treated also on a mathematically rigorous level, but on the other hand,
low dimensions have indeed turned out to be relevant in the description
of important experiments. This thesis is centered around two aspects of
low-dimensional theoretical physics: $\Zed_n$-quantum spin models and
extended conformal algebras.
\mn
Among the $\Zed_n$-quantum spin models we will mainly concentrate on a
`$\Zed_n$-chiral Potts model'. The first chiral Potts model that was
introduced in 1981 by Ostlund in order to describe incommensurate
phases of physisorbed systems $\q{\ostlund}$ was a classical 2D spin model.
The associated quantum chain Hamiltonians were obtained in 1981-82 by
Rittenberg et al.\ $\q{\rietalA,\rietalB}$.
Because this chain was not self-dual the location of the
critical manifold was difficult. In 1983, Howes, Kadanoff
and den Nijs introduced a self-dual $\Zed_3$-symmetric chiral quantum
chain $\q{\hkn}$, which however, does not correspond to a two-dimensional
model with positive Boltzmann weights. Soon afterwards, von Gehlen and
Rittenberg noticed that the remarkable property of the first gap of this model
being linear in the inverse temperature also applies to the second
gap and can be generalized to arbitrary $\Zed_n$ $\q{\gehri}$. Furthermore,
the authors of $\q{\gehri}$ showed that the Ising-like form of the eigenvalues
is related to this $\Zed_n$-Hamiltonian satisfying the Dolan-Grady
integrability condition $\q{\dogra}$ which is equivalent
$\q{\perk,\daviesA,\daviesB}$ to Onsager's algebra $\q{\onsager}$. It was then
shown by Au-Yang, Baxter, McCoy, Perk et al.\ that this integrability
property -- nowadays called `superintegrability' -- can be implemented in a 2D
classical model with Boltzmann weights defined on higher genus Riemann
surfaces that satisfy a generalized Yang-Baxter relation. In the sequel
the chiral Potts model attracted much attention because of these mathematical
aspects, i.e.\ on the one hand due to the generalized Yang-Baxter relations
$\q{\mccoyadv,\albertiniA,\albertiniB,\baxterA,\yang,\perkadv,\baxterB,
 \baxterC,\scm,\mcroan,\roanA,\tarasov}$
and on the other hand because of Onsager's algebra
$\q{\daviesA,\daviesB,\roanB,\ahn}$.
We will present new results showing that the model is also
`physically' very interesting although it is not directly related to a
realistic 2D physisorbed system.
\medskip
We start by a general discussion of quantum spin models and indicate how one
might systematically construct integrable nearest neighbour interaction
quantum spin models exclusively by algebraic means. Afterwards, we specialize
to $\Zed_n$-quantum spin chains but drop the requirement of integrability.
\medskip
One of our main interests will be the excitation spectrum
of $\Zed_n$-quantum spin chains. We will show that in the massive
phases the excitation spectrum can be explained in terms of quasiparticles.
Although this is not very surprising, a particle interpretation is not
directly incorporated into the spin chain Hamiltonian. Therefore it was not
recognized until very recently that a quasiparticle interpretation of
the complete excitation spectrum is possible. Using numerical techniques
we were able to show that the low-lying excitations in the zero momentum
sectors can be explained in terms of $n-1$ fundamental particles for
$n=3$, $4$ at general values of the parameters $\q{\weA,\gehlenph}$ and
checked for $n=3$ that this quasiparticle picture extends to general
momenta $\q{\lett}$.
For the superintegrable $\Zed_3$-chiral Potts model
McCoy et al.\ have derived a quasiparticle picture of the
complete spectrum using techniques related to the Bethe ansatz $\q{\dkcoy}$.
Recently, they argued that this quasiparticle picture should in
general be valid for the integrable $\Zed_3$-chiral Potts
quantum chain $\q{\kedem}$.
In this thesis we will show that both results can be combined
into the general statement that the massive high-temperature
phases of general $\Zed_n$ spin quantum chains have quasiparticle
spectra. In fact, this quasiparticle picture will in certain cases
give small corrections to the additivity of energy in the momentum zero
sectors observed in $\q{\weA}$.
\medskip
Another main issue will be correlation functions of the $\Zed_n$-chiral
Potts model. Since a scaling exponent for the wave vector in the
low-temperature phase of the $\Zed_3$-chiral Potts model has
been calculated in $\q{\krallm,\gehlenph}$ from level crossings
in the ground state one expects an oscillatory contribution to the
correlation functions also in the massive phases. We will demonstrate
the presence of this oscillation by computing low- and high-temperature
series and numerical evaluation of the ground state. These computations
are limited to short distances. Therefore we also use a form factor
expansion for the correlation functions which leads to a better conceptual
understanding of the oscillation.
\medskip
These $\Zed_n$-spin quantum chains have conformally invariant second order
phase transitions for particular values of the parameters. They can be
described by rational models of $\Zed_n$ parafermions $\q{\fatzamA}$ with
$c > 1$ (for $n > 4$). The $\Zed_n$ parafermion algebras have the
disadvantage that they are non-local and that each algebra only describes
precisely one rational conformal field theory. However, Virasoro-minimal
models have $c < 1$ $\q{\bpz}$ and are therefore not useful for describing
these second order phase transitions. The discovery of extended conformal
algebras $\q{\zam}$ (also called $\w$-algebras -- for a review see e.g.\
$\q{\bouwschou}$) was a major simplification because these algebras are
local and --like the Virasoro algebra-- admit infinite series of minimal
models. Nowadays the second order phase transitions of $\Zed_n$-spin
chains can conveniently be described by the first unitary minimal models
of the so-called ${\cal WA}_{n-1}$-algebras
$\q{\alcaraz,\fateev,\lykyanov,\flrep}$. Similarly, one also needs $c > 1$
for applications in string theory. Very important applications require
$N=2$ supersymmetric rational conformal fields theories with $c=9$
(see e.g.\ $\q{\stringA,\stringB}$). Consequently, the classification of
all rational conformal field theories is a natural question. Since those
with $c > 1$ are obtained from $\w$-algebras one may expect that $\w$-algebras
play an important r\^ole in this classification program. Another motivation
for the study of $\w$-algebras is that $\w$-algebras have a rich mathematical
structure, their complete classification still being an open question.
\mn
With this thesis we will also try to contribute to these issues. Once again,
we will look at the aforementioned second order phase transitions from the
point of view of representations of $\w$-algebras with non-trivial boundary
conditions. We then turn to questions relevant for the classification
problem. The classical versions of $\w$-algebras are (to some extent) more
easy to handle. We will therefore use them to examine cosets and orbifolds
after addressing some basic problems of covariance with respect to the
Virasoro subalgebra of the classical $\w$-algebra. Some puzzles
(see e.g.\ $\q{\howcl}$) and also some confusion concerning the structure
of quantum $\w$-algebras have recently been explained with the concept
of `unifying $\w$-algebras' $\q{\unilet,\unicos}$. We summarize these
results and discuss their relevance for the classification problem.
\mn
Perturbations of conformal field theories (see e.g.\ $\q{\mussardo}$) describe
the off-critical scaling region in statistical mechanics. For the
$\Zed_n$-chiral Potts models two types of perturbations are relevant:
A thermal perturbation (see e.g.\ $\q{\zamPA,\zamPB,\zamPC,\mussardo}$)
that makes the models massive and a chiral perturbation
$\q{\cardy}$ that breaks parity. Both perturbations separately are integrable
but it is still an open question if the combined perturbation is also
integrable. $\w$-algebras might also be useful for addressing problems
of this type which is however beyond the scope of this thesis.
\bigskip
The outline of this thesis is as follows. Chapter {\secSsup} contains a review
of the algebraic formulation of superintegrability and indicates how
superintegrable quantum spin models can be constructed systematically.
In chapter {\secW} we summarize basic facts about the chiral Potts model
and perturbation series. In particular, in section {\secIOPD} we present
a precise duality statement. Then, the excitation spectrum of $\Zed_n$-spin
chains is examined in chapter {\secX}. After reviewing the quasiparticle
picture for the superintegrable $\Zed_3$-chiral Potts model $\q{\dkcoy}$
we start by studying the low-lying levels with high- and low-temperature
series. These results illustrate a general quasiparticle picture
which we discuss in section {\secE}. This quasiparticle picture is strictly
valid only in the infinite lattice limit but we also discuss finite-size
corrections. Since we use perturbative arguments for the derivation of
this quasiparticle picture we also have to discuss the radius of
convergence in section {\secI}.
The various aspects of the correlation functions mentioned
above are discussed in chapter {\secY}.
\sn
The final chapter {\secZ} shifts to a different aspect of low-dimensional
theoretical physics: Extended conformal algebras. In section {\secwB}
we show how automorphisms of Casimir $\w$-algebras can be used to find
extensions of rational conformal field theories by imposing twisted
boundary conditions. The relevance of these results to $\Zed_n$-quantum
spin chains is discussed in section {\secwC}. We then turn to classical
$\w$-algebras in section {\secwCl} and discuss their $su(1,1)$ structure.
For classical $\w$-algebras one can get a good handle on orbifolds and
cosets as we show in section {\secwCO}. These results are useful also
for obtaining new insights in the structure of quantum $\w$-algebras.
Some new statements about the space of quantum $\w$-algebras are
summarized in section {\secwSQ} where in particular the notion of
`unifying $\w$-algebras' is explained.
\vfill
\eject
\noindent
This thesis is (to a large extent) a summary of already published material.
More precisely, the following sections are based on the following publications:
\mn
Sections \secIOPD., \secIOPC.\ and \secIOPE.:
\item{} N.S.\ Han, A.\ Honecker,
              {\it Low-Temperature Expansions and Correlation
              Functions of the $\Zed_3$-Chiral Potts Model},
              J.\ Phys.\ A: Math.\ Gen.\ {\bf 27} (1994) p.\ 9
\sn
Section \secD.:
\item{} G.\ von Gehlen, A.\ Honecker, {\it Multi-Particle Structure
              in the $\Zed_n$-Chiral Potts Models},
              J.\ Phys.\ A: Math.\ Gen.\ {\bf 26} (1993) p.\ 1275
\sn
Sections \secD., \secE., \secI., \secG., \secFORM.\ and appendices
{\appA}, {\appB}:
\item{} A.\ Honecker, {\it A Perturbative Approach to the
              Chiral Potts Model}, preprint BONN-TH-94-21,
              hep-th/9409122
\sn
Section \secGG.:
\item{} G.\ von Gehlen, A.\ Honecker, {\it Excitation Spectrum and
              Correlation Functions of the $\Zed_3$-Chiral Potts Quantum
              Spin Chain}, Nucl.\ Phys.\ {\bf B435} (1995) p.\ 505
\sn
Sections \secwB.\ and \secwC.:
\item{} A.\ Honecker, {\it Automorphisms of ${\cal W}$-Algebras
              and Extended Rational Conformal Field Theories},
              Nucl.\ Phys.\ {\bf B400} (1993) p.\ 574
\sn
Section \secwCO.:
\item{} J.\ de Boer, A.\ Honecker, L.\ Feh\'er,
              {\it A Class of $\w$-Algebras with Infinitely Generated
              Classical Limit},
              Nucl.~Phys.~{\bf B420} (1994) p.\ 409
\sn
Section \secwSQ.:
\item{} R.\ Blumenhagen, W.\ Eholzer, A.\ Honecker, K.\ Hornfeck,
              R.\ H\"ubel, {\it Unifying $\w$-Algebras},
              Phys.\ Lett.\ {\bf B332} (1994) p.\ 51
\item{} R.\ Blumenhagen, W.\ Eholzer, A.\ Honecker,
              K.\ Hornfeck, R.\ H\"ubel,
              {\it Coset Realization of Unifying $\w$-Algebras},
              preprint BONN-TH-94-11, DFTT-25/94, hep-th/9406203,
              to appear in Int.\ Jour.\ of Mod.\ Phys.\ {\bf A}
\mn
There are some minor extensions throughout these sections. The previously
unpublished material is concentrated mainly in sections \secSsb.\ -- \secSsd.\
and section \secwCl.
\vfill
\eject
{\helvits
{\par\noindent
\leftskip=1cm
\baselineskip=11pt
I was at the mathematical school, where the master taught his pupils after
a method scarce imaginable to us in Europe. The proposition and demonstration
were fairly written on a thin wafer, with ink composed of a cephalic
tincture. This the student was to swallow upon a fasting stomach, and for three
days following eat nothing but bread and water. As the wafer digested the
tincture mounted to the brain, bearing the proposition along with it.
\par\noindent}
\par\noindent
\rightline{Jonathan Swift in ``Gulliver's Travels''}
\par\noindent}
\mn
\chaptitle{\secSsup.\ Superintegrable quantum spin models}
\mn
By now an abundance of integrable classical spin models in two space
dimensions, or equivalently quantum spin chains in one space dimension,
are known. The quantum spin chain Hamiltonians in turn can be regarded
as evolution operators of a lattice field theory in one plus one dimensions.
This indicates that it would be desirable to know also integrable
spin models in higher dimensions in order not to be restricted to
surface physics or other planar media. Unfortunately, very little is
known so far about higher dimensions despite major efforts.
\mn
A promising idea was put forward in $\q{\dogra}$ using some algebraic
criterion for a particular kind of integrability -- the so-called
`superintegrability'. To our knowledge, no results in dimensions
greater than one have been obtained with this criterion yet either.
In this section we would like to give some indications how
one can systematically construct superintegrable quantum spin Hamiltonians.
So far, we have not yet reached the goal of finding integrable models
in higher dimensions. Nevertheless we show that the problem is in principle
amenable to computer computation and we also obtain a few related
new results
\footnote{${}^{1})$}{
Note that recently a different algebraic approach to the construction of
integrable quantum spin models in one dimension has been proposed
in $\q{\grabo}$.}.
\bn
\chapsubtitle{\secSsa.\ The general setting}
\mn
\beginddef{\DefSUPa}{Nearest neighbour interaction quantum spin model}
Let $\La$ be a lattice of dimension $d$ and $\F$ be its fundamental domain.
Furthermore, consider an associative algebra of dimension $n$
with generators
$\Om^i$ including the identity $\id$. For any finite subset
$S(\La) \subset \La$ with $p$ points define operators $\Om_\xv^i$
associated to `site' $\xv \in S(\La)$ by operators
$\id \otimes \ldots \otimes \id \otimes \Om^i
   \otimes \id \otimes \ldots \otimes \id$
in the $p$-fold tensor product of the algebra with itself.
Then, a general nearest neighbour interaction quantum spin
Hamiltonian in $d$ dimensions is given by
$$H^{(\rm general)} = \sum_{\xv \in S(\La)} \sum_{i=1}^n d_i \Om_\xv^i
        + \la \sum_{\xv \in S(\La)} \sum_{\e \in \F} \sum_{i=1}^n \sum_{j=1}^n
                v^{\e}_{i,j} \Om_\xv^i \Om_{\xv+\e}^j \, .
\eqno({\rm \secSsa.1})$$
\enddef
A Hamiltonian of this type is called `{\it isotropic}' if the coupling
constants $v^{\e}_{i,j}$ do not depend on the direction $\e$ of the
interaction, i.e.\ $v^{\e}_{i,j} = v_{i,j}$ for all $\e$.
\mn
The Hamiltonian (\secSsa.1) is of the form
$$H = H_0 + \la V  \, . \eqno({\rm \secSsa.2})$$
\eject\noindent
It was shown in $\q{\dogra,\daviesA,\daviesB}$ that Hamiltonians of this
form are integrable if
$$\eqalignno{
\lb H_0, \lb H_0, \lb H_0, V \rb \rb \rb &= c_1 \lb H_0, V \rb
 \, , &({\rm \secSsa.3a}) \cr
\lb V, \lb V, \lb V, H_0 \rb \rb \rb &= c_2 \lb V, H_0 \rb
 \, . &({\rm \secSsa.3b}) \cr
}$$
We will call a Hamiltonian of type (\secSsa.2) satisfying (\secSsa.3)
`{\it superintegrable}' (note that this terminology is not used entirely
consistent in the literature -- in contrast to us, some authors include
the generalized Yang-Baxter relations in the notion of superintegrability).
\mn
Clearly, one can modify $c_1$ and $c_2$ by rescaling the individual terms
$H_0$ and $V$ of the Hamiltonian. This is usually exploited in order to
set $c_1 = c_2 = 16$. However, we will not impose this normalization
condition but will instead normalize $H_0$ and $V$ suitably. One should
keep in mind that if $H_0 + \la V$ is superintegrable, so is $(h H_0) +
\la (v V)$ for arbitrary $h$, $v$.
\mn
The condition (\secSsa.3) is trivially satisfied if $\lb H_0, V \rb = 0$.
In this case, one can simultaneously diagonalize $H_0$ and $V$ and
in this manner diagonalize $H$ for general $\la$. Clearly, for
$\lb H_0, V \rb = 0$ all eigenvalues of $H$ are linear in $\la$.
Therefore, if $\la$ should play the r\^ole of temperature, the
Hamiltonian $H = H_0 + \la V$ gives rise to trivial thermodynamics.
Thus, the non-trivial physics must be given by the potential, usually
for $\la \to \infty$, i.e.\ the physically relevant Hamiltonian
is $H = V + \hat{\la} H_0$ with small $\hat{\la}$. In this case
the diagonalization of $V$ may still be a difficult task although
the conservation of $H_0$ can be very helpful.
\mn
For $\lb H_0, V \rb \ne 0$ one can recursively generate Onsager's algebra
$\q{\onsager}$ from $A_0 = H_0$ and $A_1 = V$
(see $\q{\dogra,\daviesA,\daviesB}$):
$$
\lb A_m, A_n \rb = 4 G_{m-n} \, , \qquad
\lb G_m, A_n \rb = 2 A_{n+m} - 2 A_{n-m} \, , \qquad
\lb G_m, G_n \rb = 0
\eqno({\rm \secSsa.4})$$
if (\secSsa.3) is satisfied with $c_1 = c_2 = 16$. From (\secSsa.4) one
obtains the following infinite set of conserved commuting charges:
$$- Q_m = {1 \over 2} \left( A_m + A_{-m} + \la A_{m+1} + \la A_{-m+1} \right)
  \, .
\eqno({\rm \secSsa.5})$$
Note that $G_{-1} = -G_{1} = {1 \over 4} \lb H_0, V \rb$. Thus, if
$\lb H_0, V \rb=0$ the recurrence relations (\secSsa.4) can be trivially
solved leading to $A_m = A_n$ if $m \equiv n$ mod $2$. In this case, one
has $Q_m = Q_n$ for $m \equiv n$ mod $2$, i.e.\ the Hamiltonian is
not necessarily integrable in the sense that it has an infinite set
of conserved commuting charges. This motivates the following
definitions:
\mn
\begindef{\DefSUPb}
A Hamiltonian of the form (\secSsa.2) is called `{\it quasi
superintegrable}' if it satisfies $\lb H_0, V \rb = 0$.
\enddef
\par\noindent
\begindef{\DefSUPc} A Hamiltonian satisfying (\secSsa.3) with
$\lb H_0, V \rb \ne 0$ is called `{\it strictly superintegrable}'.
\enddef
\bn
\chapsubtitle{\secSsb.\ First considerations}
\mn
We will be interested in associative algebras that have finite dimensional
(complex) defining representations. These algebras can be considered
as subalgebras of $gl(n,\Complex)$ for some $n$. In the Hamiltonian
(\secSsa.1) we may restrict to operators $\Omega^i \in sl(n,\Complex)$
for the following reason: The complement of $sl(n,\Complex)$ in
$gl(n,\Complex)$ is spanned by the identity $\id$. If the identity
$\id$ should appear in the potential term $V$, this part of the potential
could be absorbed in the free part $H_0$. Now, if the identity would
appear in the free part $H_0$ it would shift the ground state energy
per site $e_0$ by a constant amount and not change the excitation
spectrum. Such a modification may be neglected because of its physical
irrelevance. Thus, we assume $\Omega^i \in sl(n,\Complex)$ in (\secSsa.1).
\mn
Next we note that the conditions (\secSsa.3) are local ones. We may therefore
choose a suitable domain $S(\La)$ and suitable boundary conditions.
Furthermore note that the eigenvalues of the Hamiltonian (\secSsa.1) will
not depend on the details of the lattice $\La$ but only on the way how
neighbours are linked. We will therefore restrict to cubic lattices
in $d$ dimensions. This leads us to the following specialization of the
Hamiltonian (\secSsa.1):
$$H^{(\rm general')} = \sum_{\xv \in \zed^d \atop 1 \le x_i \le N_i}
                       \sum_{i=1}^{n^2-1} d_i \Om_\xv^i
                + \la  \sum_{\xv \in \zed^d \atop 1 \le x_i \le N_i}
                       \sum_{i=1}^d
                       \sum_{a=1}^{n^2-1} \sum_{b=1}^{n^2-1}
                       v^i_{a,b} \Om^a_\xv \Om^b_{\xv+\ei}
\eqno({\rm \secSsb.1})$$
where $n$ is now the dimension of the defining representation of
$sl(n,\Complex)$. On (\secSsb.1) we will impose periodic boundary
conditions, i.e.\ the $x_i$ will be identified if they are equal mod $N_i$.
\mn
Finally we make some observations that will be useful for treating the
problem on the computer. For a Hamiltonian of type (\secSsa.1) the
conditions (\secSsa.3) will involve at most four-fold products of generators
$\Om^i_\xv$ sitting on four neighbouring sites. Due to the locality
of the conditions (\secSsa.3) this means that we may choose all $N_i = 4$
when checking them for a general ansatz of the form (\secSsb.1).
\mn
Note that the condition $\lb H_0 , V \rb = 0$ does not involve the
dimensionality of the lattice. Therefore, for any solution $\{d_i, v_{a,b}\}$
with $\lb H_0 , V \rb = 0$ in $d=1$ we immediately obtain the corresponding
general solution in arbitrary dimensions $d$ by fixing the parameters
$\{d_i, v_{a,b}^j\}$ such that for fixed $j$ they satisfy the equations
for $d=1$ independently.
\mn
After these first general considerations the problem of solving (\secSsa.3)
is now straightforward although not necessarily very easy.
\bn
\chapsubtitle{\secSsc.\ The quasi superintegrable case}
\mn
As we have pointed out before we are mainly interested
in strictly superintegrable spin chain Hamiltonians. However, in order
to single them out among all solutions to (\secSsa.3) we have to find the
quasi superintegrable ones first.
\mn
Fortunately, this problem is comparably simple. In order to evaluate
$\lb H_0, V\rb$ for a Hamiltonian of the form (\secSsb.1) we do in fact
not need that the $\Om^a$ form an associative algebra, we need nothing
else than the Lie bracket. Introducing structure constants by
$$\lb \Om^a, \Om^b \rb = \sum_c f^{a b}_c \Om^c
\eqno({\rm \secSsc.1})$$
one obtains
$$\lb H_0, V \rb =
\sum_{\xv \in \zed^d \atop 1 \le x_i \le N_i} \sum_{i=1}^d
                       \sum_{a,b,c,d=1}^{n^2-1}
                       d_a \left\{
                       v^i_{b,c} f^{a b}_d + v^i_{d,b} f^{a b}_c \right\}
                       \Om^d_\xv \Om^c_{\xv+\ei} .
\eqno({\rm \secSsc.2})$$
Clearly, $H_0 + \la V$ will be quasi superintegrable iff
$$\sum_{a,b=1}^{n^2-1} d_a \left\{
                       v^i_{b,c} f^{a b}_d + v^i_{d,b} f^{a b}_c \right\}
      = 0 \qquad \forall c, \, d, \, i \, .
\eqno({\rm \secSsc.3})$$
The condition (\secSsc.3) is very reminiscent of an invariance condition
with respect to the generators of the Lie algebra for which $d_a \ne 0$.
Thus, it will not be surprising if the quasi superintegrable spin chain
Hamiltonians turn out to have extra continuous symmetries.
\bn
\chapsubtitle{\secSsd.\ Results}
\mn
In this section we will present some new results for quantum spin
models based on $sl(2,\Complex)$ and afterwards recall some facts
about $\Zed_n$ spin chains. After complexification $sl(2,\Complex)$ is
the same as $su(2)$. We can therefore use the standard Pauli spin matrices
$$\siz = \pmatrix{1 & 0 \cr
                0 & -1 \cr} \, , \qquad
\six = \pmatrix{0 & 1 \cr
                1 & 0 \cr} \, , \qquad
\siy = \pmatrix{0 & -i \cr
                i & 0 \cr} \, .
\eqno({\rm \secSsd.1})$$
In fact, we will not need the matrix realization given by (\secSsd.1).
Instead we will look at the associative algebra generated by $\id$,
$\siz$, $\six$, $\siy$, e.g.\ $(\siz)^2 = (\six)^2 = (\siy)^2 = \id$.
\mn
In the basis (\secSsd.1) the isotropic Ising quantum spin model on a cubic
lattice in $d$ dimensions which is one special case of (\secSsb.1) is given by:
$$H^{(\rm Ising)} = - \sum_{\xv \in \zed^d \atop 1 \le x_i \le N_i} \siz_\xv
       - \la \sum_{\xv \in \zed^d \atop 1 \le x_i \le N_i} \sum_{i=1}^d
          \six_\xv \six_{\xv+\ei} \, .
\eqno({\rm \secSsd.2})$$
Note that superintegrability is known since Onsager's days for
$H^{(\rm Ising)}$ only in $d=1$. For $d=2$ we will show
below that $H^{(\rm Ising)}$ is not superintegrable.
\mn
Let us now turn to the most general ansatz for a nearest neighbour
interaction Hamiltonian based on $sl(2,\Complex)$ in $d$ dimensions.
In addition to the previous comments we recall that $SO(3)$ acts on
$su(2)$ by inner conjugations. This can be exploited in order to fix
the basis in $sl(2,\Complex)$ and the normalization of $H_0$ such that
$H_0 = -\sum_{\xv} \siz_\xv$. Thus, the general spin-1/2 Hamiltonian
becomes
$$H = - \sum_{\xv \in \zed^d \atop 1 \le x_i \le N_i} \siz_\xv
      - \la \sum_{\xv \in \zed^d \atop 1 \le x_i \le N_i} \sum_{i=1}^d
          \sum_{a=0}^2 \sum_{b=0}^2
          v^i_{a,b} \Om^a_\xv \Om^b_{\xv+\ei}
\eqno({\rm \secSsd.3})$$
with $\Om^0 = \siz$, $\Om^1 = \six$ and $\Om^2 = \siy$. Note that there
may be some residual freedom of choice of basis corresponding to rotations
around the $z$-axis.
\mn
First we consider one-dimensional models.
In $d=1$ the general ansatz (\secSsd.3) becomes
$$H = - \sum_{x=1}^N \siz_x - \la \sum_{x=1}^N \sum_{a=0}^2 \sum_{b=0}^2
                    v_{a,b} \Om_x^a \Om_{x+1}^b
\eqno({\rm \secSsd.4})$$
With this ansatz one can show by a tedious but straightforward calculation
that there are only two different inequivalent families of solutions to
eq.\ (\secSsa.3) given by theorems {\ThSUPa} and {\ThSUPb}.
\mn
\begintheorem{\ThSUPa}
The most general quasi superintegrable Hamiltonian of type (\secSsd.4)
in $d=1$ is given by:
$$H^{(1)} = - \sum_{x=1}^N \siz_x - \la \sum_{x=1}^N \left\{
                    \six_x \six_{x+1} + \siy_x \siy_{x+1}
           + \hat{v} \left(\six_x \siy_{x+1} - \siy_x \six_{x+1}\right)
           + \tilde{v} \siz_x \siz_{x+1} \right\} \, .
\eqno({\rm \secSsd.5})$$
\endTheorem
\beginremarks{}
\item{1)}
The Hamiltonian (\secSsd.5) is invariant with respect to rotations in the
two-di\-men\-sional space spanned by $\six$, $\siy$. For
$\hat{v} = \tilde{v} = 0$
this solution contains the Hamiltonian generated by $\q{\ahn}$
with (in the notations of $\q{\ahn}$) $g_1 = g_{-1}$. Note that the
Hamiltonian eq.\ (11) of $\q{\ahn}$ is {\it not} superintegrable for
general $g_1 \ne g_{-1}$.
\item{2)} In $\q{\lacaze}$ a spin-1 Hamiltonian of the type
$H = \sum_x \vec{S}_x \cdot \vec{S}_{x+1}$ was studied numerically for up to
22 sites. This Hamiltonian is a special case of (\secSsd.5) and demonstrates
the
usefulness of conservation of $H_0$ for numerical computations.
\endremark
\mn
The second solution to (\secSsa.3) is strictly superintegrable.
For the second solution in $d=1$ we have some freedom of rotation between
$\six$ and $\siy$. After fixing this freedom we arrive at
\mn
\begintheorem{\ThSUPb}
The most general strictly superintegrable Hamiltonian of type (\secSsd.4)
in $d=1$ is given by:
$$H^{(2)} = - \sum_{x=1}^N \siz_x
            - \la \sum_{x=1}^N \left\{ \six_x \six_{x+1}
                           + v \left(\six_x \siy_{x+1}
                           -         \siy_x \six_{x+1} \right) \right\} \, .
\eqno({\rm \secSsd.6})$$
\endTheorem
\beginremark{}
The solution (\secSsd.6) contains the Ising model (\secSsd.2) in $d=1$ for
$v = 0$.
\endremark
\mn
The conditions for a general Hamiltonian (\secSsd.3) to be superintegrable
are already very complicated for spin 1/2 in $d=1$. Therefore, it does
not make sense to start with a general ansatz in $d > 1$ but one
has to restrict oneself to special cases.
\mn
A trivial $d$-dimensional strictly superintegrable Hamiltonian can be
obtained from (\secSsd.6) by setting all $v^i_{a,b} = 0$ for $i\ne k$ with
some fixed $k$. For $v^k_{a,b}$ one then simply takes the solution (\secSsd.6).
However, this solution is trivial since it consists of decoupled one
dimensional models. Therefore, we will not consider solutions of this type
any more.
\mn
First, we note that the Hamiltonian $H^{(\rm Ising)}$ given by
eq.\ (\secSsd.2) is
{\it not} superintegrable in $d=2$. Even more strongly, one can show
that
\mn
\begintheorem{\ThSUPc}
There is no strictly superintegrable $\Zed_2$ charge conserving
Hamiltonian of type (\secSsd.3) in $d=2$. Here $\Zed_2$ charge conservation
means that $v^i_{0,1} = v^i_{0,2} = v^i_{1,0} = v^i_{2,0} = 0$ for $i=1$, $2$.
\endtheorem
\mn
However, from the preceding discussions we obtain the
\mn
\begincorollary{}
The Hamiltonian
$$\eqalign{
H^{(1,d)} = - \sum_{\xv \in \zed^d \atop 1 \le x_i \le N_i} \siz_\xv
 - \la \sum_{\xv \in \zed^d \atop 1 \le x_i \le N_i} \sum_{i=1}^d
   & \left\{
  v_i \left( \six_\xv \six_{\xv+\ei} + \siy_\xv \siy_{\xv+\ei} \right)
  \right. \cr
 + & \left.
   \hat{v}_i \left( \six_\xv \siy_{\xv+\ei} - \siy_\xv \six_{\xv+\ei} \right)
 + \tilde{v}_i \siz_\xv \siz_{\xv+\ei} \right\} \cr
}  \eqno({\rm \secSsd.7})$$
is quasi superintegrable for general $d$.
\endcorollary
\mn
Current knowledge about strictly superintegrable spin quantum chains based
on $sl(n)$ with $n > 2$ is restricted to the result of von Gehlen and
Rittenberg who constructed the so-called `{\it superintegrable
$\Zed_n$-chiral Potts model}' $\q{\gehri}$. More precisely, one has
\mn
\beginttheorem{\ThSUPX}{\q{\gehri}}
The $\Zed_n$-chiral Potts spin quantum chain defined by
$$H^{(n)}_N = - \sum_{j=1}^N \left(2 \HH_j
               + \la \sum_{k=1}^{n-1}
                  \left(1 - i \cot{\pi k \over n}\right)
                   \Ga_j^k \Ga_{j+1}^{n-k} \right)
             \eqno{(\rm \secSsd.8)}$$
is strictly superintegrable.
$\HH$ and $\Ga$ are operators in the fundamental representation of
$sl(n)$. $\HH$ is the Cartan generator of the principal $sl(2)$, i.e.\
$\HH = {\rm diag}\left({n-1 \over 2}, \ldots, -{n-1 \over 2}\right)$.
Furthermore, one has
$\Ga = \left(\sum_\alpha E_\alpha \right) + E_{-\gamma}$
where the sum runs over the $n-1$ simple positive roots and $\gamma$ is
the highest root.
$E_\beta$ is the root vector associated to a root $\beta$ in a
normalization such that $E_\alpha$ for a simple root $\alpha$ has a single
non-zero entry equal to one on the first upper off-diagonal.
\endtheorem
\mn
Ahn and Shigemoto $\q{\ahn}$ realized that one can add terms of type
${1 \over 2} g_l \left(A_{m+l} + A_{-m+l}\right)$ with arbitrary parameters
$g_l$ to the conserved charges $Q_m$ in (\secSsa.5),
still ensuring the commutativity of these charges.
Furthermore, $A_{-1} = V - {1 \over 2} [H_0, V]$ is obviously still
a nearest neighbour interaction for Hamiltonians of type (\secSsa.1).
Thus, adding ${1 \over 2} \hat{\la} \left(A_{m-1} + A_{-m-1}\right)$ to
(\secSsa.5) they arrived at the following corollary to theorem {\ThSUPX}:
\mn
\begincorollary{} (\q{\ahn})
The $\Zed_n$-spin quantum chain defined by
$$H^{(n)}_N = - \sum_{j=1}^N \left[ 2 \HH_j
               + \sum_{k=1}^{n-1}
               \left(1 - i \cot{\pi k \over n}\right) \left(
                 \la \Ga_j^k \Ga_{j+1}^{n-k}
                     + \hat{\la} \Xi_j^k \Xi_{j+1}^{n-k} \right) \right]
             \eqno{(\rm \secSsd.9)}$$
is integrable for general $\la$, $\hat{\la}$, i.e.\ it gives rise to an
infinite set of commuting charges. The additional operator $\Xi$ is defined by
$\Xi = \left(\sum_\alpha E_\alpha \right) - E_{-\gamma}$.
\endcorollary
\mn
\beginremark{}
The Hamiltonian (\secSsd.9) is neither quasi nor strictly superintegrable
for general values of the parameters.
\endremark
\vfill
\eject
{\helvits
{\par\noindent
\leftskip=1cm
\baselineskip=11pt
Einstweilen wissen wir noch gar nicht, in welcher Sprache wir \"uber das
Geschehen im Atom reden k\"onnen. Wir haben zwar eine mathematische
Sprache, das hei{\ss}t ein mathematisches Schema, mit Hilfe dessen wir
die station\"aren Zust\"ande des Atoms oder \"Ubergangswahrscheinlichkeiten
von einem Zustand zu einem anderen ausrechnen k\"onnen. Aber wir wissen
noch nicht -- wenigstens noch nicht allgemein -- wie diese Sprache mit
der gew\"ohnlichen Sprache zusammenh\"angt.
\par\noindent}
\sn
\rightline{Werner Heisenberg in ``Der Teil und das Ganze'',
           Kapitel zu 1925-1926}
\par\noindent}
\bn
\chaptitle{\secW.\ The chiral Potts quantum chain}
\mn
This chapter and chapters {\secX} and {\secY} of this thesis focus on the
`$\Zed_n$-chiral Potts quantum chain' which is a generalization of
(\secSsd.8). From here on we will rely on methods that do not make use
of any particular integrability properties.
\bn
\chapsubtitle{\secB.\ Preliminaries}
\mn
This section summarizes well-known basic facts about $\Zed_n$-spin
quantum chains. We also introduce some notions that will be useful
later on. For more details see e.g.\ the review $\q{\chrihen}$.
\mn
In the following we will study the $\Zed_n$-spin quantum chain with $N$
sites which is defined by the Hamiltonian:
$$H^{(n)}_N = - \sum_{j=1}^N \sum_{k=1}^{n-1} \left\{ \ab_k \si_j^k
                 + \la \a_k \Ga_j^k \Ga_{j+1}^{n-k} \right\} \, .
             \eqno{(\rm \secB.1)}$$
For the low-temperature phase (large $\la$) it is more convenient to
use a different normalization than (\secB.1):
$$\tilde{H}^{(n)}_N = \tilde{\la} H^{(n)}_N
            = - \sum_{j=1}^N \sum_{k=1}^{n-1} \left\{ \tilde{\la} \ab_k \si_j^k
              + \a_k \Ga_j^k \Ga_{j+1}^{n-k} \right\}
\eqno({\rm \secB.2})$$
with $\tilde{\la} = \la^{-1}$.
\mn
$\si_j$ and $\Ga_j$ freely generate a finite dimensional
associative algebra with involution by the following relations
($1 \le j,l \le N$):
$$\eqalign{
\si_j \si_l &= \si_l \si_j \ , \qquad
\Ga_j \Ga_l = \Ga_l \Ga_j \ , \qquad
\si_j \Ga_l = \Ga_l \si_j \om^{\delta_{j,l}}_{} \ , \qquad
\si_j^n = \Ga_j^n = \id \ , \cr
\si_j^{+} &= \si_{n}^{n-1} \ , \qquad
\Ga_j^{+} = \Ga_{j}^{n-1} \cr
} \eqno{(\rm \secB.3)}$$
where $\om$ is the primitive $n$th root of unity $\om = e^{2 \pi i \over
n}_{}$.
\sn
For the Hamiltonian (\secB.1) we also have to specify boundary conditions,
i.e.\ we have to define $\Ga_{N+1}$. The following choices will be considered
(in the terminology we follow Cardy $\q{\cardyB}$):
\item{1)} {\it Free} boundary conditions
$$\Ga_{N+1} = 0 \, . \eqno{(\rm \secB.4a)}$$
\item{2)} {\it Toroidal} boundary conditions, i.e.\ we identify the $N+1$st
          site with the 1st site.
\itemitem{a)} {\it Cyclic} boundary conditions
$$\Ga_{N+1} = \om^{-R} \Ga_1 \, .  \eqno{(\rm \secB.4b)}$$
\itemitem{b)} {\it Periodic} boundary conditions which are the $R = 0$
              special case of cyclic boundary conditions
$$\Ga_{N+1} = \Ga_1 \, .  \eqno{(\rm \secB.4c)}$$
\itemitem{c)} {\it Twisted} boundary conditions
$$\Ga_{N+1} = \om^{-R} \Ga_1^{+} \, .  \eqno{(\rm \secB.4d)}$$
\par\noindent
We will mainly focus on periodic boundary conditions.
\mn
The Hamiltonian (\secB.1) contains $2n-1$ parameters: The temperature-like
parameter $\la$ that we choose to be real and the complex
constants $\ab_k$ and $\a_k$. $H^{(n)}_N$ is hermitean iff
$\ab_k = \ab_{n-k}^{*}$ and $\a_k = \a_{n-k}^{*}$.
\medskip
The algebra (\secB.3) is conveniently represented in
$$\HN := \underbrace{
\Complex^n \otimes \Complex^n \otimes \ldots \otimes \Complex^n}_{
N \phantom{l} {\rm times}}              \eqno{(\rm \secB.5)}$$
labeling the standard basis of $\Complex^n$ by $\{e_0, \ldots, e_{n-1}\}$.
Then a basis for (\secB.5) is given by:
$$\state{i_1 \ldots i_N} := e_{i_1} \otimes \ldots \otimes e_{i_N}
\ , \qquad 0 \le i_j \le n-1.   \eqno{(\rm \secB.6)}$$
Now the following operation in the space (\secB.5) is a faithful
irreducible representation $r$ of the algebra (\secB.3):
$$\eqalign{
r(\si_j) \state{i_1 \ldots i_j \ldots i_N}
            &= \om^{i_j} \state{i_1 \ldots i_j \ldots i_N} \, , \cr
r(\Ga_j) \state{i_1 \ldots i_j \ldots i_N}
            &= \state{i_1 \ldots (i_j+1 \mod n) \ldots i_N} \, . \cr
}         \eqno{(\rm \secB.7)}$$
The involution is the adjoint operation with
respect to the standard scalar product in the tensor product of $\Complex^n$.
For low-temperature expansions of (\secB.2)
it is more convenient to consider a representation $\rt$ different from $r$:
$$\eqalign{
\rt(\Ga_j) \state{i_1 \ldots i_j \ldots i_N}
            &= \om^{i_j} \state{i_1 \ldots i_j \ldots i_N} \cr
\rt(\si_j) \state{i_1 \ldots i_j \ldots i_N}
            &= \state{i_1 \ldots (i_j-1 \mod n) \ldots i_N}. \cr
}         \eqno{(\rm \secB.8)}$$
\indent
The Hamiltonian (\secB.1) commutes with the $\Zed_n$ charge operator
$\hat{Q} := \prod_{j=1}^N \si_j$ acting on the vectors (\secB.6) as
$$r(\hat{Q}) \state{i_1 \dots i_N} = \om^{\left(\sum_{j=1}^N i_j\right)}_{}
     \state{i_1 \dots i_N}     \eqno{(\rm \secB.9)}$$
which shows that the eigenvalues of $\hat{Q}$
have the form $\om^Q$ with $Q$ integer. Thus, $H^{(n)}_N$ has $n$ charge
sectors which we shall refer to by $Q=0,$ $\ldots,$ $n-1$.
\medskip
$H^{(n)}_N$ also commutes with the translation operator $T_N$ that
acts on the basis vectors (\secB.6) in the following way:
$$r(T_N) \state{i_1 i_2 \dots i_{N}} = \state{i_2  \ldots i_N i_1}.
     \eqno{(\rm \secB.10)}$$
The eigenvalues of $T_N$ are $N$th roots of unity. We label them by
$e^{i P}$ and call $P$ the `momentum'. We choose $0 \le P < 2 \pi$
corresponding to the first Brillouin zone
and have $P \in \{0, {2 \pi \over N}, \ldots, {2 \pi (N-1) \over N} \}$.
Note that the states
$$\eqalign{
\pstate{i_1 i_2 \ldots i_{N-1} i_{N}}_P^{} :=
  {1 \over \sqrt{{\cal N}}} {\bigg (} &\state{i_1 i_2 \ldots i_{N-1} i_{N}}
       + e^{i P} \state{i_N i_1 i_2 \ldots i_{N-1}}
       + \ldots \cr
      &+ e^{i P (N-1)} \state{i_2 \ldots i_{N-1} i_N i_1} {\bigg )} \cr
}\eqno{(\rm \secB.11)}$$
are eigenstates of $T_N$ with eigenvalue $e^{i P}$.
${\cal N}$ is a suitable normalization constant.
If the state $\state{i_1 \ldots i_{N} }$ has no symmetry
(i.e.\ $T_N^k \state{i_1 \ldots i_{N} }
\ne \state{i_1 \ldots i_{N} }$ for all $0<k<N$), one has ${\cal N} = N$.
This will apply to most cases below where we need (\secB.11).
For a definition of the momentum eigenstates (\secB.11) in case
of cyclic boundary conditions $R \ne 0$ see e.g.\ $\q{\gehritell,\chrihen}$
\bigskip
We will frequently use the following parameterization of the constants
$\a_k$ and $\ab_k$, fixing their dependence on $k$:
$$\a_k = {e^{i \phi ({2 k \over n}-1)} \over \sin{\pi k \over n} } \ , \qquad
\ab_k = {e^{i \vphi ({2 k \over n}-1)} \over \sin{\pi k \over n} }.
    \eqno{(\rm \secB.12)}$$
This is a suitable choice because it includes a large class of
interesting models.
\mn
For $\phi = \vphi = 0$ one obtains real
$\a_k = \ab_k = {1 \over \sin{\pi k \over n}}$. This leads to
models with a second order phase transition at $\lambda=1$ which
can be described by a parafermionic conformal field theory in the limit
$N \to \infty$ at criticality $\q{\fatzamA,\alcaraz}$.
These so-called Fateev-Zamolodchikov-models $\q{\fateev}$
lead to extended conformal algebras ${\cal WA}_{n-1}$ where the simple
fields have conformal dimension $2, \ldots, n$ $\q{\lykyanov}$.
The spectrum of
the Hamiltonian (\secB.1) can be described by the first unitary minimal
model of the algebra ${\cal WA}_{n-1}$. For $n=3$ the symmetry algebra
is Zamolodchikov's well-known spin-three extended conformal algebra
$\q{\zam}$ at $c={4 \over 5}$.
\mn
Choosing $\phi = \vphi = {\pi \over 2}$ in (\secB.12) for the Hamiltonian
(\secB.1) one recovers the superintegrable $\Zed_n$-chiral Potts model
(\secSsd.8).
\mn
The parameterization (\secB.12) also includes the family of integrable models
discovered in refs.\
$\q{\mccoyadv,\albertiniA,\albertiniB,\baxterA,\yang,\perkadv}$
which interpolates between the integrable cases at $\phi = \vphi = 0$,
$\la = 1$ and $\phi = \vphi = {\pi \over 2}$. The Hamiltonian
(\secB.1) is integrable if one imposes the additional constraint
$$\cos \vphi = \la \cos \phi
    \eqno{(\rm \secB.13)}$$
(or equivalently $\tilde{\la} \cos \vphi = \cos \phi$ for the low-temperature
phase) on the parameterization (\secB.12).
For $\phi = \vphi = 0$ this yields $\la=1$ -- the conformally invariant
critical points. At $\phi = \vphi = 0$, the Hamiltonian
is self-dual, i.e.\ it is invariant under a duality-transformation such
that $H^{(n)}_N(\la) \cong \la H^{(n)}_N(\la^{-1})$. The Hamiltonian is also
self-dual on the superintegrable line $\phi=\vphi={\pi \over 2}$.
$H^{(n)}_N$ with the choices (\secB.12), (\secB.13) is in general
not self-dual any more
whereas particular choices yield a self-dual Hamiltonian. If we choose
for (\secB.12) $\phi=\vphi$ and neglect (\secB.13) $H^{(n)}_N$ will be
self-dual again. Therefore we choose to refer to (\secB.1) with (\secB.12) as
the general `chiral Potts model'. We will not consider the integrable case
where the additional constraint (\secB.13) is satisfied in detail.
\mn
\centerline{\psfig{figure=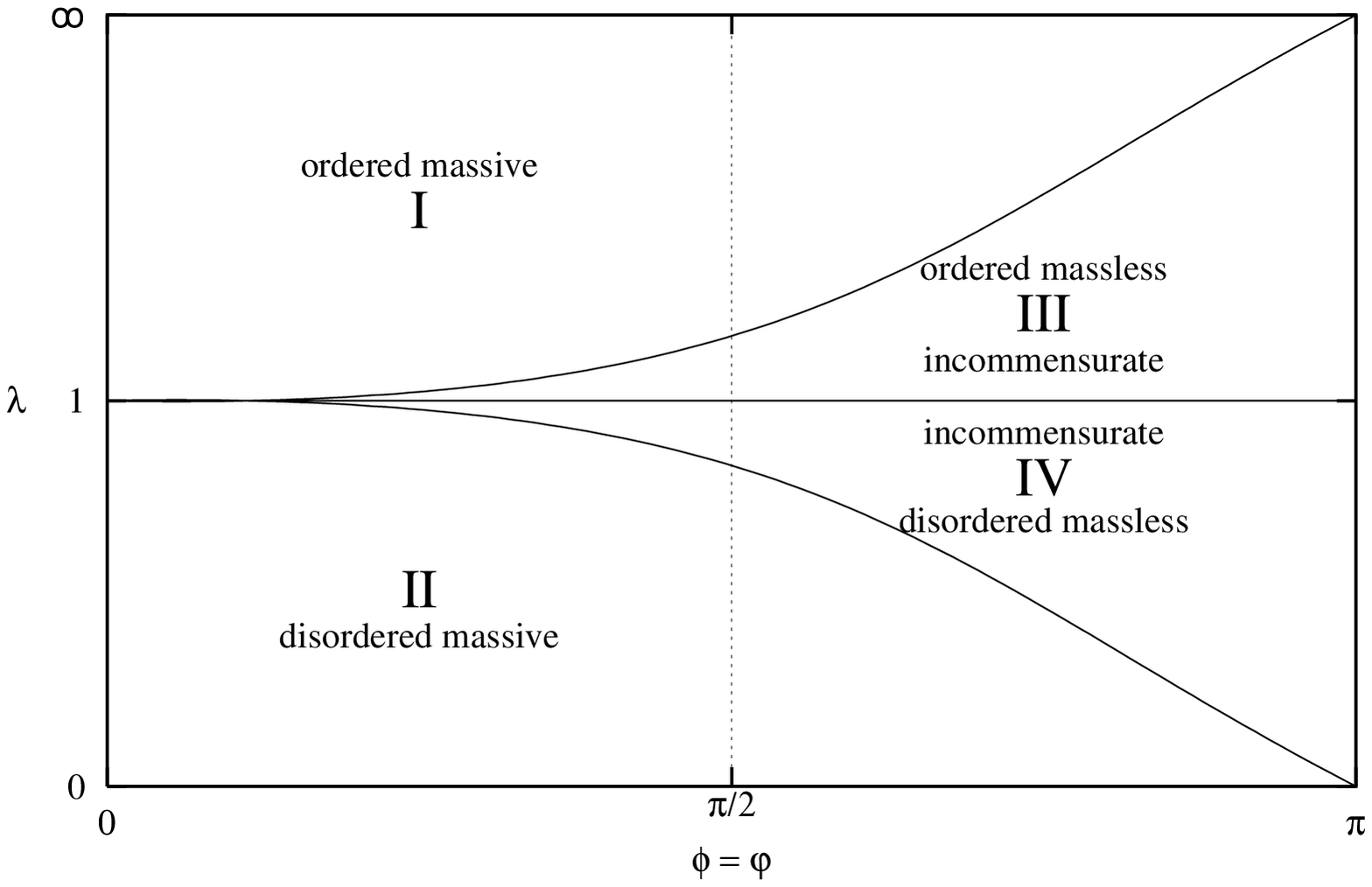}}
{\par\noindent\figindents
{\bf Fig.\ 1:}
Schematic phase diagram of the self-dual chiral Potts model for $n=3$.
\par\noindent}
\mn
The main features of the phase diagram of the $\Zed_n$-chiral Potts chain
are already present for $n=3$.
The $\Zed_3$ version of (\secB.1) is known to have four phases
$\q{\mccoyadv,\albertiniB,\gehlenph}$: Two massive and two
massless phases (see Fig.\ 1). One of the massive phases is ordered and the
other massive phase is disordered. In the following we will mainly be
interested in the massive phases. At $\phi = \vphi = {\pi \over 2}$
the massive high-temperature phase (small $\la$) appears in the range $0 \le
\la
\le 0.901292\ldots$ and the massive low-temperature phase appears
in the range $1/0.901292\ldots \le \la \le \infty$ $\q{\mccoyadv,\albertiniB}$.
There is a second order phase transition at the point $\la = 1$ for
$\phi = \vphi = 0$ and general $\Zed_n$. This conformally invariant point
is described by the models of Fateev and Zamolodchikov with
${\cal WA}_{n-1}$-symmetry $\q{\fatzamA,\fateev,\alcaraz,\lykyanov}$.
The conformal field theories can be perturbed by a thermal operator
(see e.g.\ $\q{\fatzamB,\zamPA,\zamPB,\zamPC,\mussardo}$). This perturbation is
integrable and corresponds to going to $\la \ne 1$ on the line
$\phi = \vphi = 0$ of the $\Zed_n$-quantum spin chain.
One can also introduce an integrable chiral perturbation of the conformal
field theories $\q{\cardy}$. The $\Zed_n$-spin quantum chains describe
both this chiral perturbation as well as the integrable thermal perturbations.
\bigskip
Our main interest is the spectrum in the limit $N \to \infty$ of $H_N^{(n)}$.
Of course, we have to specify how the limit is to be taken. One possible
definition was given in $\q{\perturb}$. Here, we want to omit the technical
details. We just mention that the basic idea is first to identify momentum
eigenstates for chains of different lengths and then to consider the
weak limit of $T_N$ and the excitation spectrum operator
$$\Delta H_N^{(n)} := H_N^{(n)} - E_N^0 \id
\eqno({\rm \secB.14})$$
where $E_N^0$ is the groundstate energy for $N$ sites.
\bn
\chapsubtitle{\secIOPD.\ Duality of spectra}
\sn
There is a first statement about the spectra of the Hamiltonian (\secB.1)
which allows us to restrict to the high-temperature phase $\la < 1$ for
the discussion of the spectrum: The spectra in the high-temperature phase
at $\la$ are {\it dual} to those in the low-temperature phase
at $\la=\tilde{\la}$ if we interchange $\a_k$ and $\ab_k$.
This statement for $\Zed_n$-quantum spin chains has been known
for a long time $\q{\elitzur,\horn}$ and was also used
in $\q{\hkn}$. However, special attention has to be
paid to the boundary conditions when performing duality transformations.
It has been observed in $\q{\gehri}$ that the duality transformation
interchanges the r\^ole of the charge $Q$ and boundary conditions $R$.
The precise statement is given by the following
\mn
\beginttheorem{{\ThA}}{Duality}
Denote the Hamiltonian (\secB.1) by
$H_N^{(n)}(\la,R^{ht},\ab_k^{ht},\a_k^{ht})$ and (\secB.2)
by $\tilde{H}_N^{(n)}(\tilde{\la},R^{lt},\ab_k^{lt},\a_k^{lt})$,
including in both cases explicitly the corresponding parameters.
Furthermore, abbreviate the space with charge $Q^{ht}$ in the
high-temperature phase by $\H^{Q^{ht}}$ and the eigenspace of $\rt(\hat{Q})$
to eigenvalue $\om^{Q^{lt}}$ by $\tilde{\H}^{Q^{lt}}$. Then
$H_N^{(n)}(\la,R^{ht},\ab_k^{ht},\a_k^{ht})$ restricted
to $\H^{Q^{ht}}$ and
$\tilde{H}_N^{(n)}(\tilde{\la},R^{lt},\ab_k^{lt},\a_k^{lt})$ restricted to
$\tilde{\H}^{Q^{lt}}$ have the same spectra if
$$\eqalign{
Q^{lt} &= R^{ht} \ , \qquad  R^{lt} = Q^{ht} \ , \cr
\ab_k^{lt} &= \a_k^{ht} \ , \qquad  \a_k^{lt} = \ab_k^{ht} \ , \qquad
\tilde{\la} =  \la. \cr
}  \eqno{(\rm \secIOPD.1)}$$
\endTheorem
\mn
\beginproof{}
We present a simple non-standard proof of theorem {\ThA} $\q{\han}$ that
is different from the approach of e.g.\ $\q{\elitzur,\horn}$.
We derive duality by comparing the representation $r$ (\secB.7) to
the representation $\rt$ (\secB.8). Working with the states has the
advantage that it is straightforward to take care of the boundary conditions.
Note that $\rt(\Ga_j) = r(\si_j)$ and $\rt(\si_j) = r(\Ga_j^{+})$ and
that the representations $\rt$ and $r$ are unitarily equivalent.
Now fix the state $\GS{Q^{lt}}$ to be the ground state
in $\tilde{\H}^{Q^{lt}}$:
$$\GS{Q} := {1 \over \sqrt{n}} \sum_{l=0}^{n-1} \om^{l \cdot Q}
  \state{l \ldots l} \, .    \eqno{(\rm \secIOPD.2)}$$
Then the following states are a basis for $\tilde{\H}^{Q^{lt}}$:
$$\state{Q; i_2 \ldots i_N} := \rt(\si_2^{-i_2}) \ldots
 \rt(\si_N^{-i_N}) \GS{Q^{lt}}.              \eqno{(\rm \secIOPD.3)}$$
Note that this implies:
$$\rt(\si_1) \state{Q; i_2 \ldots i_N} =
\om^{Q^{lt}} \state{Q; (i_2+1) \ldots (i_N+1)}.       \eqno{(\rm \secIOPD.4)}$$
Now consider the following intertwining isomorphism $I$:
$$I \state{Q; i_2 \ldots i_N} := \state{(-i_2) (i_2 -i_3) \ldots
                                     (i_{N-1} - i_N) (i_N + R^{lt}) }.
  \eqno{(\rm \secIOPD.5)}$$
It is now straightforward to check using the basis (\secIOPD.3) that
$$I \rt(\si_{(j+1 \mod N)}) = r(\Ga_j \Ga_{j+1}^{+}) I \ , \qquad
I \rt(\Ga_j \Ga_{j+1}^{+}) = r(\si_j) I \ .
  \eqno{(\rm \secIOPD.6)}$$
The observation that $I$ is a unitary map and $r$ and $\rt$ are unitarily
equivalent in conjunction with (\secIOPD.6) concludes the proof.
\endproof
\mn
\beginremarks
\item{1)}
The momentum decomposition can be applied alike in the high- and
low-temperature phase. Thus, theorem {\ThA} is also valid if we further
restrict to eigenspaces with momentum $P$.
\item{2)}
Duality (\secIOPD.1) preserves the condition (\secB.13).
Thus, each integrable chiral Potts model is dual to exactly one
integrable chiral Potts model.
\endremark
\bn
\chapsubtitle{\secC.\ Generalities about perturbation theory}
\mn
Some of the results that will be reported later on rely on perturbation
theory. Therefore, we now review the general outline for perturbation theory
to all orders as presented in $\q{\baym}$ which directly applies to the
degenerate case as well. It should be clear to the reader that this method
is not in all cases the most powerful one available. Nevertheless, unlike
e.g.\ cluster expansions it can be directly applied to any problem we are
interested in (including higher excitations and correlation functions).
Note also that we do not intend to study the massless regimes III and IV
of Fig.\ 1 for which perturbation expansions are certainly not well suited.
\mn
The Hamiltonian (\secB.1) can be written as $H = H_0 + \la V$
with $H_0 = - \sum_{j,k} \ab_k \si_j^k$,
$V = - \sum_{j,k} \a_k \Ga_j^k \Ga_{j+1}^{n-k}$.
The eigenstates for $H_0$ are obvious, thus we have solved:
$$H_0 \state{a} = E_{\state{a}}^{(0)} \state{a}.      \eqno({\rm \secC.1})$$
Now one can solve $H \state{a(\la)} = E_{\state{a}} \state{a(\la)}$
for small $\la$ as follows:
Let $q_{\state{a}}$ be the projector onto the eigenspace of $H_0$ with
eigenvalue $E_{\state{a}}^{(0)}$. We can treat non-degenerate and
degenerate perturbation theory alike if we choose $\state{a}$ such
that
$$q_{\state{a}} V \state{a} = E_{\state{a}}^{(1)} \state{a}
     \eqno({\rm \secC.2})$$
with a constant $E_{\state{a}}^{(1)}$,
i.e.\ $q_{\state{a}} V q_{\state{a}}$ is to be chosen diagonal.
One also needs a regularized resolvent $g(z)$ of $H_0$:
$$g(z) := \left( \id - q_{\state{a}} \right) \left(z-H_0\right)^{-1}.
      \eqno({\rm \secC.3})$$
Then, the Wigner-Brillouin perturbation series
$$E_{\state{a}} = \sum_{\nu=0}^{\infty} \la^{\nu} E_{\state{a}}^{(\nu)}
\ , \qquad
\state{a(\la)} = \sum_{\nu = 0}^{\infty} \la^{\nu} \state{a, \nu}
      \eqno({\rm \secC.4})$$
are given by the following recurrence relations $\q{\baym}$:
$$\eqalign{
\state{a, 0} &= \state{a} \cr
\state{a, \nu} &= g(E_{\state{a}}^{(0)}) \left\{
       V \state{a, \nu-1} - \sum_{\mu = 1}^{\nu-1}
       \state{a, \nu-\mu} E_{\state{a}}^{(\mu)}
       \right\}, \cr
E_{\state{a}}^{(\nu+1)} &= \astate{a} V \state{a, \nu}. \cr
}       \eqno({\rm \secC.5})$$
Note that neither $\state{a(\la)}$ nor $\state{a, \nu}$ are in general
normalized although $\state{a}$ must be normalized to one.
Observe that the derivation of (\secC.5) does not rely on $H$
being hermitean. Therefore, (\secC.5) may also be applied
to diagonalizable but non-hermitean $H$.
\sn
The radius of convergence of the series (\secC.4) can be more
easily discussed in a different framework. Therefore, we postpone such a
discussion to section \secI.
\bigskip
There is one observation that makes explicit evaluation of high orders
for the $\Zed_n$-Hamiltonian (\secB.1) possible.
The energy-eigenvalues $E_{\state{a}}$ of $H_N^{(n)}$ do depend on the
chain length $N$. However, for the low lying gaps $\Delta E_{\state{a}}$
\footnote{${}^{2})$}{This will apply precisely to the fundamental
quasiparticle states to be discussed below.}
of $\Delta H_N^{(n)}$ (see (\secB.14) ) the coefficients for powers
of $\la$ become independent of $N$ up to order $\la^{N-2}$
(see e.g.\ $\q{\tang}$). Intuitively, this can be inferred from the
fact that (\secB.1) shows only nearest neighbour interaction and thus we
need $N-1$ powers in $V$ to bring us around a chain of length $N$.
Smaller powers in $V$ (or $\la$) do not feel that the length of the chain
is finite.
\vfill
\eject
{\helvits
{\par\noindent
\leftskip=1cm
\baselineskip=11pt
Paa den anden Side m{\o}der vi Vanskeligheder af en saa dybtliggende Natur,
at vi ikke aner Vejen til deres L{\o}sning; efter min personlige Opfattelse
er disse Vanskeligheder af en saadan Art, at de n{\ae}ppe lader os haabe
indenfor Atomernes Verden at gennemf{\o}re en Beskrivelse i Rum og Tid af den
Art, der svarer til vore Saedvanlige Sansebilleder.
\par\noindent}
\sn
\rightline{Niels Bohr, Letter to H{\o}ffding in 1922}
\par\noindent}
\bn
\chaptitle{\secX.\ Excitation spectrum}
\mn
In this chapter we will study the excitation spectrum of the
$\Zed_n$-chiral Potts model. For this end we denote the $k$th gap
in charge sector $Q$ by $\Delta E_{Q,k}$ starting to count with
$k=0$. More precisely, we define
$$\Delta E_{Q,k}(P) := E_{Q,k}(P) - E_{0,0}(0)
     \eqno({\rm \secX.1})$$
where $E_{Q,k}(P)$ is the $k$th energy level with momentum $P$
in charge sector $Q$. In the massive phases the groundstate
has $Q = P = 0$, i.e.\ $E_{0,0}(0)$ is the energy of the groundstate.
We will mainly be interested in the thermodynamic limit, i.e.\ in the
limit $N \to \infty$ of (\secX.1). Thus, if we write down energy gaps
$\Delta E_{Q,k}(P)$ we will usually think of sufficiently large lattices.
It will be clear from the context when gaps are evaluated for particular
finite chain length $N$.
\sn
In general we have the following symmetry under charge conjugation
$$\Delta E_{n-Q,k}(P, \phi, \vphi) = \Delta E_{Q,k}(P, -\phi, -\vphi).
     \eqno({\rm \secX.2})$$
Therefore, it is sufficient to restrict to those charge sectors with
$Q \le {n \over 2}$ in all explicit calculations where we keep the
dependence on $\phi$ and $\vphi$.
\bn
\chapsubtitle{\secF.\ Analytic results for the superintegrable $\Zed_3$-chain}
\mn
The results in $\q{\dkcoy,\kedem}$ and $\q{\weA,\lett,\perturb}$
suggest that we may expect a quasiparticle spectrum in the complete
massive high-temperature phase of the general $\Zed_n$-Hamiltonian (\secB.1).
At $\phi=\vphi={\pi \over 2}$ the spectrum of the Hamiltonian (\secB.1)
has been determined analytically by McCoy et al.\ $\q{\dkcoy}$ for $n=3$.
Their result is that the spectrum is a perfect quasiparticle spectrum.
We formulate their precise result in this section.
\mn
\beginddef{\DefA}{$\q{\dkcoy}$} Let
$$\eqalign{
\EE_1(P_r) = e_r &= 2 \mid 1 - \la \mid + {3 \over \pi}
\int_1^{\mid{1+\la \over 1-\la}\mid^{2 \over 3}} {\rm d}t
{v_r (2 v_r t + 1) \over v_r^2 t^2 + v_r t + 1}
\sqrt{ {4 \la \over t^3 -1} - (1-\la)^2 } \, , \cr
\EE_2(P_{2s}) = e_{2s} &= 4 \mid 1 - \la \mid + {3 \over \pi}
\int_1^{\mid{1+\la \over 1-\la}\mid^{2 \over 3}} {\rm d}t
{v_{2s} (4 v_{2s}^2 t^2 - v_{2s} t  + 1) \over v_{2s}^3 t^3 + 1}
\sqrt{ {4 \la \over t^3 -1} - (1-\la)^2 } \cr
}    \eqno({\rm \secF.1})$$
with the auxiliary variables $v_r$ and $v_{2s}$ related to the momenta by:
$$\eqalign{
v_r &=
{e^{i P_r}-1 \over \om (\om e^{i P_r}-1)}
\ , \quad \qquad 0 \le P_r < 2 \pi \, ,\cr
v_{2s} &=
{e^{i P_{2s}}-1 \over e^{-i {\pi \over3}} e^{i P_{2s}}- e^{i {\pi \over3}} }
\ , \quad \qquad \ {2 \pi \over 3} \le P_{2s} \le 2 \pi \, . \cr
}    \eqno({\rm \secF.2})$$
\enddef
Note that the Brillouin zone of the $Q=2$ excitation is only the interval
$[{2 \pi \over 3}, 2 \pi]$.
\mn
It is very difficult to evaluate the integrals in (\secF.1) explicitly
except for a few special values of the momentum. One can check that
$$\eqalignno{
P_r = 0 \qquad \Rightarrow \qquad v_r = & 0 \qquad \Rightarrow
 \qquad e_r = 2 \abs{1 - \la} \, , &({\rm \secF.3a})\cr
P_{2s} = 0 \qquad \Rightarrow \qquad v_{2s} = & 0 \qquad \Rightarrow
 \qquad e_{2s} = 4 \abs{1 - \la} \, , &({\rm \secF.3b})\cr
P_r \to {\textstyle {4 \pi \over 3}} \qquad
             \Rightarrow \qquad v_r \to & \pm \infty \qquad \Rightarrow
 \qquad e_r = 2 (1 + \la) \, , &({\rm \secF.3c})\cr
P_{2s} \to {\textstyle {2 \pi \over 3}} \qquad
             \Rightarrow \qquad v_{2s} \to & \infty \qquad \Rightarrow
 \qquad e_{2s} = 4 (1 + \la) \, . &({\rm \secF.3d})\cr
}$$
This shows that even for infinite values of the rapidities the energy as
a function of the physical variable momentum $P$ is well-defined. Thus,
as a function of $P$ the $Q=1$ excitation has {\it no} singularity at
$P = {4 \pi \over 3}$ and also $e_{2s}$ is continuous at both boundaries
including $P = {2 \pi \over 3}$.
For other values of the momentum than in (\secF.3) one can perform
numerical integration where special attention has to be paid to the
singularities in the integrand at $t=1$.
\mn
\beginpprop{{\PropA}}{\q{\dkcoy}}
All excitations above
the groundstate can be composed in the large chain limit of the fundamental
excitations $\EE_1(P) := e_r(P)$ and $\EE_2(P) := e_{2s}(P)$
according to the following rules:
$$
\Delta E_{Q,r}(P, \phi, \vphi) = \sum_{k=1}^{m_r} \EE_{Q^{(k)}}(P^{(k)})
\, , \quad
P = \sum_{k=1}^{m_r} P^{(k)} \hbox{ mod } 2 \pi \, , \quad
Q = \sum_{k=1}^{m_r} Q^{(k)} \hbox{ mod } n .
     \eqno({\rm \secF.4})$$
subject to the Pauli principle,
i.e.\ $Q^{(i)} = Q^{(j)}$ implies $P^{(i)} \ne P^{(j)}$.
\endprop
\mn
It was argued in $\q{\kedem}$ by examining the zeroes of the transfer matrix
that the same statement should also apply to the integrable three states chain
although here the relation between rapidities and momentum is not known.
\mn
For a detailed comparison with numerical methods and the perturbative
approach to be presented below see $\q{\lett}$.
\bn
\chapsubtitle{\secD.\ High-temperature expansions}
\mn
In order to generalize the quasiparticle picture of the previous section we
first study the low lying levels in the spectrum of the chiral Potts
model perturbatively in this section. In particular, for $n=3$ and $n=4$
we calculate the dispersion relations of the lowest excitations in the
charge sectors $Q \ne 0$. Some first results in this direction
have been presented in $\q{\weA}$ for the self-dual version of
these models. In this section we derive higher orders and admit general
$\phi \ne \vphi$. We also present some explicit results on the next
excitations.
\sn
In $\q{\hkn}$ high-temperature perturbation series were computed
for the disorder operator (or magnetization $m$) and the first
energy gap in the momentum zero sector of the superintegrable
$\Zed_3$-chiral Potts model leading to exact conjectures for
both of them. After the superintegrable chiral Potts model
had been generalized to general $n$ $\q{\gehri}$ perturbation
series for the ground state energy, energy gap in the momentum
zero sector, magnetization and susceptibility of this superintegrable
$\Zed_n$-chiral Potts model were presented in $\q{\hela}$.
At the same time elaborate expansions of the ground state energy
and some excitations of the superintegrable chiral Potts model
for $n \in \{ 3, 4, 5 \}$ and in particular perturbation series
for the order parameters with general $n$ were calculated in
$\q{\tang}$. First perturbative results for the energy gaps
at more general values of the angles $\phi$, $\vphi$ were
obtained in $\q{\weA}$ where second order high-temperature
expansions for the translationally invariant energy gaps in
each charge sector of the general self-dual $\Zed_3$- and
$\Zed_4$-chiral Potts models as well as a first order expansion
for the dispersion relations for general $n$ was presented.
\medskip
For arbitrary $n$, $N$ the groundstate of the Hamiltonian (\secB.1)
in the limit $\la \to 0$ is given by:
$$\state{{\rm GS}} :=
\state{0 \ldots 0} \eqno({\rm \secD.1})$$
provided that $-{\pi \over 2} \leq \vphi \leq {\pi \over 2}$. For $n=3$
(\secD.1) will be the groundstate for the larger range
$-\pi \leq \vphi \leq \pi$ and for $n=4$ (\secD.1) is the groundstate for
$-{5 \pi \over 6} \leq \vphi \leq {5 \pi \over 6}$.
\mn
The first excited states at $\la = 0$ for $Q > 0$ and arbitrary $P$
are the states
$$\pstate{s^Q}_P := \pstate{Q 0 \ldots 0}_P^{}   \eqno({\rm \secD.2})$$
in the range $-{\pi \over 2} \leq \vphi \leq {\pi \over 2}$.
According to the definition of the Hilbert space $\H$ in $\q{\perturb}$,
the states (\secD.2) give rise to proper eigenstates in
the limit of $\Delta H^{(n)}$. Thus, the corresponding gaps
$\Delta E_{Q,0} (P)$ belong to the point spectrum of $\Delta H^{(n)}$.
\mn
More generally, we wish to argue later on that the complete spectrum
can be explained in terms of quasiparticles. At $\la = 0$, a single-particle
excitation corresponds to just one flipped spin (\secD.2). Due to the absence
of interactions $k$-particle states have $k$ flipped spins at $\la = 0$.
Therefore, one can think in terms of states keeping in mind that for
$\la > 0$ one has to take the interactions into account using perturbation
theory. Note, however, that except for the single-particle states and
certain two-particle states we cannot directly compute the perturbation
series for multi-particle states and check that they have the correct
properties. Nevertheless, such a picture may be suggestive and is in fact
traditional (see e.g.\ $\q{\camp}$ for a transfer-matrix approach to the
spectrum of the Ising model in $d$ dimensions).
\medskip
First, one can derive a first order expansion for the dispersion relations
of the lowest excitations in the charge sectors $0 < Q < n$ using
(\secD.1) and (\secD.2) $\q{\weA}$:
$$\Delta E_{Q,0}(P, \ab_k, \phi) =
   \left( \sum_{k=1}^n \ab_k \left(1 - \om^{Q \, k} \right) \right)
    - {2 \la \over \bsin{\pi Q \over n}}
        \bcos{P - \left(1 - {2 Q \over n}\right) \phi}
    + \O(\la^2).
     \eqno{(\rm \secD.3)}$$
{}From (\secD.3) we can immediately read off that in the limit $\la \to 0$
the minimum of the dispersion relations $P_{{\rm min},Q}$ is located at
$$P_{{\rm min},Q} = \left(1 - {2 Q \over n}\right) \phi
     \eqno{(\rm \secD.4)}$$
which demonstrates that for $\phi \ne 0$ parity is not conserved.
\medskip
Let us now specialize to the $\Zed_3$-chain.  We will use the abbreviations:
$$\eqalign{
\cab := \bcos{{\vphi \over 3}} \ , \qquad
\chab := \bcos{{\pi - \vphi \over 3}} \ , & \qquad
\Rab := 1 - 4 \cab^2 \ , \qquad
\Uab := 7 \Rab - 4 \ , \cr
\cnab{r} :&= \bcos{{r \phi \over 3}}. \cr
}     \eqno{(\rm \secD.5)}$$
For $n=3$ we can calculate the groundstate energy per site $e_0$
which is defined by $E_N^0 = N e_0$ perturbatively:
$$\eqalign{
e_0 = &-{4 \over \sqrt{3}} \cab
      -{2 \la^2 \over 3 \sqrt{3} \cab}
      -{\bcos{\phi} \la^3 \over 9 \sqrt{3} \cab^2}
       + {\sqrt{3} \over 81 \cab} \left \{ {1 \over 2 \cab^2 }
             +{4 \over \Rab } \right \} \la^4
       + {\sqrt{3} \bcos{\phi} \over 81 \cab^2} \left \{ {3 \over 4 \cab^2 }
             +{4 \over \Rab } \right \} \la^5 \cr
      &+{1 \over \sqrt{3}} \Biggl[
       {\bcos{2 \phi} \over 324 \cab^3} \left\{
       {6 \over \Rab}
       +{1 \over \cab^2}
        \right\}
      + {23 \over 3888 \cab^5} \cr
      & \qquad + {1 \over 9 \Uab} \left\{
        -{40 \over 81 \cab \Rab}
        +{8 \over \cab \Rab^2}
        -{32 \over 27 \cab \Rab^3}
        -{11 \over 27 \cab^3 \Rab}
        -{28 \over 27 \cab^3 \Rab^2}
        +{55 \over 27 \cab^3} \right\} \Biggr] \la^6 \cr
       &+{\bcos{\phi} \over \sqrt{3}} \left[
        -{1 \over 324 \cab^6}
        +{1 \over 729 \Uab^2} \left\{
           {63976 \over 9 \cab^2 \Rab}
          -{47972 \over 9 \cab^2 \Rab^2}
          +{3104 \over 3 \cab^2 \Rab^3}
          -{64 \over \cab^2 \Rab^4}
          +{6664 \over 9 \cab^2}
 \right.\right. \cr
      &\qquad \left.\left.
          -{2326 \over 3 \cab^4 \Rab}
          +{284 \over \cab^4 \Rab^2}
          +{3529 \over 12 \cab^4}
           \right\}
         \right] \la^7
          + \O(\la^8).  \cr
}  \eqno{(\rm \secD.6)}$$
For $\vphi = \phi = {\pi \over 2}$ (\secD.6) reproduces the result of
$\q{\hela}$. The orders $e_0^{(k)}$ of the free energy per site $e_0$
are independent of $N$ if $N > k$.
Comments on the accuracy of a truncation to 5th order of (\secD.6)
can be found in $\q{\han,\perturb}$.
\mn
In principle, one can derive a critical exponent $\alpha$ for the specific
heat ${{\rm d}^2 e_0 \over {\rm d}\la^2}$ from a perturbation
expansion of the ground state energy $e_0$. In fact,
$\alpha$ has been estimated using a 13th order
expansion of $e_0$ in $\q{\hkn}$ for the self-dual case $\phi = \vphi$
and for the superintegrable case $\phi = \vphi = {\pi \over 2}$
the Ising-like form of the eigenvalues has been exploited to calculate
even higher orders of $e_0$ in $\q{\tang}$. The results in $\q{\hkn,\tang}$
indicate $\alpha = {1 \over 3}$ independent of the angles $\phi$, $\vphi$.
\medskip
We omit the results presented for the lowest translationally invariant
gaps of the three states chain presented in $\q{\perturb}$.
For $n=3$ and general $P$ we obtain from the states (\secD.2) the
perturbation expansion (\secD.7) below for the dispersion relation
of the lowest $Q=1$ excitation
and the lowest $Q=2$ excitation is given by (\secX.2)
\footnote{${}^{3})$}{The fourth order was independently obtained by
G.\ von Gehlen using the slightly more compact notation of $\q{\lett}$.}.
In $\q{\lett,\perturb}$ we restricted to a 3rd order version of
(\secD.7) because of the complicatedness of the 4th order which we
present here just as an illustration. It is also not difficult to
compute the 5th order on a computer but we do not want
to present this to the reader here either.
$$\eqalign{
\EE_{1}&(P) :=
\Delta E_{1,0}(P, \phi, \vphi) = 4 \bsin{{\pi - \vphi \over 3}}
                - \la {4 \over \sqrt{3}} \bcos{P-{\phi \over 3}} \cr
  &- \la^2 {2 \over 3 \sqrt{3}} {\Biggl \{}
               { \bcos{{P+{2 \phi \over 3} }} + 1\over \chab }
             + { \bcos{{2 P-{2 \phi \over 3} }} - 2\over \cab }
                 {\Biggr \}} \cr
  &+ \la^3 {1 \over 9 \sqrt{3}} {\Biggl \{}
             - {2 \bcos{2 P + {\phi \over 3}} - 3 \bcos{P - {\phi \over 3}}
                  + 2 \bcos{3 P - \phi} - 2 \bcos{\phi}
                        \over \cab^2 } \cr
  & \quad    -{2 \bcos{2 P + {\phi \over 3}} + 2 \bcos{P - {\phi \over 3}}
                   \over \cab \chab }
         +{\bcos{2 P + {\phi \over 3}} + 2 \bcos{P - {\phi \over 3}}+
                   \bcos{\phi}
                   \over \chab^2 }
             {\Biggr \} } \cr
& + \la^4 \left[ {1 \over 54 \sqrt{3}} \left\{
\left({6 \over \cab^3} + {4 \over \cab \chab^2} - {2 \over \chab^3}\right)
\bcos{2 P - {2 \phi \over 3}}
-\left({4 \over \cab^3} - {3 \over \cab \chab^2} + {1 \over \chab^3} \right)
\bcos{3 P} \right. \right. \cr
& \qquad \quad
- \left({1 \over \cab^3} + {2 \over \cab \chab^2}
+ {1 \over \cab^2 \chab}\right) \bcos{2 P + {4 \phi \over 3}}
\cr
& \qquad \quad
+ \left({2 \over \cab^3} + {1 \over \cab \chab^2}
  - {1 \over \chab^3} - {1 \over \cab^2 \chab} \right)
\bcos{P - {4 \phi \over 3}}
- {5 \over \cab^3} \bcos{4 P - {4 \phi \over 3}} \cr
\cr
& \qquad \quad \left.
+ \left({6 \over \cab^3} + {16 \over \cab \chab^2}
  - {2 \over \chab^3} + {20 \over \cab^2 \chab} \right)
\bcos{P + {2 \phi \over 3}}
 - {5 \over \cab^3}
 + {18 \over \cab \chab^2}
 -{2 \over \chab^3}
 +{18 \over \cab^2 \chab} \right\} \cr
&\qquad
-{1 \over 9 \bsin{{\pi + \vphi} \over 3}} \left\{
\left({1 \over \cab^2} + {2 \over \cab \chab} +  {1 \over \chab^2}\right)
\bcos{P + {2 \phi \over 3}}
+{2 \over 3 \cab^2} + {2 \over \cab \chab} + {1 \over \chab^2} \right. \cr
& \qquad \quad \left.\left.
+ {2 \over 9} \left({1 \over \cab^2} + {1 \over \cab \chab}\right)
\left(\bcos{2 P - {2 \phi \over 3}} + \bcos{3 P}\right)
 \right\}
-{ 2 \bcos{3 P} - 3 \over 81 \bsin{{\pi - \vphi} \over 3} \cab^2}
    \right]
              +\O(\la^5) \, . \cr
}     \eqno({\rm \secD.7})$$
\sn
Note that the agreement even of the truncations to 3rd order of (\secD.7)
with the results of a numerical diagonalization is usually
good as was discussed in detail in $\q{\lett}$ and in $\q{\perturb}$
for $\phi = \vphi = P = 0$.
\sn
We would like to mention that after specializing (\secD.7) to
$\phi = \vphi = 0$ it becomes much simpler and it even makes sense to
also present the 5th order $\q{\perturb}$. For $\phi = \vphi = 0$
one can read off from (\secD.7) that for $\la$ small, a Klein-Gordon
dispersion relation is a good approximation. It was shown in $\q{\albCONF}$
that for $\phi = \vphi = 0$ and in the limit $\la = 1$ one obtains
$\E_1(P) = 6\abs{\bsin{{P \over 2}}}$ which is also of the Klein-Gordon
form on the lattice. However, one can show that for general $\la$
and $\phi = \vphi = 0$ a Klein-Gordon relation is nothing but a very
good approximation to $\EE_1(P)$ $\q{\perturb}$. Also certain generalizations
of Klein-Gordon dispersion relations can be ruled out $\q{\perturb}$.
\sn
In section {\secC} we mentioned that the $k$th orders of (\secD.5) --
(\secD.7) are independent of $N$ if $N \ge k+2$. In particular, this
implies the existence of the limits $N \to \infty$ of (\secD.6) and
(\secD.7) if the perturbation series converge at all.
\mn
In the derivation of (\secD.7) we have not assumed that
the Hamiltonian (\secB.1) is hermitean. Thus, we may admit
$\phi \in \Complex$. We have checked in a few cases that results
of a numerical diagonalization at $N=12$ sites are still in good
agreement with (\secD.7) also for complex $\phi$.
\medskip
For the $\Zed_4$-chain one obtains up to second order in $\la$:
$$\eqalign{
\Delta & E_{1,0}(P, \phi, \vphi) =
      2\left(1 + 2 \bsin{{\pi - 2 \vphi} \over 4}\right)
      - \la 2 \sqrt{2} \bcos{P - {\phi \over 2}} \cr
     &- \la^2 \left\{
        {2 \sqrt{2} \bcos{P + {\phi \over 2}} + 3 \over
                4 \bcos{{2 \vphi + \pi \over 4}} }
        + {\bcos{2 P - \phi} - 2 \over 1 + \sqrt{2} \bcos{{\vphi \over 2}} }
        - {1 \over 4 \sqrt{2} \bcos{{\vphi \over 2}} }
        \right\}
     + \O(\la^3) \, ,\cr
\Delta & E_{2,0}(P, \phi, \vphi) =
      4 \sqrt{2}  \bcos{\vphi \over 2}
      - \la 2 \bcos{P} \cr
     &- \la^2 \left\{
        {\bcos{P - \phi} + 1 \over 1 - \sqrt{2} \bsin{{\vphi \over 2}} }
      + {\bcos{P + \phi} + 1 \over 1 + \sqrt{2} \bsin{{\vphi \over 2}} }
      - {2 \over 1 + \sqrt{2} \bcos{{\vphi \over 2}} }
      + {\bcos{2 P} - 1 \over 4 \sqrt{2} \bcos{{\vphi \over 2}} }
        \right\}
     + \O(\la^3) \, . \cr
}     \eqno({\rm \secD.8})$$
Again, $\Delta E_{3,0}$ is given by (\secX.2). The special case
$\phi = \vphi$ and $P = 0$ of (\secD.8) was presented in $\q{\weA}$.
\medskip
Returning to the $\Zed_3$-chain we observe that
for $\phi=\vphi = {\pi \over 2}$ we have to use degenerate
perturbation theory. This was not done before $\q{\perturb}$ but can easily
implemented for $P=0$ with the method of section $\secC$ and was carried out
in $\q{\perturb}$. It is straightforward to verify the exact result
(\secF.3b) which was first shown by $\q{\baxterB}$
$\Delta E_{2,0}\left({\textstyle {\pi \over 2}, {\pi \over 2}}\right)
 = 4 (1 - \la)$ up to order $\la^8$. Unfortunately, $P \ne 0$
is not yet accessible to perturbative computations because the lowest
eigenvector of the potential has a very complicated $P$-dependence and
becomes quite simple just for $P=0$. Nevertheless, the fact that we are
able to compute fairly high orders even in degenerate cases
demonstrates the universality of the approach to perturbation expansions
outlined in section \secC.
\medskip
Also for the higher excitations we must apply degenerate
perturbation theory. The next simplest case are the states where
two spins are different from zero. For general $n$, $P$,
$-{\pi \over 2} < \vphi < {\pi \over 2}$ the space of the excitation
with one spin flipped into charge sector $Q_1$ and another one flipped
into charge sector $Q_2$ is spanned by the states
$$\pstate{t^{Q_1,Q_2}_j}_P :=
\pstate{Q_1 \underbrace{0 \ldots 0}_{j-1 \ {\rm times}} Q_2 0 \ldots 0}_P \ ,
\qquad 1 \le j \le
\cases{N-1, & if $Q_1 \ne Q_2$;\cr
   \left\lb{N \over 2}\right\rb, & if $Q_1 = Q_2$.\cr}
 \eqno({\rm \secD.9})$$
Obviously, we will have to consider two cases: $Q_1 \ne Q_2$ and
$Q_1 = Q_2$.
\sn
For $Q_1 \ne Q_2$ it is not obvious even for $n=3$ how to diagonalize $V$ in
the space (\secD.9) in closed form although one can easily diagonalize it
numerically for comparably long chains -- for details see appendix {\appA}.
\mn
In the second case $Q_1 = Q_2$ one can diagonalize the potential $V$
in the space (\secD.9) explicitly exploiting a connection
to graph theory (see e.g.\ $\q{\jones}$). We shift the details of the
computation to appendix {\appA} and now present just the result.
\sn
It turns out that in the case of two identical excitations one has to
distinguish between even and odd momenta in terms of lattice
sites. It is convenient to introduce a further abbreviation $\delta_P^N$
encoding this distinction:
$$\delta_P^N := 0\ , \quad {\rm if} \ {P N \over 2 \pi} \ {\rm odd};
\qquad \qquad
\delta_P^N := 1\ , \quad {\rm if} \ {P N \over 2 \pi} \ {\rm even}.
  \eqno({\rm \secD.10})$$
\medskip
The result of the calculation in appendix {\appA} for the eigenvectors of
the potential $V$ is:
$$\eqalign{
\pstate{\tau^{Q,Q}_k}_P := {2 \over \sqrt{N}}
\biggl\{& \sum_{j=1}^{\lb{N \over 2}\rb - 1}
      \bsin{{(2 k - \delta_P^N) \, j \, \pi \over N}}
      e^{-i \Ph (j-1)} \pstate{t^{Q,Q}_j}_P \cr
&+    {\sqrt{2} \over \sqrt{3 + (-1)^N }}
      \bsin{{(2 k - \delta_P^N) \, \lb{N \over 2}\rb \, \pi \over N}}
      e^{-i \Ph (\lb{N \over 2}\rb-1)} \pstate{t^{Q,Q}_{\lb{N \over 2}\rb}}_P
\biggr\}. \cr
}  \eqno({\rm \secD.11})$$
\mn
The final result for the first order expansion of the energy for these
excitations is for $N \ge 3$:
$$\eqalign{
\Delta E_{2 Q,k}(P,\phi,\vphi) = &
       2 \left(\sum_{k=1}^{n-1} \ab_k (1 - \om^{Q k}) \right)
        - 4 \la {\bcos{\Ph - \left(1-{2 Q \over n}\right) \phi}
                    \bcos{{(2 k-\delta_P^N) \pi \over N}}
                 \over \bsin{\pi Q \over n} } \cr
       &+ \O(\la^2) \ ,
\qquad  1 \le k \le \left\lb{N+\delta_P^N-1 \over 2}\right\rb \cr
}  \eqno({\rm \secD.12})$$
For a few remarks on the second order see the end of appendix {\appA}.
\bn
\chapsubtitle{\secIOPC.\ Spectrum in the low-temperature phase}
\sn
In this section we calculate the ground state energy
and the lowest excitations in the low-temperature phase using
perturbative expansions according to $\q{\baym}$ around
$\tilde{\la} = 0$ of the Hamiltonian (\secB.2).
We restrict once again to the $\Zed_3$-version
of the chiral Potts model (\secB.2) with periodic boundary conditions
$R=0$ but impose no restrictions on
the angles $\phi$, $\vphi$. We present low-temperature
expansions for the ground state energy and in particular the first
translationally invariant energy gaps that before $\q{\han}$ had not
been treated by perturbative methods because of high degeneracies.
\medskip
In each charge sector $Q$ of the low-temperature phase there is
one unique ground state. For arbitrary $n$ it is given by
(\secIOPD.2) provided that $-{\pi \over 2} \leq \phi \leq {\pi \over 2}$.
For $n=3$ (\secIOPD.2) is the ground state if $-\pi \leq \phi \leq \pi$ and for
$n=4$ (\secIOPD.2) is the ground state for
$-{5 \pi \over 6} \leq \phi \leq {5 \pi \over 6}$.
The excited states are more complicated and highly degenerate. The
space of the first excitation is spanned by those states which
have precisely two blocks of different spins.
Furthermore, the values of the spins in these two blocks must
have difference one. For fixed
$P$, $Q$ and $-{\pi \over 2} \leq \phi \leq {\pi \over 2}$
we can choose the following basis for the space of the first excitation:
$$\state{a_k^Q} := {1 \over \sqrt{n}} \sum_{l=0}^{n-1} \om^{l \cdot Q}
  \pstate{\underbrace{(l+1 \mod n) \ldots (l+1 \mod n)}_{k \phantom{l}
                                     {\rm times}} l \ldots l}_P^{}.
  \eqno{(\rm \secIOPC.1)}$$
In order to perform explicit calculations we now specialize to
$n=3$ with $P = 0$. In $\q{\han}$ a 5th order low-temperature expansion
of the groundstate energy per site $e_0$ was computed. Upon interchanging
$\phi$ and $\vphi$ it agrees with (\secD.6) as predicted by duality (theorem
{\ThA}). It is noteworthy that the expansion in powers of $\tilde{\la}$
of $e_0$ does not depend on the charge sector for large $N$.
More precisely, the order $\tilde{\la}^k e_0^{(k)}$
of the free energy does not depend on the charge if $N > k$.
However, ground state level crossings have been observed in
$\q{\krallm,\gehlenph}$ which in the perturbative
approach are due to the fact that for short chain length $N$
high orders $\tilde{\la}^k$ ($k \ge N$) do depend on the charge $Q$.
These level crossings at fixed $\tilde{\la}$ in the massive phase
were exploited in $\q{\gehlenph}$ to compute a critical exponent
of the wave vectors by the argument introduced in $\q{\hoeger}$.
For more details on the relation to perturbation series compare
$\q{\han}$.
\bigskip
The calculation of the smallest gap $\Delta E_{Q,1}$ is more difficult.
Let $q$ be the projector onto the space spanned by the states
$\state{a_k^Q}$ (\secIOPC.1). In this space, the potential acts as follows:
$$\eqalign{
q \rt(V) \state{a_1^Q} &= -{2 \over \sqrt{3}}
   (2 e^{{i \vphi \over 3}} \state{a_2^Q}
   + e^{{i \vphi \over 3}} \om^Q \state{a_{N-1}^Q}) \cr
q \rt(V) \state{a_{k}^Q} &= -{2 \over \sqrt{3}}
   (2 e^{{i \vphi \over 3}} \state{a_{k+1}^Q}
   + 2 e^{-{i \vphi \over 3}} \state{a_{k-1}^Q} )
   \qquad 1 < k < N-1 \cr
q \rt(V) \state{a_{N-1}^Q} &= -{2 \over \sqrt{3}}
   (2 e^{-{i \vphi \over 3}} \state{a_{N-2}^Q}
   + e^{-{i \vphi \over 3}} \om^{2 Q} \state{a_1^Q}). \cr
}  \eqno{(\rm \secIOPC.2)}$$
In the limit $N \to \infty$ the eigenvector with lowest eigenvalue
converges to
$${1 \over \sqrt{N-1}} \sum_{k=1}^{N-1} \state{a_k^Q}.
  \eqno{(\rm \secIOPC.3)}$$
Using (\secIOPC.2) and (\secIOPC.3) we can calculate that for $N \to \infty$
and $\phi < {\pi \over 2}$
$$\eqalign{
\lim_{N \to \infty} \Delta E_{Q,1} = &
   4 \sqrt{3} \bcos{\phi \over 3}
   - \tilde{\la} {8 \over \sqrt{3}} \bcos{{\vphi \over 3}} \cr
  &+ \tilde{\la}^2 \left\{ {8\bcos{\phi \over 3}
                 \left(1+\bcos{{2\vphi \over 3}}\right)
                 \over 3 \sqrt{3} \left(3 - 4 \bcos{\phi \over 3}^2\right)}
   + {4 \left(2 - \bcos{{2 \vphi \over 3}} \right)
                 \over 3 \sqrt{3} \bcos{\phi \over 3} }
        \right\}
   + \O(\tilde{\la}^3). \cr
}  \eqno{(\rm \secIOPC.4)}$$
Comparing (\secIOPC.4) with the corresponding high-temperature expansion
(\secD.7) shows that up to the order calculated it coincides with
$\Delta E_{1,0}(\la) + \Delta E_{2,0}(\la) $ at the dual point
$\la = \tilde{\la}$ with $\phi$, $\vphi$ interchanged. Furthermore, one
can argue that the finite-size corrections to (\secIOPC.4) are of
order $N^{-2}$ $\q{\han}$. This further supports the identification of
(\secIOPC.4) with the dual of a two-particle state.
Of course, (\secIOPC.4) holds only for $\phi < {\pi \over 2}$.
\sn
For $\phi \ge {\pi \over 2}$, the states (\secIOPC.1) are not the first excited
states any more. Now we have to consider the following states:
$$\eqalign{
{1 \over \sqrt{n}} {\bigg (} &
   \pstate{1 \ldots 1 \ 2 \ldots 2 \ldots 0 \ldots 0 }_P^{}
   + \om^Q \pstate{2 \ldots 2 \ldots 1 \ldots 1 }_P^{} \cr
 & + \ldots
   + \om^{Q (n-1)} \pstate{0 \ldots 0 \ldots (n-1) \ldots (n-1)}_P^{}
          {\bigg )}.        \cr
}  \eqno{(\rm \secIOPC.5)}$$
Going through the same steps as before we find:
$$\eqalign{
\lim_{N \to \infty} \Delta E_{Q,1} = &
   12 \bsin{{\pi - \phi \over 3}}
   - \tilde{\la} 4 \sqrt{3} \bcos{{\vphi \over 3}} \cr
  &- \tilde{\la}^2 {2 \over \sqrt{3} } \left\{
      {\bcos{{2 \vphi \over 3}} - 2 \over \bcos{\phi \over 3} }
   +  {\bcos{{2 \vphi \over 3}} + 1 \over \bcos{{\pi - \phi \over 3}} }
      \right\}
   + \O(\tilde{\la}^3),
\quad {\pi \over 2} \le \phi < \pi. \cr
}  \eqno{(\rm \secIOPC.6)}$$
(\secIOPC.6) coincides with $3 \Delta E_{1,0}(\la)$ at the dual point
$\la = \tilde{\la}$ with interchanged $\phi$, $\vphi$ in the corresponding
high-temperature expansion (\secD.7) up to the order calculated.
\sn
For $\phi= {\pi  \over 2}$ the states (\secIOPC.1) and (\secIOPC.5) are
degenerate. However, for large $N$
the dominant contribution comes from the states (\secIOPC.5)
such that (\secIOPC.6) is valid for $\phi = {\pi \over 2}$ as well.
At $\phi = \vphi = {\pi \over 2}$ this is in agreement with the exact
result of $\q{\baxterB}$:
$$\lim_{N \to \infty} \Delta E_{Q,1} =
   6 (1 - \tilde{\la})
  \qquad {\rm for} \phantom{x} \phi = \vphi = {\pi \over 2}.
  \eqno{(\rm \secIOPC.7)}$$
\bn
\chapsubtitle{\secE.\ Evidence for quasiparticle spectrum}
\mn
Due to the studies of $\q{\dkcoy}$ (see proposition {\PropA}) and $\q{\kedem}$
one expects a quasiparticle picture for the {\it integrable} $\Zed_3$-chiral
Potts model. Extensive numerical studies $\q{\weA,\gehlenph,\lett}$ showed
that this quasiparticle picture is neither linked to integrability nor
restricted to the $\Zed_3$-chain. Furthermore, we were able to argue in
$\q{\perturb}$ that this quasiparticle picture is to be expected in a
quite general situation. Here, we do not want to go into technical details.
We first recall the main statement, then discuss its origin and its
consequences.
\mn
By a $\Zed_n$-quasiparticle spectrum we understand the obvious generalization
of (\secF.4):
\mn
\beginddef{\DefSpec}{$\Zed_n$-quasiparticle spectrum}
A family of spin chain Hamiltonians $H_N$ has a $\Zed_n$-quasiparticle
spectrum iff in the limit $N \to \infty$ the spectrum of the excitation
operator (\secB.14) can be obtained by the following rules:
$$\Delta E_{Q,r}(P, \phi, \vphi) = \sum_{k=1}^{m_r} \EE_{k}(P^{(k)})
\, , \quad
P = \sum_{k=1}^{m_r} P^{(k)} \hbox{ mod } 2 \pi \, , \quad
Q = \sum_{k=1}^{m_r} Q^{(k)} \hbox{ mod } n
     \eqno({\rm \secE.1})$$
with $m$ fundamental quasiparticles carrying energy
$\EE_1(P)$ to $\EE_{m}(P)$ and charge $Q^{(1)}$ to $Q^{(m)}$.
\enddef
\mn
Now the precise statement is:
\mn
\beginprop{\PropSpec}
Let $H_N$ be a nearest neighbour interaction spin chain Hamiltonian
of the form $H_N = H_0 + \la V$ such that $H_N$ conserves $\Zed_n$-charge
for all $\la$. Assume that $H_0$ has a $\Zed_n$-quasiparticle spectrum
with $n-1$ fundamental particles of charge
$Q^{(j)} = j$ ($j \ne 0$) and Brillouin zones $[0, 2\pi]$. Assume
furthermore that the perturbation series
of the fundamental particles converge for $\la < \la_0$.
\sn
Then for $\la < \la_0$ the limit $N \to \infty$ of the complete excitation
spectrum of $H_N$ exists and it is a $\Zed_n$-quasiparticle spectrum with
$n-1$ fundamental particles of charge $Q^{(j)} = j$ ($j \ne 0$) and
Brillouin zones $[0, 2\pi]$.
\endprop
\mn
\beginremarks{}
\item{1)}
For the $\Zed_n$-chiral Potts model (\secB.1) with the parameterization
(\secB.12) the assumptions of proposition {\PropSpec} can be easily
checked except for a non-trivial radius of convergence which we shall
check for the $\Zed_3$-chain below in section {\secI}. However, proposition
{\PropSpec} also covers more general cases, e.g.\ also the Hamiltonians
(\secSsd.5), (\secSsd.6) and (\secSsd.9).
\item{2)}
Due to the assumptions the only overlap between proposition {\PropA}
and proposition {\PropSpec} is the trivial point $\la = 0$ on the
superintegrable line of the $\Zed_3$-chain.
\endremark
\medskip
Recall that one may expect to interpret the energy bands as continuous spectrum
in the weak limit of the Hamiltonian $\q{\lett,\perturb}$ and that
the single-particle excitations (\secD.2) lead to point spectrum.
The normalization factors ${2 \over \sqrt{N}}$ for the
two-particle states in (\secD.11) demonstrate that these
states tend to zero for $N \to \infty$ and will therefore
not give rise to proper eigenvectors. Furthermore,
comparing (\secD.12) with the first order expansion for the single-particle
states (\secD.3) one observes that this first order expansion
of the two-particle excitations is in agreement with the quasiparticle rule
(\secE.1). Up to first order the composite particle states satisfy
either $2 \Delta E_{Q,0}(P, \phi, \vphi) <
\Delta E_{2Q,k} (2 P, \phi, \vphi) <
2 \Delta E_{Q,0}(P+2 \pi, \phi, \vphi)$ or
$2 \Delta E_{Q,0}(P, \phi, \vphi) >
\Delta E_{2Q,k} (2 P, \phi, \vphi) >
2 \Delta E_{Q,0}(P+2 \pi, \phi, \vphi)$ depending on which one of the
single-particle energies is larger. Thus, the two-particle states
do indeed lie inside the energy band of two single-particle states
and the boundaries are not included. Even more, we can see from
(\secD.12) that the two-particle states become dense
in this energy band for $N \to \infty$.
\medskip
The proof of proposition {\PropSpec} relies on
the interaction in the Hamiltonian (\secB.1) being very short
ranged -- in fact, only among nearest neighbours. In the massive
high-temperature phase there is no spontaneous order and the correlation
length is finite. Thus, if one puts two excitations of `short' chains with
a sufficient separation on a longer chain, the interaction will be
negligible. In pictures:
$$
\overbrace{\bull\mskp\hr\mskp\bull\cdots\cdots\bull\mskp\hr\mskp\bull}^{
\displaystyle N}_{\displaystyle E_1, P_1, Q_1}
\ \otimes \
\overbrace{\bull\mskp\hr\mskp\bull\cdots\cdots\bull\mskp\hr\mskp\bull}^{
\displaystyle M}_{\displaystyle E_2, P_2, Q_2}
\ \longTO_{N,M \to \infty} \
\overbrace{\bull\mskp\hr\mskp\bull\cdots\cdots\bull\mskp\hr\mskp\bull}^{
\displaystyle N+M}_{\displaystyle E_1 + E_2, P_1 + P_2, Q_1+Q_2} \, .
    \eqno({\rm \secE.2})$$
For example, putting one single-particle
excitation one the left half of the chain and another on the right
half will approximate a two-particle excitation.
\mn
In order to make the derivation rigorous one has to show the vanishing
of boundary terms in (\secE.2). This can indeed be done using e.g.\
perturbative arguments $\q{\perturb}$. Because this is very technical
we do not want to recall the complete argument here. Just note that
the crucial point is that
the momentum eigenstates have normalization factors $N^{-{1 \over 2}}$,
$M^{-{1 \over 2}}$. Any operator acting only at boundaries yields only
a finite part of these states in contrast to the operators $T_N$ and
$\Delta H_{N}^{(n)}$ which act
on the complete chain and yield complete momentum eigenstates.
The finite pieces of momentum eigenstates are suppressed by the
normalization factors $N^{-{1 \over 2}}$ in the infinite chain length
limit.
\mn
Note that boundary terms are substantial for conformally
invariant systems with long ranged correlations. Also in the low-temperature
phase boundary terms play an important r\^ole because the free part of the
Hamiltonian depends on the difference of neighbouring spins $\q{\han}$.
Thus, proposition {\PropSpec} applies neither to critical points where
one might have conformal invariance nor to the low-temperature phase.
\sn
We have not assumed the Hamiltonian to be hermitean. In particular,
the quasiparticle picture should also be valid for $\phi \in \Complex$
as long as the single-particle excitations exist and converge.
This is indeed supported by numerical calculations $\q{\yildirim}$.
\bigskip
The argument proving the quasiparticle structure can be refined in order
to give an upper estimate for the rate of convergence in $N$ of the
energy of a $k$-particle state. As an approximation to a $k$-particle state
for $k N$ sites, total energy $E_{\rm tot}$ and total momentum $P$
we may take the $k$-fold tensor product of single-particle states
$$\pstate{k N; E_{\rm tot}}_{P}
  := \pstate{N; E_1}_{P_1} \otimes \ldots \otimes \pstate{N; E_k}_{P_k}
    \eqno({\rm \secE.3})$$
with $E_{\rm tot} = \sum_{l=1}^k E_l$, $P = \sum_{l=1}^k P_l$.
Now, the deviation from the limit $N \to \infty$ is given by:
$${}_{P}\apstate{k N; E_{\rm tot}} \Delta H_{k N}^{(n)}
  \pstate{k N; E_{\rm tot}}_{P} - E_{\rm tot}
= \prod_{l=1}^k {}_{P_l} \apstate{N; E_l} \O(\Delta H_N)
   \pstate{N; E_l}_{P_l}
= \O(N^{-k})
    \eqno({\rm \secE.4})$$
where $\O(\Delta H_N)$ is some operator that acts only at sites $1$ and $N$.
The first equality simply uses the definition of the scalar product in
tensor products. The last equality is more profound and
due to the fact that operators acting only at boundaries
of the chain are suppressed by $N^{-1}$ due to the normalization factor in
the finite fourier transformation for momentum eigenstates.
This general argument is confirmed by our results for the two-particle
states. Expanding $\cos(x) = 1 - {1 \over 2} x^2 + \O(x^4)$ we can read
off from (\secD.12) that the first order correction of the
$r$th two-particle state with respect to the boundary of the energy
band behaves as $N^{-2}$ which agrees with (\secE.4).
\mn
In summary, we have the following results on the finite-size effects:
\item{1)} Single-particle states converge exponentially in $N$
          as long as the corresponding perturbation series converge
          since the $r$th orders are independent of $N$ if $N > r+1$.
\item{2)} The deviation of the energy of a $k$-particle state
          ($k > 1$) from the limit is at most of order $N^{-k}$ for
          $N \to \infty$. Note that consequently, the $N$-dependence
          of some energy eigenvalue yields only a lower bound on the
          number $k$ of particles involved.
\smallskip
In particular, the {\it energy} of any state remains unchanged to
order ${1 \over N}$ in a finite-size system. Thus, the only
modification in (\secE.1) at order ${1 \over N}$ in the massive high
temperature phase is a discretization of the momentum (and possible
minor modifications of the Brillouin zones and selection rules
$\q{\kedem}$).
\bigskip
Note that the proof of the vanishing of boundary terms as sketched above
and presented in detail in $\q{\perturb}$ also directly applies to the
Hamiltonian (\secB.1) itself. So far, we have restricted ourselves to
periodic boundary conditions but one could also use cyclic, twisted or even
free boundary conditions. Our argument shows
that all these different choices lead to the same spectrum in the
limit $N \to \infty$. In particular, our results are valid for all
choices of boundary conditions and one is free to choose those which
seem most appropriate, e.g.\ one can leave the ends of the chain open
instead of the unnatural end-identification for a realistic physical
system.
\sn
Again, this observation for the massive high-temperature phase is to
be contrasted with other situations. In particular, at the second order
phase transition $\phi = \vphi = 0$, $\la=1$ the correlation length
becomes infinite and the boundary terms are very important
$\q{\cardyB,\schuetz,\automos}$. Even in the massive low-temperature phase one
observes long range order and boundary terms cannot be neglected $\q{\han}$.
\sn
Similarly one can directly argue that in the low-temperature regime all charge
sectors are degenerate for $N \to \infty$, at least if the perturbation
expansion converges $\q{\han}$. That means that in the massive low-temperature
regime the thermodynamic limit does not depend on $Q$. The independence of the
spectrum of the charge $Q$ in the low-temperature regime is also expected due
to duality (theorem {\ThA}) because in the massive high-temperature phase it
does not depend on the boundary conditions $R$.
\bigskip
So far we have not addressed the question of whether the fundamental
particles satisfy a Pauli principle or not -- note that the above discussion
is intrinsically insensitive to a Pauli principle because the limit
was defined such that the spectrum forms a closed set. Nevertheless, for
the special case $n=3$ and $\phi = \vphi = {\pi \over 2}$, eq.\ (\secE.1)
was obtained in $\q{\dkcoy}$ supplemented with the Pauli principle.
Fortunately, due to (\secD.12), we have some control over the
finite-size dependence of the scattering states of two identical particles in
the general case. Up to first order in $\la$ these finite-size effects
do essentially neither depend on the charge $Q$ nor on the number of states
$n$ for periodic boundary conditions. Therefore, the nature of the fundamental
excitations can be determined by looking at one particular choice of $Q$ and
$n$. However, for $n=2$ one obtains the Ising model where it is well-known
that the excitation spectrum can be explained in terms of one fundamental
{\it fermion} (see e.g.\ $\q{\kogut}$). This indicates that the fundamental
excitations for general $n$ should behave similar to fermions. In particular,
for a scattering state of two identical excitations $i$ and $j$ the momenta
must satisfy $P_i \ne P_j$. In a scattering state of two {\it different}
fundamental particles these two fundamental particles can easily be
distinguished because they carry different $\Zed_n$-charges. Therefore,
two different particles should not be subject to a Pauli principle (like
it is the case for two different non-interacting fermions).
\smallskip
{}From theorem {\ThA} we conclude that the
quasiparticle interpretation for the high-tem\-per\-a\-ture
phase of the general $\Zed_n$-chiral Potts quantum chain given by
proposition {\PropSpec}
can be pulled over to the low-temperature phase. The duality
transformation interchanges charge sector $Q$ and boundary conditions
$R$. Thus, the ground state of the high-temperature
phase is mapped to periodic boundary conditions $R=0$
in the low-temperature phase. However, the fundamental excitations
are mapped to different boundary conditions corresponding to
the charge sectors $R \in \{1, \ldots n-1\}$. Therefore we
observed only composite particle states in section \secIOPC.
\bn
\chapsubtitle{\secI.\ Convergence of single-particle excitations}
\mn
As far as the quasiparticle picture of proposition {\PropSpec} is concerned
the main open question is the convergence of the single-particle states, or
equivalently the existence of the limits $N \to \infty$ of the
corresponding eigenvalues of the Hamiltonian. We have argued in section
{\secD} that convergence of the perturbation expansions
is sufficient to guarantee the existence of the limits
$N \to \infty$. Therefore we will discuss the radius of convergence
for the perturbation expansion of the single-particle
excitations in this section.
\mn
The potential for $\Delta H_N^{(n)}$ as defined in (\secB.1) and (\secB.14)
is unbounded if $N$ is not fixed. Thus, we have to apply the Kato-Rellich
theory of regular perturbations. Reviews of this subject can be found e.g.\
in the monographs $\q{\reedsimon,\katobook}$. The main
results we are going to use were originally published in
$\q{\rellich,\kato}$. The theory of Kato and Rellich applies
in particular to operators of the form  $H(\la) = H_0 + \la V$.
\medskip
Suppose that the single-particle eigenvalues $\Delta E$ have a non-zero
distance from the scattering eigenvalues (the continuous
spectrum) at $\la = 0$. Then it is clear from the discussion
in the previous sections that these eigenvalues are
non-degenerate and isolated. In particular, the resolvent
$(\Delta H_N^{(n)}(\la) - z)^{-1}$ is bounded for $\abs{\Delta E - z} > 0$.
Restricting to the hermitean case,
this is sufficient to guarantee that the $\Delta H_N^{(n)}(\la)$
are an analytic family in the sense of Kato. In this case,
the Kato-Rellich theorem ($\q{\reedsimon}$ Theorem XII.8)
may be used to guarantee a non-zero
radius of convergence $r_0 > 0$ for the single-particle
eigenvalues of $\Delta H_N^{(n)}(\la)$.
\mn
In order to obtain explicit estimates of the radius of convergence
one needs the inequality
$$\norm{ V \state{a} } \le {\cal V} \norm{H_0 \state{a} }
                         + {\cal K} \norm{ \state{a} }
   \eqno{\rm (\secI.1)}$$
on ${\cal D}(H_0)$ which in our case is dense in the the complete
Hilbert space $\H$. Then, the isolated point eigenvalues of $H(\la)$ are
convergent at least for
$$\la < r_1 := {\cal V}^{-1}   \eqno{\rm (\secI.2)}$$
as long as these eigenvalues {\it do not come in contact with
continuous spectrum} $\q{\kato}$. On the one hand this criterion
is very simple, on the other hand one must estimate not
only the constant ${\cal V}$ but also examine the level
crossings between single-particle excitations and scattering
states. There is another estimate $r_2$ that guarantees the
separation of eigenvalues as
well but gives smaller radii of convergence. For self-adjoint
$H_0$ with isolated eigenvalue $E^{(0)}_0$ where the
nearest eigenvalue $E^{(0)}_1$ has distance
$\epsilon := \abs{E^{(0)}_1 - E^{(0)}_0}$
($\epsilon^{-1} = \norm{g(E^{(0)}_0)}$) the perturbation
expansion of $E_0(\la)$ is convergent for
$$\la < r_2 := {\epsilon \over 2
           \left( {\cal K} + {\cal V} (\abs{E^{(0)}_0} + \epsilon) \right) }
   \eqno{\rm (\secI.3)}$$
and there are no crossings with neighbouring levels. In order to
compare the estimates (\secI.2) and (\secI.3) let us assume
${\cal K} = 0$ and $\abs{E^{(0)}_0} = \epsilon$. For this
almost optimal case one has $r_1 = 4 r_2$ showing that the criterion
(\secI.3) is much more restrictive.
\medskip
Let us now apply these general results to the present case of
$\Zed_n$-spin quantum chains. For non-degenerate single-particle
eigenvalues the Kato-Rellich theorem can be applied to guarantee
a positive radius of convergence $r_0$. Then we know from section
{\secE} that the spectrum of $\Delta H_N^{(n)}(\la)$
is a quasiparticle spectrum for $\la < r_0$. This fact can be used
to calculate the constant ${\cal V}$ and obtain explicit
estimates $r_1$ (where level crossings still have to be discussed)
or $r_2$. One can obtain the estimate (\secI.1) with ${\cal K}=0$
using Schwarz' inequality:
$${\cal V} := \sup_{\state{a} \in \H}
{\astate{a} \Delta V \state{a} \over \norm{\Delta H_{N,0}^{(n)} \state{a}}}.
   \eqno{\rm (\secI.4)}$$
In general, this supremum need not be finite but then it is very
difficult to ensure convergence at all. In our case,
the important observation is that due to the quasiparticle
picture we can evaluate (\secI.4) exclusively
from the single-particle excitations. To see this one performs
a first order expansion in $\la$ for any composite particle state,
compares coefficients and uses the quasiparticle property to
expand the expectation values of $\Delta H_{N,0}^{(n)}$ and $\Delta V$
in single-particle excitations. Thus, ${\cal V}$ can be calculated
as
$${\cal V} = \max_{Q,P}
     {{}_P\apstate{s^Q} \Delta V \pstate{s^Q}_P
          \over \norm{\Delta H_{N,0}^{(n)} \pstate{s^Q}_P } }.
   \eqno{\rm (\secI.5)}$$
\medskip
In order to implement this program explicitly we specialize to
the case of $\Zed_3$ with the parameterization (\secB.12).
At $\la = 0$ both single-particle eigenvalues are isolated
for $-{\pi \over 2} < \vphi < {\pi \over 2}$. This guarantees
a non-zero radius of convergence $r_0$.
\mn
The simplest case is the parity conserving case $\phi = \vphi =0$.
Here, the maxima are located at zero momentum $P=0$ and
both charge sectors are degenerate. Furthermore, we have
$\norm{\Delta H_{N,0}^{(3)} \pstate{s^Q}_0} = \epsilon = E_0^{(0)}$.
{}From (\secD.7) at $\phi = \vphi = 0$ we can therefore read off
${\cal V} = {2 \over 3}$, or in terms of radii of convergence
$$r_1 = {3 \over 2}, \qquad r_2 = {3 \over 8}, \qquad
{\rm for} \quad n=3, \ \phi = \vphi = 0 .
   \eqno{\rm (\secI.6)}$$
$r_2 = 0.375$ is certainly too small (for more details see $\q{\perturb}$).
\mn
For general angles $0 \le \vphi < \pi$, the free part of
the Hamiltonian $\norm{\Delta H_{N,0}^{(3)} \pstate{s^Q}_0}$ is minimized
for $Q=1$ and the potential
${}_P\apstate{s^Q} \Delta V \pstate{s^Q}_P$ is maximal
for $P = {\phi \over 3}$. Thus, we read off from (\secD.7)
${\cal V} = \left( \sqrt{3} \bsin{{\pi - \vphi \over 3}} \right)^{-1}$.
The relevant distance $\epsilon$ in the spectrum at $\la = 0$ is
given by the distance from the $Q=2$ fundamental particle to a state
of $k$ particles with $Q=1$ where $k$ depends on $\vphi$. Note that
$k$ must be equal $2=Q$ mod 3. Thus, one has $\epsilon = 4 \abs{
k \bsin{{\pi - \vphi \over 3}}- \bsin{{\pi + \vphi \over 3}} }$ and
$E_0^{(0)} = 4 \bsin{{\pi - \vphi \over 3}}$.
This amounts to the following radii
$$r_1 = \sqrt{3} \bsin{{\pi - \vphi \over 3}} , \qquad
 r_2 = \mathop{\rm min}\limits_k {\sqrt{3} \bsin{{\pi - \vphi \over 3}} \abs{
 k \bsin{{\pi - \vphi \over 3}}- \bsin{{\pi + \vphi \over 3}} }
\over
2  \left(
\bsin{{\pi - \vphi \over 3}} + \abs{
 k \bsin{{\pi - \vphi \over 3}}- \bsin{{\pi + \vphi \over 3}} }
\right) }
   \eqno{\rm (\secI.7)}$$
for $n=3$, $0 \le \vphi < \pi$. For $0 \le \vphi \le {\pi \over 2}$
the closest state is a two-particle state, i.e.\ $k=2$.
\sn
For $\vphi \to {\pi \over 2}$ the situation is contrary to that at
$\phi = \vphi = 0$. The $Q=2$ particle state becomes degenerate with
two $Q=1$ scattering states at $\vphi = {\pi \over 2}$ such that
the radius of convergence must tend to zero for $\vphi \to {\pi \over 2}$.
Whereas $r_2$ has precisely this property, $r_1$ tends to $0.866\ldots$
which is certainly too large.
\medskip
Because for small $\vphi$ we would prefer the large radius of
convergence $r_1$ but at $\vphi \approx {\pi \over 2}$ this is much
too large and $r_2$ seems more appropriate we have to enhance the
estimate given by $r_1$ by a discussion of level crossings between
single-particle states and scattering states.
For $0 \le \vphi < {\pi \over 2}$
the first level crossing of this kind will take place between
the $Q=2$ single-particle excitation and a two $Q=1$ particles
scattering state.
\sn
It is very difficult
to determine those values of $\la$ explicitly and precisely
where they take place. Therefore, we will use the first order
approximation of the perturbation expansion.
We are looking for those values of $\la$ where a single point
$P$ exists such that $x(k) := k \Delta E_{1,0}({P \over k},\phi,\vphi) -
\Delta E_{2,0}(P,\phi,\vphi)$ (with $k \equiv 2$ mod $3$) vanishes.
The fact that we are looking
for no real crossings but $x(k)=0$ implies ${{\rm d} x(k) \over {\rm d}P} = 0$.
Inserting (\secD.7) and (\secX.2) up to first order leads to the
condition
$$\bsin{{P \over k} - {\phi \over 3}} = \bsin{P + {\phi \over 3}}
    \eqno({\rm \secI.8})$$
Eq.\ (\secI.8) has a solution
$$P = {k \pi \over k+1}    \eqno({\rm \secI.9})$$
that does not depend on $\phi$. Now we can solve the linear equation
$x(k)\mid_{\la_0}=0$ for the value $\la_0$. One obtains
$$\la_0(k) = \sqrt{3} {\bsin{{\pi + \vphi \over 3}}
                     - k \bsin{{\pi - \vphi \over 3}}
            \over
           \bcos{{k \pi \over k+1} + {\phi \over 3}} - k \bcos{{\pi \over k+1}
                                                        - {\phi \over 3}} }
    \eqno({\rm \secI.10})$$
The $k=2$ special case of (\secI.7) -- (\secI.10) was discussed in
$\q{\perturb}$.
\mn
Fig.\ 2 shows a plot of the estimates (\secI.7) and (\secI.10) for
the self-dual case $\vphi = \phi$. Note that $r_1$ and $r_2$ are independent
of $\phi$. However, we have assumed that the Hamiltonian is hermitean and
therefore $\phi$ must be real.
\mn
For completeness we have also included an estimate for the boundary
of the massive high-temperature phase in Fig.\ 2. At this boundary,
levels of the $Q=1$ particle with generically non-zero momentum
cross with the ground state. Its explicit location has been
obtained estimating the minimum of the dispersion relation (\secD.7)
with $P = {\phi \over 3}$ and solving the second order approximation
$\Delta E_{1,0} ({\phi \over 3}, \phi, \phi) = 0$ for $\la$.
This is a slightly modified version of the estimate already proposed in
$\q{\hkn}$ where we have used the second order because it gives good
agreement with exactly known points.
\mn
For a detailed discussion of Fig.\ 2 in the interval $0 \le \vphi
\le {\pi \over 2}$ see $\q{\perturb}$.
\mn
The level crossings transition $\la_0$ divides the massive high-temperature
phase of the $\Zed_3$-chiral Potts model into two parts
which we label IIa and IIb. Note that above the superintegrable line
$\vphi = {\pi \over 2}$ there are infinitely many `windows' of type IIa
in which the $Q=2$ quasiparticle exists in the entire Brillouin
zone $[0, 2 \pi]$. At those points where the $Q=2$ fundamental particle
crosses $k$ $Q=1$ particle states ($k \equiv 2$ mod 3) already for $\la = 0$,
the radii of convergence become zero. The first crossing of this kind
takes places at $\vphi = {\pi \over 2}$ with a two-particle state.
The crossings and consequently also the `windows' accumulate for
$\vphi \to \pi$.
Note that for $\vphi \ge {\pi \over 2}$
the estimates $r_2$ and $\la_0$ follow each other closely, i.e.\
the estimate $r_2$ is already close to optimum.
\eject
\centerline{\psfig{figure=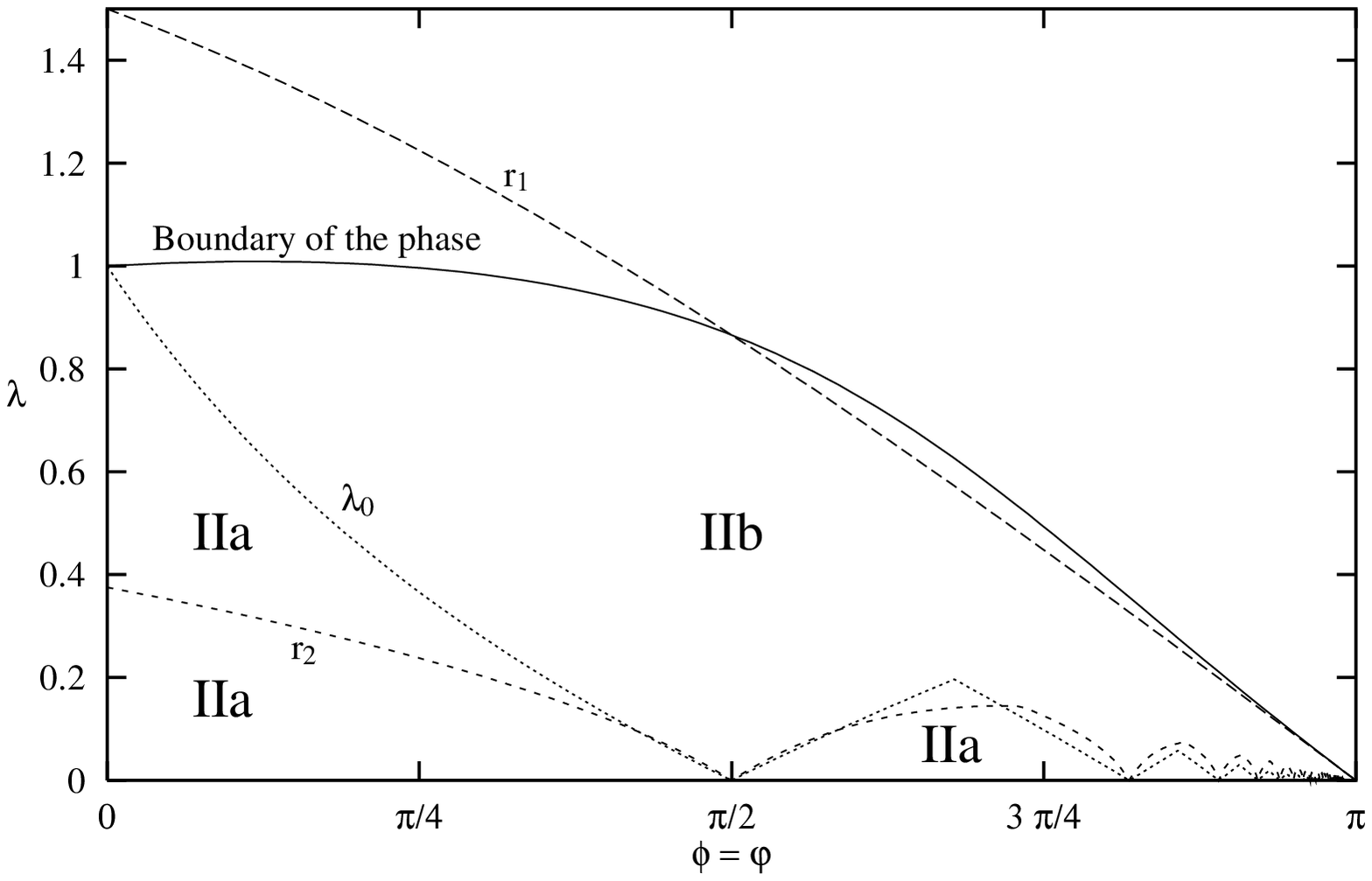}}
{\par\noindent\figindents
{\bf Fig.\ 2:}
Radii of convergence and boundary of the massive high-temperature phase
for the hermitean $\Zed_3$-chain. $r_1$ is an estimate ensuring
convergence if {\it no level crossings between point and continuous
spectrum occur}. The estimate $r_2$ also ensures the absence of
level crossings. The perturbation series are definitely
convergent for $\la < r_2$ although the true radius of convergence is
larger. It extends until the value $\la_0$ where the first
level crossings between fundamental quasiparticles and scattering
states occur.
\par\noindent
Note that $r_1$ and $r_2$ are independent of $\phi$
whereas for $\la_0$ we put $\vphi = \phi$.
\par\noindent}
\mn
In part IIa the derivation of the quasiparticle
picture is rigorous. Thus, in regime IIa the spectrum is
a quasiparticle spectrum with two fundamental particles existing
for all momenta. In $\q{\lett}$ we have presented numerical evidence that
regime IIb probably also exhibits a quasiparticle spectrum with two fundamental
particles where the $Q=2$ particle has the unusual property that it exists
only in a limited range of the momentum $P$. At $\vphi=\phi={\pi \over 2}$
this statement is given by proposition {\PropA}. We expect
that the idea to approximate multi-particle states by putting single-particle
states of `small' chains with a sufficient separation on a longer chain and
to use the finite correlation length in order to ensure vanishing of boundary
terms (which we cannot show directly like in $\q{\perturb}$ any more)
will apply also in regime IIb for general angles $\phi$, $\vphi$.
However, in contrast to section {\secE} we loose control over the fundamental
$Q=2$ excitation because the perturbation series does not converge any more
and there is no guarantee for the completeness of this construction.
At least it is plausible to still expect a quasiparticle spectrum in regime
IIb with two fundamental particles of which the $Q=2$ particle may have a
Brillouin zone that is smaller than the interval $\lbrack 0, 2 \pi \rbrack$.
\vfill
\eject
{\helvits
{\par\noindent
\leftskip=1cm
\baselineskip=11pt
Tout cela est d'une grande beaut\'e et d'une extr\^eme importance, mais
malheureusement nous ne le comprenons pas. Nous ne comprenons ni l'hypoth\`ese
de Planck sur les vibrateurs, ni l'exclusion des orbits non stationnaires,
et nous ne voyons pas, dans la th\'eorie de Bohr, comment, en fin de
compte, la lumi\`ere est produite. Car, il faut bien l'avouer, la
m\'ecanique des quanta, la m\'ecanique des discontinuit\'es, doit encore
\^etre faite.
\par\noindent}
\sn
\rightline{H.A.\ Lorentz in ``L'ancienne et la nouvelle m\'ecanique'' (1925)}
\par\noindent}
\bn
\chaptitle{\secY.\ Correlation functions}
\mn
In recent papers a systematic investigations of the correlation
functions of the $\Zed_3$-chiral Potts model in the massive phases
has been started. First, a non-vanishing wave vector has been predicted
in $\q{\gehlenph,\krallm}$ for the massive low-temperature phase and
its critical exponent was calculated from level crossings.
Below we wish to present the direct computations of the correlation
functions of $\q{\han,\lett,\perturb}$.
Note also that for the massless phases around $\la \sim 1$ of the
$\Zed_3$-chain correlation functions have been derived in $\q{\albcoy}$
using results from conformal field theory.
\bn
\chapsubtitle{\secG.\ Correlation functions of the high-temperature phase}
\mn
In this section we study correlation functions for the $\Zed_3$-chiral
Potts model perturbatively.
Before defining correlation functions, we first note that the
two-point functions are translationally invariant because
the groundstate $\vac$ is translationally invariant:
$$\eqalign{
\avac \Ga_{x+r}^{+} \Ga_r \vac &= \avac \Ga_{x+1}^{+} \Ga_1 \vac \ , \cr
\avac \si_{x+r}^{+} \si_r \vac &= \avac \si_{x+1}^{+} \si_1 \vac . \cr
}    \eqno({\rm \secG.1})$$
Thus, it makes sense to define the correlation function for an operator
$\Xi$ by the following expression:
$$C_{\Xi}(x) := {\avac \Xi_{x+1}^{+} \Xi_1 \vac \over \normvac}
              - {\avac \Xi_{x+1}^{+} \vac \avac \Xi_1 \vac \over
                   \normvac^2}
    \qquad 0 \le x < {N \over 2}    \eqno({\rm \secG.2})$$
where $\vac$ is the eigenvector of the Hamiltonian to lowest energy.
Here, we do not assume that $\vac$ is normalized to one and have therefore
included the proper normalization factors in (\secG.2).
\mn
The operator $\Xi$ for the $\Zed_3$-chiral Potts model can be
either $\Gamma$ or $\sigma$. (For $n>3$ also different powers of
these operators may be interesting.) The correlation
functions of the operators $\Ga_x$ and $\si_x$ have the property
$$\eqalign{
C_{\Ga}(-x) &= C_{\Ga}(x)^{*} \ , \cr
C_{\si}(-x) &= C_{\si}(x)^{*} = C_{\si}(x) \cr
}    \eqno({\rm \secG.3})$$
such that it makes sense to restrict to positive $x$. Note that
(\secG.3) follows by complex conjugation using (\secG.1).
Explicit calculations show the validity of (\secG.1) and (\secG.3) as well.
\bigskip
Let us now turn to the explicit computation of correlation functions
for the $\Zed_3$-chain.
In order to be able to calculate the correlation functions we need
to know the groundstate $\vac$. We will calculate it from
the free ground state $\state{{\rm GS}}$ using the perturbation
expansion (\secC.5).
We should stress again that although we assume the free groundstate
$\state{{\rm GS}}$ to be normalized to
$1$, this is not necessarily true for the complete state $\vac$.
The expansion of the groundstate $\vac$ provides us with an
expansion for the correlation functions in powers of $\la$
$$C_{\Xi}(x) = \sum_{\nu=0}^{\infty} \la^{\nu} C_{\Xi}^{(\nu)}(x) \, .
    \eqno({\rm \secG.4})$$
Note that according to (\secC.5) a $k$th order
expansion of the groundstate yields a $k+1$th order expansion
of the groundstate energy as a byproduct.
\sn
The operator $\Gamma_x$ creates charge such that charge conservation implies
$\avac \Ga_x^{+} \vac = \avac \Ga_x \vac = 0$ for all $x$. Thus
$$C_{\Ga}(x) =  {\avac \Ga^{+}_{x+1} \Ga_1 \vac \over \normvac}.
    \eqno({\rm \secG.5})$$
\medskip
Here we omit the intermediate steps and
present just the final results of the computations. The interested
reader may find some more details in $\q{\perturb}$.
\sn
The final result for the expansion of $C_{\si}(x)$ is
$$\eqalign{
C_{\si}^{(0)}(x) =&
C_{\si}^{(1)}(x) = 0 \, , \cr
C_{\si}^{(2)}(x) =& {1 \over 3 \cab^2} \left \{
                     \delta_{x,0}
                   + {\delta_{x,1} \over 4}
                    \right \} \, , \qquad
C_{\si}^{(3)}(x) = {\bcos{\phi} \over 9 \cab^3} \left \{
                     \delta_{x,0}
                   + {\delta_{x,1} \over 4}
                    \right \} \, , \cr
C_{\si}^{(4)}(x) =& {1 \over 27 \cab^2} {\Biggl \{}
                   - \delta_{x,0} \left (
                           {2 (1 - 10 \cab^2)
                               \over \Rab^2 }
                         + {3 \over 2 \cab^2}
                                  \right )
                   + \delta_{x,1} \left (
                           {1 + 20 \cab^2
                               \over 3 \Rab^2 }
                         - {1 \over \cab^2}
                                  \right ) \cr
                  & \phantom{{1 \over 27 \cab^2} {\Biggl \{} }
                   + \delta_{x,2} \left (
                           {2( 1 + 2 \cab^2)
                               \over 3 \Rab^2 }
                         + {1 \over 16 \cab^2}
                                  \right )
                    {\Biggr \}}. \cr
}    \eqno({\rm \secG.6})$$
For the other correlation function $C_{\Ga}(x)$ one finds the following
final result
$$\eqalign{
C_{\Ga}^{(0)}(x) =& \delta_{x,0} \ , \qquad
C_{\Ga}^{(1)}(x) =  \delta_{x,1}
                     {e^{i {\phi \over 3}} \over 3 \cab } \ , \qquad
C_{\Ga}^{(2)}(x) = {1 \over 6 \cab^2} {\Biggl \{}
                     \delta_{x,1} {e^{- i {2 \phi \over 3}} \over 2}
                   + \delta_{x,2} e^{i {2 \phi \over 3}}
                    {\Biggr \}} \ , \cr
C_{\Ga}^{(3)}(x) =& {1 \over 54 \cab} {\Biggl \{}
                   - \delta_{x,1} \ e^{i {\phi \over 3}} \left(
                           {1 \over \cab^2}
                         + {8 \over \Rab}
                            \right)
                   + \delta_{x,2} \ e^{-i {\phi \over 3}} \left(
                           {2 \over \cab^2}
                         - {8 \over \Rab}
                            \right)
                   + \delta_{x,3} {5 e^{i \phi} \over \cab^2}
                    {\Biggr \}} \cr
C_{\Ga}^{(4)}(x) =& {1 \over 81 \cab^2} {\Biggl \{}
            - \delta_{x,1} \left( e^{i {4 \phi \over 3}}
             + 4 e^{-i {2 \phi \over 3}} \right)
                  \left( {9 \over 16 \cab^2}
           + {3 \over \Rab}
                  \right) \cr
          &+ \delta_{x,2} \left( {8 e^{i {2 \phi \over 3}}
          (19 \cab^2 - 4) \over 3 \Rab^2 }
           + {3 e^{-i {4 \phi \over 3}} - 20 e^{i {2 \phi \over 3}}
                  \over 8 \cab^2}
           - {3 e^{-i {4 \phi \over 3}} \over \Rab}
                  \right) \cr
          &+ \delta_{x,3} \left(
           {40 \cab^2 - 7 \over \Rab^2 }
               + {9 \over 4 \cab^2 }
                  \right)
           + \delta_{x,4} {35 e^{i {4 \phi \over 3}} \over 8 \cab^2}
            {\Biggr \}} \cr
}    \eqno({\rm \secG.7})$$
It is easy to see that $C_{\Ga}^{(k)}(x)$ and
$C_{\si}^{(k)}(x)$ are independent of $N$ if $N > 2 k$ and $x \le k$.
\medskip
$C_{\si}(x)$ is real and positive for all values of $\phi$ and $\vphi$
up to the order calculated. However, it is not easy to read off from
(\secG.6) what might be the form for large $x$. Thus, we specialize
to $\phi = \phi = {\pi \over 2}$ and calculate two further orders for
$C_{\si}(x)$:
$$\eqalign{
C_{\si}^{(0)}(x) =& C_{\si}^{(1)}(x) = C_{\si}^{(3)}(x)
                 = C_{\si}^{(5)}(x)= 0 \ , \cr
C_{\si}^{(2)}(x) =& {1 \over 9} \left \{
                     4 \delta_{x,0} + \delta_{x,1} \right \} \ , \cr
C_{\si}^{(4)}(x) =& {1 \over 81} \left \{
                     5 \delta_{x,0} + 2 \delta_{x,2} \right \} \ , \cr
C_{\si}^{(6)}(x) =& {1 \over 6561} \left \{
                     190 \delta_{x,0} - 13 \delta_{x,1} + 38 \delta_{x,2}
                     + 60 \delta_{x,3} \right \}. \cr
}    \eqno({\rm \secG.8})$$
\indent
$C_{\Ga}(x)$ in general has a non-vanishing imaginary part and therefore
is worth while being considered in more detail. Thus, we specialize
again to the superintegrable case $\phi = \vphi = {\pi \over 2}$
\footnote{${}^{4})$}{Another natural specialization would be the standard
Potts case $\phi = \vphi = 0$. This was discussed in detail in $\q{\perturb}$.
There, a deviation from the well-known relation (see e.g.\ $\q{\kogut}$)
$\xi_\Ga \sim m(\la)^{-1}$ was observed which is clearly due to lattice
artifacts.}
and obtain after calculating two further orders
$$\eqalign{
C_{\Ga}^{(0)}(x) =& \delta_{x,0} \ , \qquad \qquad
C_{\Ga}^{(1)}(x) =  \delta_{x,1} \left({1 \over 3} + i {\sqrt{3} \over 9}
            \right) \ , \cr
C_{\Ga}^{(2)}(x) =& {1 \over 18} \left\{\delta_{x,1} +2 \delta_{x,2} \right\}
      + i {\sqrt{3} \over 18} \left\{ -\delta_{x,1} + 2 \delta_{x,2}
      \right\} \ , \cr
C_{\Ga}^{(3)}(x) =& {1 \over 81} \left\{ 4 \delta_{x,1} + 10 \delta_{x,2}
            \right\}
      + i {\sqrt{3} \over 243} \left\{ 4 \delta_{x,1} - 10 \delta_{x,2}
             + 20 \delta_{x,3} \right\} \ , \cr
C_{\Ga}^{(4)}(x) =& {1 \over 1458} \left\{ 27 \delta_{x,1} + 18 \delta_{x,2}
              + 210 \delta_{x,3} - 70 \delta_{x,4}\right\}
      + i {\sqrt{3} \over 1458} \left\{ -27 \delta_{x,1} + 18 \delta_{x,2}
             + 70 \delta_{x,4} \right\}
      \hskip 2pt ,  \cr
C_{\Ga}^{(5)}(x) =& {1 \over 2187} \left\{ 45 \delta_{x,1} + 108 \delta_{x,2}
         + 252 \delta_{x,4}- 126 \delta_{x,5} \right\} \cr
      &+ i {\sqrt{3} \over 2187} \left\{ 15 \delta_{x,1} - 36 \delta_{x,2}
             - 14 \delta_{x,3} + 84 \delta_{x,4} + 42 \delta_{x,5} \right\}
      \ , \cr
C_{\Ga}^{(6)}(x) =& {1 \over 39366} \left\{381 \delta_{x,1} +214 \delta_{x,2}
              + 2314 \delta_{x,3} + 784 \delta_{x,4} + 2310 \delta_{x,5}
              - 1848 \delta_{x,6} \right\} \cr
      &+ i {\sqrt{3} \over 39366} \left\{-381 \delta_{x,1} +214 \delta_{x,2}
             - 784 \delta_{x,4} + 2310 \delta_{x,5} \right\}. \cr
}    \eqno({\rm \secG.9})$$
Of course, we still have to calculate the sum (\secG.4). Thus, changes
of signs in individual orders need not necessarily turn up in the final
result. In fact, it turns out that the imaginary part of $C_{\Ga}(x)$
is always positive up to order 6 because the smallest orders are positive
and they dominate the others. However, for sufficiently small $\la$ the
real part does indeed change signs around $x=4$.
We fit (\secG.9) by a complex exponential function:
$$\eqalignno{
C_{\Ga}(x) &= a \ e^{\left({2 \pi i \over L} - {1 \over \xi_{\Ga}}\right) x}
            + (1-a) \delta_{x,0} \ , &({\rm \secG.10a})\cr
C_{\si}(x) &= p \ e^{- {x \over \xi_{\si}} } + q \delta_{x,0} \, .
             &({\rm \secG.10b})\cr
}$$
In (\secG.10) we have also taken into account that from (\secG.8)
${C_{\si}(0) \over C_{\si}(1)} \approx 4$ independent of the correlation
length $\xi_{\si}$.
\sn
If (\secG.10a) is the correct form for $C_{\Ga}(x)$ we infer from (\secG.9)
that $L$ is about $14$ for small $\la$. We can also see from
the higher orders that $L$ increases with increasing $\la$ such that
it might well be singular at $\la = 1$. The correlation length $\xi$ tends
to zero as $\la \to 0$. This implies that -- after proper re-normalization
of the Hamiltonian -- the mass gap becomes infinite at $\la = 0$. It has
already been observed in $\q{\weA}$ that there are physical
reasons to divide (\secB.1)  by $\sqrt{\la}$ which would have exactly the
effect of infinite mass at $\la = 0$. Fits to (\secG.10) for
$\la \in \{ {1 \over 4}, {1 \over 2}, {3 \over 4} \}$ in the superintegrable
case are given by the values in table 1.
\mn
\centerline{\vbox{
\hbox{
\vrule \hskip 1pt
\vbox{ \offinterlineskip
\def\tablespace{height2pt&\omit&\vl&\omit&&\omit&&\omit&\vl&\omit&&
                          \omit&&\omit&\vl&\omit&&\omit&\cr}
\def\tablerule{ \tablespace
                \noalign{\hrule}
                \tablespace        }
\hrule
\halign{&\vrule#&
  \strut\hskip 4pt\hfil#\hfil\hskip 4pt\cr
\tablespace
\tablespace
& $\la$  &\vl& $\xi_{\Ga}$ && $a$       && $L$      &\vl&
          $\xi_{\si}$ && $p$       && $q$       &\vl&
            $P_{\min}$ && ${L P_{\min} \over 2 \pi}$ &
                                           \cr \tablespace \tablerule
& $0.25$ &\vl& $0.55(3)$   && $0.55(5)$ && $14.3(2)$ &\vl&
          $0.25(2)$   && $0.35(4)$ && $0.32(4)$ &\vl&
            $0.471$     && $1.07(2)$ & \cr \tablespace
& $0.50$ &\vl& $0.9(1)$    && $0.59(3)$ && $16.5(8)$ &\vl&
          $0.38(4)$   && $0.35(3)$ && $0.24(3)$ &\vl&
            $0.401$     && $1.05(5)$ & \cr \tablespace
& $0.75$ &\vl& $1.5(6)$    && $0.64(3)$ && $18.3(8)$ &\vl&
          $0.55(6)$   && $0.36(2)$ && $0.09(2)$ &\vl&
            $0.308$     && $0.90(4)$ & \cr \tablespace
}
\hrule}\hskip 1pt \vrule}
\hbox{Table 1: Parameters for the correlation functions
            (\secG.18) at $\phi = \vphi = {\pi \over 2}$. }}
}
\mn
The estimates in table 1 have been obtained as follows.
First, $\xi_{\Ga}$ has been estimated by calculating
$\re(\ln({C_{\Ga}(x) \over C_{\Ga}(x+1) }))^{-1}$ and averaging
over $x$. Next, the zero of $\re(e^{x \over \xi_{\Ga}} C_{\Ga}(x))$
has been estimated by linear interpolation for two neighbouring
values and $L / 4$ was obtained by averaging. Finally,
$a$ was estimated such that the difference
$$\re(C_{\Ga}(x)) -
a e^{-{x \over \xi_{\Ga}}} \bcos{{2 \pi x \over L}}
    \eqno({\rm \secG.11})$$
is minimal for $x=1,2$. That this procedure yields reasonable
fits is demonstrated by Fig.~3 which shows the stretched correlation
function $e^{x \over \xi_{\Ga}} C_{\Ga}(x)$ in comparison to the
fits. The `error bars' are not really error bars but given by
$a e^{6 - x \over \xi_{\Ga}}$ which gives an idea how much the
values have actually been stretched and what might be the contribution
of the next orders in the perturbation expansion. The agreement
for all $x$ not only in the real part but also in the imaginary part
is convincing.
\mn
Let us now discuss the implications of (\secG.7) under the
assumption that (\secG.10a) is the correct form for general values
of the chiral angles.
{}From the leading orders in (\secG.7) we read off the following identity
for the ratio of $C_{\Ga}(1)$ and $C_{\Ga}(2)$:
$${C_{\Ga}(2) \over C_{\Ga}(1)} = {
{e^{i{2 \phi \over 3}} \over 6 \cab^2} \la^2 + \O(\la^3) \over
{e^{i{\phi \over 3}} \over 3 \cab} \la + \O(\la^2) }
= {e^{i{\phi \over 3}} \over 2 \cab} \la + \O(\la^2).
    \eqno({\rm \secG.12a})$$
On the other hand we immediately obtain from (\secG.10a)
$${C_{\Ga}(2) \over C_{\Ga}(1)}
= e^{-{1 \over \xi_{\Ga}}} e^{{2 \pi i \over L}}.
    \eqno({\rm \secG.12b})$$
\mn
\centerline{\psfig{figure=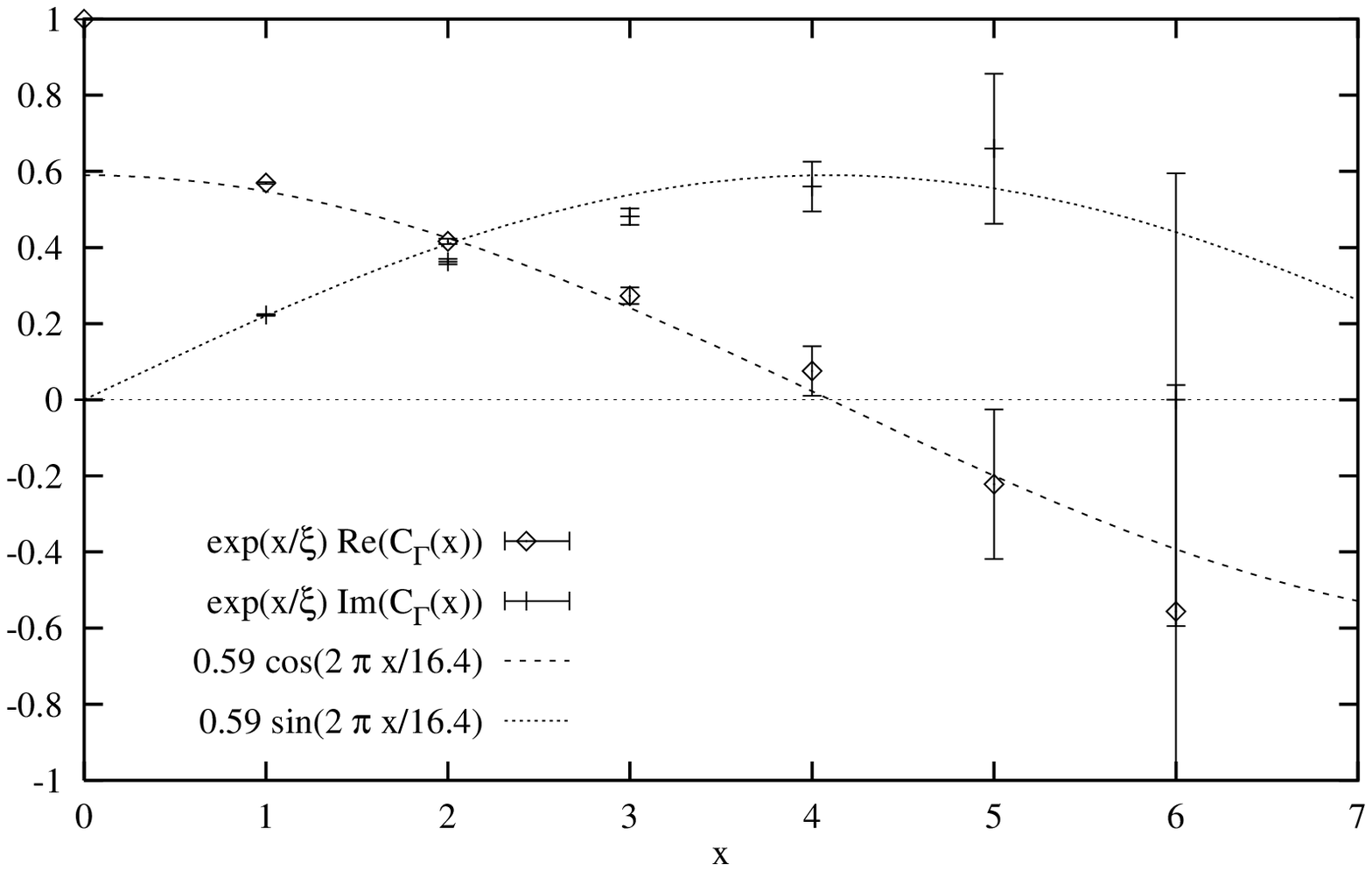}}
{\par\noindent\figindents
{\bf Fig.\ 3:}
Correlation function $C_{\Ga}(x)$ stretched by $e^{x \over \xi_{\Ga}}$
in comparison to the fit (\secG.10a) at $\phi = \vphi = {\pi \over 2}$,
$\la = {1 \over 2}$. The `error bars' are given by
$a e^{6 - x \over \xi_{\Ga}}$ which conveys an idea how much the
values have actually been stretched. The oscillatory contribution
to $C_{\Ga}(x)$ is clearly visible.
\par\noindent}
\mn
Comparison of (\secG.12a) and (\secG.12b) leads to
$$L = {6 \pi \over {\rm Re}(\phi)} \ , \qquad \qquad
\xi_{\Ga} = -{1 \over \ln{\left(
{\la \over 2 \bcos{{\vphi \over 3}} }
\right)}  - {{\rm Im}(\phi) \over 3} }
    \eqno({\rm \secG.13})$$
for small values of $\la$. It is noteworthy that we obtain the same
result for the oscillation length if we apply a similar argument to
${C_{\Ga}(x_1) \over C_{\Ga}(x_2)}$ in lowest non-vanishing order with
$x_1,x_2 \in \{1,2,3,4\}$. At $\phi = {\pi \over 2}$ (\secG.13) yields
the approximations $L = 12$, $\xi_{\Ga} = 0.52$, $0.80$, $1.2$
for $\la = 0.25$, $0.50$, $0.75$. The agreement with the numbers of table 1
is very good. Thus, for very high temperatures the oscillation length $L$
is proportional to the inverse chiral angle $\phi^{-1}$. In particular,
the oscillation vanishes smoothly for $\phi \to 0$.
According to (\secD.4) for very high temperatures the minimum of the
dispersion relation of the fundamental particles is also proportional to
$\phi$ -- for $n=3$ and $Q=1$ we have $P_{\min} = {{\rm Re}(\phi) \over 3}$.
Thus, we obtain from (\secG.13) for very high temperatures
$$P_{\min} L\mid_{\la \to 0} = 2 \pi \qquad \forall \phi, \vphi.
     \eqno({\rm \secG.14})$$
Furthermore, the second order in (\secD.7) shows that the minimal
momentum $P_{\min}$ decreases with increasing $\la$ (compare also
$\q{\lett}$). Similarly,
we read off from (\secG.7) that the inverse oscillation length $L^{-1}$ also
decreases with increasing inverse temperature $\la$. Thus, (\secG.14) has
a chance to be valid for all values of $\la$ in the massive
high-temperature phase. Indeed, using the values of $P_{\min}$
given in table 8 of $\q{\weA}$ we see that $P_{\min} L = 2 \pi$
holds quite accurately for $\la = 0.25, 0.5, 0.75$ at
$\phi = \vphi = {\pi \over 2}$ (compare table 1).
We suspect that in general this relation is not exact but
an excellent approximation.
\medskip
Note that even at $\phi = \vphi = {\pi \over 2}$ the correlation lengths
$\xi_{\Ga}$ and $\xi_{\si}$ are clearly different. Furthermore,
$\xi_{\si}$ coincides with its dual in the low-temperature phase
whereas $\xi_{\Ga}$ does not (see $\q{\han}$). This can be understood
in terms of the form factor decomposition to be presented in section
{\secFORM} (see $\q{\perturb}$).
\bn
\chapsubtitle{\secGG.\ Numerical computation of the correlation functions}
\mn
In this section we study the correlation function for the
operator $\Ga$ of the $\Zed_3$-chain numerically.
We will in particular check the relation (\secG.14) numerically.
\mn
In order to evaluate the correlation function (\secG.5) numerically
one needs the groundstate of the Hamiltonian (\secB.1).
The groundstate of the $\Zed_3$-chain is easily obtained for
general values of the parameters (even in the non-hermitian case
$\phi \in \Complex$)
using vector iteration up to $N=13$ sites. The more difficult
point is to calculate the matrix elements of $\Ga_{x+1}^{+} \Ga_1$
because this operator does not conserve momentum and thus does
not leave a space of momentum eigenstates invariant. It is
crucial to really work with $\Ga_{x+1}^{+} \Ga_1$ and {\it not}
to replace it by ${1 \over N} \sum_{r=1}^{N} \Ga_{x+r}^{+} \Ga_r$
because this would destroy the oscillation. Table 2 shows
the correlation function $C_\Ga(x)$ obtained in this
manner for $N=12$ and $N=13$ sites at $\phi = \vphi = {\pi \over 2}$,
$\la = {1 \over 2}$ which corresponds to Fig.\ 3.
This table also contains the perturbative results for the correlation
function $C_\Ga^{\rm pert.}(x)$ which were obtained in section {\secG}.
\def\sc{\scriptstyle}
\mn
\centerline{\vbox{
\hbox{
\vrule \hskip 1pt
\vbox{ \offinterlineskip
\def\tablespace{height2pt&\omit&\vl&\omit&&\omit&&\omit&&\omit&&
                          \omit&&\omit&&\omit&\cr}
\def\tablerule{ \tablespace
                \noalign{\hrule}
                \tablespace        }
\hrule
\halign{&\vrule#&
  \strut\hskip 2pt\hfil#\hfil\hskip 2pt\cr
\tablespace
& $x$  &\vl& \hskip 2pt 0 \hskip 2pt && 1 && 2 && 3 && 4 && 5 && 6 &
                                                      \cr \tablerule
&$C_{\Ga}^{\rm pert.}(x)$  &\vl& $1$ &&
          $\sc .18868+.07384 i$ &&
          $\sc .04561+.03980 i$ &&
          $\sc .00992+.01747 i$ &&
          $\sc .00091+.00674 i$ &&
          $\sc -.00088+.00263 i$ &&
          $\sc -.00074$ & \cr \tablerule
&$C_{\Ga}^{\rm 12,num.}(x)$   &\vl& $1$ &&
          $\sc .18881+.07385 i$ &&
          $\sc .04587+.03967 i$ &&
          $\sc .01004+.01737 i$ &&
          $\sc .00126+.00679 i$ &&
          $\sc -.00056+.00224 i$ &&
          $\sc -.00080$ & \cr \tablerule
&$C_{\Ga}^{\rm 13,num.}(x)$   &\vl& $1$ &&
          $\sc .18882+.07385 i$ &&
          $\sc .04588+.03967 i$ &&
          $\sc .01007+.01738 i$ &&
          $\sc .00132+.00684 i$ &&
          $\sc -.00043+.00242 i$ &&
          $\sc -.00063+.00058 i$ & \cr \tablespace
}
\hrule}\hskip 1pt \vrule}
\hbox{\quad Table 2: Perturbative results and numerical results
for $N=12$ and $N=13$ sites for the correlation}
\hbox{\quad \phantom{Table 2} function $C_{\Ga}(x)$
at $\phi = \vphi = {\pi \over 2}$, $\la = {1 \over 2}$.}}
}
\mn
The agreement between the results of both methods and also between
$N=12$ and $N=13$ sites is good. This
shows that on the one hand higher orders are indeed negligible in
(\secG.9) for $x < 7$ and on the other hand that the finite chain
length does not considerably affect the correlation function $C_{\Ga}(x)$.
Thus, one can take the values obtained e.g.\ at $N=12$ sites
as an approximation for the infinite chain limit as long as $x \le 6$.
\mn
Table 3 collects results of numerical calculations
at $N=12$ sites for the self-dual values $\phi=\vphi \in \{ {3 \pi \over 4},
{\pi \over 2}, {3 \pi \over 8} \}$ and a few integrable cases.
\mn
\centerline{\vbox{
\hbox{
\vrule \hskip 1pt
\vbox{ \offinterlineskip
\def\tablespace{height2pt&\omit&&\omit&&\omit&\vl&\omit&&\omit&&\omit&\vl&
                          \omit&&\omit&\cr}
\def\tablesp{height1pt&\omit&&\omit&&\omit&\vl&\omit&&\omit&&\omit&\vl&
                          \omit&&\omit&\cr}
\def\tablerule{ \tablespace
                \noalign{\hrule}
                \tablespace        }
\def\tablerul{ \tablespace
                \noalign{\hrule} \tablesp \noalign{\hrule}
                \tablespace        }
\hrule
\halign{&\vrule#&
  \strut\hskip 4pt\hfil#\hfil\hskip 4pt\cr
\tablespace
\tablespace
& $\vphi$ && $\phi$    && $\la$         &\vl&
             $\xi_{\Ga}$ && $L$     && $a$      &\vl&
               $P_{\min}$ && ${L P_{\min} \over 2 \pi}$
                                             & \cr \tablespace \tablerul
& ${3 \pi \over 4}$ && ${3 \pi \over 4}$ && ${1 \over 4}$ &\vl&
             $0.62(8)$   && $10(1)$ && $0.4(2)$ &\vl&
               $0.746$       && $1.18 \pm 0.12$ & \cr \tablespace
& ${3 \pi \over 4}$ && ${3 \pi \over 4}$ && ${1 \over 2}$ &\vl&
             $1.2(6)$    && $12(1)$ && $0.4(1)$ &\vl&
               $0.685^{\dag}$       && $1.34 \pm 0.14$ & \cr \tablespace
& ${3 \pi \over 4}$ && ${3 \pi \over 4}$ && ${3 \over 4}$ &\vl&
             $1.4(1)$    && $16(1)$ && $0.65(3)$ &\vl&
               $0.602^{\dag}$       && $1.54(8)$ & \cr \tablerule
& ${\pi \over 2}$   && ${\pi \over 2}$   && ${1 \over 4}$ &\vl&
             $0.6(1)$    && $14.05(6)$ && $0.44(8)$ &\vl&
               $0.470$       && $1.051(4)$ & \cr \tablespace
& ${\pi \over 2}$   && ${\pi \over 2}$   && ${1 \over 2}$ &\vl&
             $0.60(5)$   && $16(1)$ && $0.59(3)$ &\vl&
               $0.401$       && $1.07(8)$ & \cr \tablespace
& ${\pi \over 2}$   && ${\pi \over 2}$   && ${3 \over 4}$ &\vl&
             $1.3(2)$    && $29(6)^{*}$ && \omit &\vl&
               $0.189$       && $0.9(2)$ & \cr \tablerule
& ${3 \pi \over 8}$ && ${3 \pi \over 8}$ && ${1 \over 4}$ &\vl&
             $0.53(3)$   && $18(2)$ && $0.62(1)$ &\vl&
               $0.341$       && $1.0(1)$ & \cr \tablespace
& ${3 \pi \over 8}$ && ${3 \pi \over 8}$ && ${1 \over 2}$ &\vl&
             $0.8(2)$    && $25(1)^{*}$ && \omit &\vl&
               $0.283$       && $1.14(6)$ & \cr \tablespace
& ${3 \pi \over 8}$ && ${3 \pi \over 8}$ && ${3 \over 4}$ &\vl&
             $1.8(6)$    && $50 \pm 23^{*}$ && \omit &\vl&
               $0.211$       && $1.7(8)$ & \cr \tablerul
& ${3 \pi \over 8}$ && $-i \, 0.98942$   && ${1 \over 4}$ &\vl&
             $0.7(2)$   && $\infty$ && \omit &\vl&
               $0$           && \omit & \cr \tablerule
& ${3 \pi \over 8}$ && $0.69919$         && ${1 \over 2}$ &\vl&
             $1.1(3)$    && $49(7)^{*}$ && \omit &\vl&
               $0.159$       && $1.2(2)$ & \cr \tablerule
& ${3 \pi \over 4}$ && $\pi + i \, 1.70004$ && ${1 \over 4}$ &\vl&
             $1.0(5)$    && $5.9(1)$ && $0.3(1)$ &\vl&
               ${\pi \over 3}$ && $0.99(2)$ & \cr \tablespace
}
\hrule}\hskip 1pt \vrule}
\hbox{\quad Table 3: Parameters for the correlation function
            (\secG.10a)}
\hbox{\quad \phantom{Table 3:}
            calculated numerically for $N=12$ sites. }}
}
\mn
The real part of the correlation function $C_\Ga$ has no zero in the interval
[0,6] for the entries marked with a~`${}^{*}$'. Therefore, the procedure
of section {\secG} could not be used for the computation of $L$. Here $L$ has
instead been obtained by the formula
$$L={1 \over 4} \sum_{x=1}^4 {2 \pi \over {\rm Im}\left(\ln\left(
          {C_\Ga(x) \over C_\Ga(x+1)}
               \right)\right)} \,  .
 \eqno({\rm \secGG.1})$$
The values of $P_{\min}$ in table 3 have been obtained by
first calculating $\EE_{1}(P)$ numerically at $N=12$
sites and afterwards minimizing the finite fourier decomposition
of $\left( \EE_{1}(P) \right)^2$ numerically.
This method for estimating $P_{\min}$ failed for those
values marked with a `${}^{\dag}$' because here $\EE_{1}(P)$
partly crosses two-particle scattering states.
$\EE_{1}(P)$ becomes even negative for
$\phi = \vphi = {3 \pi \over 4}$, $\la = {3 \over 4}$. This implies
that the groundstate does not belong to the $Q=P=0$ sector and
this point in the parameter space lies in the massless
incommensurate phase. However, we have ignored this fact and used
the lowest energy state in the $Q=P=0$ sector for evaluation
of the correlation functions. This is justified because it yields
results that are similar to those for the other values of the
parameters.
The values of $P_{\min}$ marked with a `${}^{\dag}$' have been determined
by first determining the smallest energy gap in the $Q=1$ sector at
{\it finite} $N  = 8, \ldots, 12$. For $8 \le N \le 12$ the minimum
of the discretized dispersion relation is located at $P={2 \pi \over N}$.
Next, a polynomial interpolation between these five values for the
energy gap was minimized numerically.
\mn
That the above procedures yield reasonable fits was demonstrated by
a figure in $\q{\lett}$.
\mn
We have also included a few values for integrable points at the end of
table 3, i.e.\ points where the parameters satisfy
$\cos \vphi = \lambda \cos \phi$. For
$\vphi = {3 \pi \over 8}$, $\la = {1 \over 2}$, the integrable case is
hermitean and behaves precisely like the self-dual case $\phi = \vphi$.
For small values of $\la$ one has to choose one of the angles $\vphi$,
$\phi$ complex in order to satisfy the integrability-condition (\secB.13).
We choose $\phi$ complex: For $\vphi < {\pi \over 2}$ we
choose purely complex $\phi \in i \Real$ whereas for $\vphi > {\pi \over 2}$
we take $\phi - \pi \in i \Real$ (compare also $\q{\yildirim}$). For complex
choices of $\phi$ the Hamiltonian becomes non-hermitean such that the
excitation energies become complex. Thus, $P_{\min}$ is not strictly the
minimum of the dispersion relations but instead a point of symmetry -- compare
$\q{\yildirim,\perturb,\lett}$ and in particular section {\secFORM}
below.
\mn
The agreement with the prediction $L P_{\min} = 2 \pi$ is good for all
values in table 3 bearing in mind that we have ignored systematic errors.
This applies also to the point $\phi = \vphi =
{3 \pi \over 4}$, $\la = {3 \over 4}$ in the massless incommensurate
phase because here the estimate for the error of $L$ is probably too small.
\bn
\chapsubtitle{\secIOPE.\ Correlation functions in the low-temperature regime}
\sn
In this section we apply the method explained in section {\secG}
to the correlation functions in the low-temperature phase of the
$\Zed_3$-chiral Potts quantum chain. Note that the duality argument
of section {\secIOPD} applies only to the Hamiltonian and not to other
operators. Thus, quantities like e.g.\ correlation lengths may be different
in these two phases.
\sn
We study again the correlation functions $C_{\Ga}(x)$ and $C_{\si}(x)$
as defined by (\secG.2) and (\secG.5)
using a low-temperature expansion for the ground state
from the state $\GS{Q}$ (\secIOPD.2).
The expansion of the ground state
in powers of $\tilde{\la}$ leads to an expansion of the correlation
functions in powers of $\tilde{\la}$:
$$C_{\Xi}(x) = \sum_{k=0}^{\infty} \tilde{\la}^k \Ct_{\Xi}^{(k)}(x).
    \eqno({\rm \secIOPE.1})$$
\indent
Below, we will first give the final results for general angles
$\phi$, $\vphi$.
We then specialize to the superintegrable case
$\phi = \vphi = {\pi \over 2}$ and calculate even higher orders.
By looking for a good fit we try to guess the structure of the
correlation functions -- as before. With this experience we turn back
to the general case and discuss how the correlation functions
should change for general $\phi$, $\vphi$.
\medskip
In order to save space we present only the final results
for the correlation functions. For $C_{\Ga}(x)$ one obtains,
using the abbreviations $\cabt = \bcos{{\phi \over 3}}$,
$\Rabt = 1 - 4 \cabt^2$:
$$\eqalign{
\Ct_{\Ga}^{(0)}(x) =& 1 \ , \qquad \qquad \qquad \qquad \qquad
\Ct_{\Ga}^{(1)}(x) = 0 \ , \cr
\Ct_{\Ga}^{(2)}(x) =& {1 \over 6 \cabt^2} \left\{ \delta_{x,0} - 1 \right\}
          \ , \qquad \qquad
\Ct_{\Ga}^{(3)}(x) = {\cos{\vphi} \over 18 \cabt^3}
           \left\{ \delta_{x,0} - 1 \right\} \ , \cr
\Ct_{\Ga}^{(4)}(x) =& \cr
       {1 \over 27 \cabt^2} {\Biggl\{} &
           (1 - \delta_{x,0}) \left( { 2 ( 1 - 16 \cabt^2) \over 3 \Rabt^2}
                  + {1 \over 16 \cabt^2} \right)
                   + \delta_{x,1} \left(
   {1 + 2 \cabt^2 - 3 \ i \ \bsin{{2 \phi \over 3}} \over 3 \Rabt^2}
                    + {5 \over 16 \cabt^2} \right)
                      {\Biggr\}}. \cr
}   \eqno({\rm \secIOPE.2})$$
The first orders of the correlation function $C_{\si}(x)$ read
as follows:
$$\eqalign{
\Ct_{\si}^{(0)}(x) =& \delta_{x,0} \ , \qquad \qquad \qquad \qquad \qquad
                                     \qquad \quad \hskip 5pt
\Ct_{\si}^{(1)}(x) = 0 \ , \cr
\Ct_{\si}^{(2)}(x) =& -{1 \over 9} {\Biggl\{} {\delta_{x,0} \over \cabt^2}
                               + \delta_{x,1} \left(
                           {2 \over \Rabt}
                         + {1 \over 2 \cabt^2} \right) {\Biggr\} } \ , \quad
\Ct_{\si}^{(3)}(x) = -{\cos{\vphi} \over 9 \cabt } {\Biggl\{}
                           {\delta_{x,0}  \over 2 \cabt^2}
                      + {\delta_{x,1} \over \Rabt} {\Biggr\} } \ , \cr
\Ct_{\si}^{(4)}(x) =&  \cr
   {1 \over 81 \cabt^2 } {\Biggl\{} \delta_{x,0} & \left(
                              {7 \over 16 \cabt^2 }
                             +{8 \over  \Rabt} \right)
                         +  \delta_{x,1} \left( {15 \over 16 \cabt^2 }
                            + {29 \over 12 \Rabt }
                            - {16  \cabt^4 \over \Rabt^3} \right)
                         + \delta_{x,2} \left(
                            - {1 \over 8 \cabt^2 }
                            + {7 + 44 \cabt^2 \over 12 \Rabt^2}
                        \right) {\Biggr\} }. \cr
}   \eqno({\rm \secIOPE.3})$$
Note that the correlation functions (\secIOPE.2) and (\secIOPE.3) do not
depend on the charge sector.
\medskip
Eq.\ (\secIOPE.2) suggests that the correlation function $C_{\Ga}(x)$ tends
to a non-zero constant for large distances $x$ -- in contrast to the
correlation functions in the high-temperature phase so
that that the low-temperature phase is ordered over long ranges.
However, beyond this general conclusion, it is difficult to guess from
(\secIOPE.2) or (\secIOPE.3) what might be the behaviour even for small $x$.
Thus, we set $\phi = \vphi = {\pi \over 2}$, calculate four
further orders and obtain
$$\eqalign{
\Ct_{\Ga}^{(0)}(x) =& 1 \ , \qquad \qquad \qquad \qquad
\Ct_{\Ga}^{(1)}(x) =
\Ct_{\Ga}^{(3)}(x) =
\Ct_{\Ga}^{(5)}(x) =
\Ct_{\Ga}^{(7)}(x) = 0 \ , \cr
\Ct_{\Ga}^{(2)}(x) =& {2 \over 9} \left( \delta_{x,0} - 1 \right) \ , \cr
\Ct_{\Ga}^{(4)}(x) =& {7 \over 81} \left( \delta_{x,0} - 1 \right)
                   + \delta_{x,1} {5 - i \sqrt{3} \over 162} \ , \cr
\Ct_{\Ga}^{(6)}(x) =& {1 \over 6561} \left\{
                        336 \left(\delta_{x,0} - 1 \right)
                   + 160 \delta_{x,1} + 60 \delta_{x,2} \right\}
                   - i {\sqrt{3} \over 6561} \left\{
                     26 \delta_{x,1} + 20 \delta_{x,2} \right\} \ , \cr
\Ct_{\Ga}^{(8)}(x) =& {1 \over 354294} \left\{
                        12600 \left(\delta_{x,0} - 1 \right)
                   + 6852 \delta_{x,1} + 3521 \delta_{x,2}
                   + 1225 \delta_{x,3} \right\} \cr
                  &- i {\sqrt{3} \over 354294} \left\{
                     960 \delta_{x,1} + 995 \delta_{x,2}
                   + 525 \delta_{x,3} \right\} \cr
}   \eqno({\rm \secIOPE.4})$$
and
$$\eqalign{
\Ct_{\si}^{(0)}(x) =& \delta_{x,0} \ , \qquad \qquad \qquad \qquad
\Ct_{\si}^{(1)}(x) =
\Ct_{\si}^{(3)}(x) =
\Ct_{\si}^{(5)}(x) =
\Ct_{\si}^{(7)}(x) = 0 \ , \cr
\Ct_{\si}^{(2)}(x) =& {1 \over 27} \left\{ -4 \delta_{x,0}
                               + \delta_{x,1} \right\} \ , \cr
\Ct_{\si}^{(4)}(x) =& {1 \over 729}\left\{
                        - 41 \delta_{x,0} + 14 \delta_{x,1} + 8 \delta_{x,2}
                    \right\} \ , \cr
\Ct_{\si}^{(6)}(x) =& {1 \over 19683}\left\{
                - 586 \delta_{x,0} + 147 \delta_{x,1} + 126 \delta_{x,2}
                     + 80 \delta_{x,3} \right\} \ , \cr
\Ct_{\si}^{(8)}(x) =& {1 \over 531441}\left\{
                - 9927 \delta_{x,0} + 2130 \delta_{x,1} + 1721 \delta_{x,2}
                   + 1728 \delta_{x,3}  + 910 \delta_{x,4} \right\} . \cr
}   \eqno({\rm \secIOPE.5})$$
Up to the order calculated, $C_{\si}(x)$ is real for all values
of the parameters $\phi$, $\vphi$, $\tilde{\la}$.
$C_{\Ga}(x)$, in contrast, has a non-vanishing imaginary part.
By analogy to the high-temperature regime and from the results
in $\q{\krallm,\gehlenph}$ we expect that $C_{\Ga}(x)$ is
oscillating. Indeed, the correlation functions in the superintegrable case
$\phi = \vphi = {\pi \over 2}$ can nicely be fitted by
$$\eqalignno{
C_{\si}(x) &= a \delta_{x,0} + b e^{-{x \over \xi_{\si}}} \ ,
                                             &({\rm \secIOPE.6a})\cr
C_{\Ga}(x) &= m^2 +
      p e^{-\left({1 \over \xi_{\Ga}} + {2 \pi i \over L}\right) x}.
                                             &({\rm \secIOPE.6b})\cr
}$$
For $\tilde{\la} \in \left\{ {1 \over 4}, {1 \over 2}, {3 \over 4} \right\}$
good fits to (\secIOPE.4) and (\secIOPE.5) using (\secIOPE.6) are given by
the values in table 4.
\mn
\centerline{\vbox{
\hbox{
\vrule \hskip 1pt
\vbox{ \offinterlineskip
\def\tablespace{height2pt&\omit&\vl&\omit&&\omit&&\omit&\vl&\omit&&
                          \omit&&\omit&&\omit&\vl&\omit&&\omit&\cr}
\def\tablerule{ \tablespace
                \noalign{\hrule}
                \tablespace        }
\hrule
\halign{&\vrule#&
  \strut\hskip 2pt\hfil#\hfil\hskip 2pt\cr
\tablespace
\tablespace
&$\tilde{\la}$ &\vl& $\xi_{\si}$ && $a$       && $b$      &\vl&
            $\xi_{\Ga}$ && $m^2$       && $L$  && $p$ &\vl&
                          $\xi_{\Ga}^{{\rm num.}}$ && $L^{{\rm num.}}$
                          & \cr \tablespace \tablerule
& $0.25$       &\vl& $0.26(1)$   && $0.891(9)$ && $0.099(9)$ &\vl&
            $0.24(3)$   && $0.9857605$ && $30 \pm 18$ &&
                      $0.011(3)$ &\vl& $0.26(3)$ && $27 \pm 7$
                          & \cr \tablespace
& $0.50$       &\vl& $0.41(2)$   && $0.841(7)$ && $0.118(7)$ &\vl&
            $0.37(5)$   && $0.9381$    && $30 \pm 15$ &&
                      $0.05(1)$  &\vl& $0.42(6)$ && $30 \pm 8$
                          & \cr \tablespace
& $0.75$       &\vl& $0.58(5)$   && $0.74(1)$ && $0.16(1)$ &\vl&
            $0.50(7)$   && $0.832$     && $30 \pm 13$ &&
                      $0.15(2)$  &\vl& $0.6(1)$  && $28 \pm 4$
                          & \cr \tablespace
}
\hrule}\hskip 1pt \vrule}
\hbox{\quad Table 4: Parameters for the correlation functions
            (\secIOPE.6) at $\phi = \vphi = {\pi \over 2}$. }}
}
\mn
First, we remark that the correlation lengths satisfy
$\xi_{\si} = \xi_{\Ga} =: \xi$ for all values
of $\tilde{\la}$ within the numerical accuracy. In fact, one
expects this equality because the correlation lengths should be
the inverses of some mass scale, and there is only one mass scale
in our problem because all three charge sectors are degenerate.
Furthermore, we observe that our data is
compatible with an oscillating correlation function for
the operator $\Ga$. The oscillation length $L$ (or wave vector)
is around $30$ sites in a major part of the low-temperature phase.
In $\q{\krallm,\gehlenph}$ it has been predicted that $L$
should diverge as $\tilde{\la}$ crosses the phase boundary
and approaches $\tilde{\la} = 1$ where the critical exponent
is expected to equal ${2 \over 3}$. Our results are compatible
with a divergent oscillation length at $\tilde{\la} = 1$
although due to the large errors we do not even see that
$L$ increases with $\tilde{\la}$.
\sn
In addition to the perturbation expansions we can also perform numerical
evaluations of the correlation function $C_{\Ga}(x)$. Since we can use
vector iteration for up
to $N=13$ sites implying a distance of up to $x=6$ one might hope to obtain
more accurate results with this method. That this is not really the case is
demonstrated by the values $\xi_{\Ga}^{{\rm num.}}$ and $L^{{\rm num.}}$
which we have included into table 4. The main problem with numerical
calculations is the numerical accuracy: In particular, the evaluation of
the constant contribution $m^2$ is not free of errors. On the other hand
the short correlation lengths strongly damp the non-constant part of the
correlation functions: For example $C_{\Ga}(x) - m^2$ has already
decreased by 10 orders of magnitude at $x=6$ for $\tilde{\la} = {1 \over 4}$.
Therefore, for our estimates we have used only $x \le 4$ for $\tilde{\la}
= {1 \over 4}$ and $x \le 5 $ for $\tilde{\la} = {1 \over 2}$ and
$\tilde{\la} = {3 \over 4}$.
\sn
Thus, it will be very
difficult to obtain more precise results from approximative
arguments and an exact expression for $C_{\Ga}(x)$ is
probably needed in order to decide whether (\secIOPE.6b) really
is the correct form and to determine the wave vector $L$
accurately.
\sn
Before we conclude the discussion of the correlation functions
for the superintegrable chiral Potts model, we mention that a conjecture
for the form of $C_{\Ga}(x)$ has been formulated in $\q{\mccoyadv}$:
$C_{\Ga}(x) = m^2 + \O ( e^{-{x \over \xi_{\Ga}}})$
where $m$ is the order parameter. Our result (\secIOPE.6) is compatible
with this from. In $\q{\hkn}$ the conjecture for the order parameter
$$m = {\avact \Ga_x \vact \over \normvact} =
      \left( 1 - \tilde{\la}^2 \right)^{1 \over 9}
   \eqno({\rm \secIOPE.7})$$
has been formulated, but (\secIOPE.7) has not been proven yet.
The constant term in (\secIOPE.4) is in exact agreement with
(\secIOPE.6) and (\secIOPE.7) up to the order calculated,
such that we may assume that at least the constant term of $C_{\Ga}(x)$
is now known exactly.
\medskip
A few remarks on the choice of ground state (\secIOPD.2) are
in place because  in $\q{\hkn,\yang}$ (\secIOPE.7) has actually been
derived considering an expectation value of the operator $\Ga_x$.
We have already pointed out that the one point functions of $\Ga_x$
vanish identically due to charge conservation if one uses the charge
eigenstates (\secIOPD.2). However, if one uses instead non-charge
eigenstates like $\state{0 \ldots 0}$ for a perturbative expansion
of $\vact$ they do not vanish. Indeed, using an expansion for $\vact$
from $\state{0 \ldots 0}$ we once again verified equality of this
one point function with the order parameter $m$. If we redefine
$\bar{C}_{\Ga}(x)$ by replacing $\vac$ by $\vact$ and
subtracting the contribution from the one point functions, this
is in fact the only change, i.e.\ $\bar{C}_{\Ga}(x)=
C_{\Ga}(x)-m^2$. $C_{\si}(x)$ remains unchanged under this redefinition.
\medskip
For more general values of the angles $\phi$, $\vphi$ one
expects the correlation functions to be also of the form (\secIOPE.6) --
of course with different values of the parameters. We can
see from the constant term $m^2$ of the correlation function $C_{\Ga}(x)$
(\secIOPE.2) that it will not be of the form (\secIOPE.7) for general
$\phi \ne {\pi \over 2} \ne \vphi$. In general, the coefficient
of $\tilde{\la}^3$ for the constant term does not vanish and
$m^2$ does not even have an expansion in powers of $\tilde{\la}^2$.
Among the powers that we have calculated for the general case
only the fourth order in (\secIOPE.2) has a non-vanishing imaginary
part at $x=1$. Under the assumption that (\secIOPE.6b) is the general
form we would expect the imaginary part at $x=1$ to be proportional
to $\bsin{{2 \pi \over L}}$ for very small temperatures $\tilde{\la}$.
Thus, we expect for very low temperatures $\tilde{\la}$ the relation
$L^{-1} \sim \phi$. On the one hand, this explains the conjectured
presence of a second length scale $L$ in addition to the correlation
length $\xi$. The oscillation length $L$ just comes from the chiral
angle $\phi$ and thus these two scales must be related to each other.
On the other hand, the oscillation (should it really be present)
will vanish smoothly as the parity conserving Potts
case $\phi = \vphi = 0$ is approached.
\bn
\chapsubtitle{\secFORM.\ Form factors and correlation functions}
\mn
In this section we show how to use form factors in order to determine the
oscillation length $L$ exactly using symmetries of the Hamiltonian
arising for particular values of the parameters.
\mn
One can use the quasiparticle picture which we have discussed earlier
in order to rewrite a correlation function $C_{\Xi}(x)$ as follows:
$$\eqalign{
C_{\Xi}(x)&={\sum_{n=0}^{\infty} \int_0^{2 \pi}
               \left(\prod_{i=1}^n {\rm d}p_i\right) \,
               \avac \Xi_{x+1}^{+} \state{p_1,\ldots,p_n}
               \astate{p_1,\ldots,p_n} \Xi_1 \vac
               \over \normvac}
         - {\abs{\avac \Xi_1 \vac}^2 \over \normvac^2} \cr
    &= \sum_{n=1}^{\infty} \int_0^{2 \pi}
        \left(\prod_{i=1}^n {\rm d}p_i\right) \,
        e^{i x \left(\left(\sum_{j=1}^n p_j \right) - P_{\vac}\right)} \,
                 { \abs{\astate{p_1,\ldots,p_n} \Xi_1 \vac}^2
                 \over \normvac} \cr
}    \eqno({\rm \secFORM.1})$$
where we have inserted a complete set of normalized $n$-particle states
$\state{p_1,\ldots,p_n}$. Representations similar to (\secFORM.1) have been
used in quantum field theory for a long time (see e.g.\ $\q{\lehmann}$)
and are well-known to be useful for the evaluation of correlation functions
of statistical models (see e.g.\ $\q{\mussardo}$). According to (\secFORM.1)
one could compute the correlation function $C_{\Xi}(x)$ by computing its
`{\it form factors}' $\astate{p_1,\ldots,p_n} \Xi_1 \vac$, but one can even
derive interesting results without doing so.
Clearly, if the groundstate $\vac$ has non-zero momentum $P_{\vac} \ne 0$ we
expect an oscillatory contribution to the correlation function.
However, one can read off from (\secFORM.1) that an oscillatory contribution
is also to be expected if $P_{\vac} = 0$ but the model breaks parity which
precisely applies to the massive high-temperature phase of the
chiral Potts model. The correlation functions of massive models
in general have an exponential decay, i.e.\
$C_{\Xi}(x) = e^{-{x \over \xi}} f_{\Xi}(x)$ where $f_{\Xi}(x)$ is
some bounded function. According to (\secFORM.1) we also expect an
oscillatory contribution of the form $e^{i {2 \pi x \over L}}$.
In summary, we expect correlation functions of the approximate form
$$C_{\Xi}(x) \sim  e^{-{x \over \xi}+i {2 \pi x \over L}}.
    \eqno({\rm \secFORM.2})$$
$\xi$ is called `correlation length' and $L$ is the `oscillation length'
($L^{-1}$ is the `wave vector'). Eq.\ (\secFORM.2) explains the fits
(\secG.10) and (\secIOPE.6) that we have used earlier.
More precisely, for the $\Zed_n$-chiral Potts model the operator
$\Ga_1$ creates $Q=1$-single-particle excitations from the groundstate.
The dispersion relations of these particles clearly violate parity.
Therefore we expect that $C_{\Ga}(x)$ is of the form (\secFORM.2).
The action of the operator $\si_1$ is much less spectacular.
In particular, it leaves the charge sector $Q=0$ invariant and thus
it need not necessarily have an oscillatory contribution. In fact,
from (\secG.3) we see that $C_{\si}(x)$ should be real which in
view of (\secFORM.2) implies the absence of oscillations.
\mn
For massive perturbations of conformal field theories it is well-known that
form factor expansions already for few particles yield excellent
approximations (see e.g.\ $\q{\mussardo}$). In order to gain some
intuition we perform some simple computations for the form factors of the
operator $\Ga$ for the $\Zed_3$-chain.
\sn
First, considering perturbation series for the states one sees that
$$\astate{p_1, \ldots, p_r} \Ga_1 \vac = \O(\la^{[{r \over 2}]}).
      \eqno({\rm \secFORM.3})$$
Next we compute the states $\vac$ and $\state{p_1}$ from perturbation
series of (\secD.1) and (\secD.2) respectively. One finds that up to
order $\la$ the states are already normalized properly, i.e.\
$$\normvac = 1 + \O(\la^2) \, , \qquad \qquad
  \normstate{p_1} = 1 + \O(\la^2) \, .
      \eqno({\rm \secFORM.4})$$
Finally one finds the single-particle form factor for $n=3$ up to order
$\la$:
$$\astate{p_1} \Ga_1 \vac = {1 \over \sqrt{\int 1 {\rm d} p}} \left\{
          1 + \la {\bcos{p_1 - {\phi \over 3}} \over 3 \bcos{{\vphi \over 3}} }
            + \O(\la^2) \right\} \, .
      \eqno({\rm \secFORM.5})$$
Comparing (\secFORM.5) with (\secD.7) up to order $\la$ one sees that
the maximum of the form factor is obtained when $\EE_1(p_1)$ has a minimum
and vice versa. Furthermore, multi-particle states have higher energy
than single-particle states and due to (\secFORM.3) the contributions of
the form factors decrease with the number of particles for small $\la$.
In summary, one obtains the qualitative statement that the contributions
of the form factors decrease with increasing energy.
\sn
According to (\secFORM.3) the single-particle form factor (\secFORM.5)
is the only one contributing to $C_{\Ga}(x)$ up to order $\la$. Indeed upon
inserting (\secFORM.5) into (\secFORM.1) one recovers (\secG.7) up
to first order in $\la$ -- as it should be. Note however that the direct
computation of (\secG.7) is simpler than the one using a form factor
expansion.
\medskip
Symmetries of the Hamiltonian translate into symmetries of the form
factors. In certain cases these symmetries are already sufficient to
compute the oscillation length $L$ as we will see now.
\mn
\beginddef{\DefB}{Parity operator}
The parity operator $\Parity$ is defined by the following action on the
states (\secB.6):
$$r(\Parity) \state{i_1 \ldots \ldots i_N}
            = \state{i_1 i_N i_{N-1} \ldots \ldots i_2} \, .
  \eqno{(\rm \secFORM.6)}$$
\enddef
\beginremark{}
$\Parity \si_{1+x} \Parity = \si_{1-x}$, $\Parity \Ga_{1+x} \Parity
= \Ga_{1-x}$.
\endremark
\mn
The following lemma is based on results obtained via numerical
diagonalization in $\q{\yildirim}$ which showed that the Hamiltonian
(\secB.1) has exact symmetries for particular values of the parameters:
\mn
\beginlemma{\LemA}
Denote the restriction of the Hamiltonian $H_N^{(n)}$ in eq.\ (\secB.1)
to the spaces with momentum $P$ and charge $Q$ by `$H_N^{(n)}(P,Q)$'.
Then one has the following identities (see also \q{\yildirim}):
$$\eqalign{
\a_k = \a_{n-k}
   \quad & \Rightarrow \quad \Parity H_N^{(n)}(P,Q) \Parity
       = H_N^{(n)}(-P,Q) \, ,\cr
\ab_k^{*} = \ab_{n-k} \hbox{ and } \a_k \in \Real
   \quad & \Rightarrow \quad \Parity H_N^{(n)}(P,Q) \Parity =
                  \left(H_N^{(n)}(-P,Q)\right)^{+} \, , \cr
\ab_k^{*} = \ab_{n-k} \hbox{ and } \a_k^{*} = e^{-{2 \pi i z k}} \a_k
   \quad & \Rightarrow \quad \Parity H_N^{(n)}(P_{{\rm m},Q} + P,Q) \Parity =
                  \left(H_N^{(n)}(P_{{\rm m},Q}-P,Q)\right)^{+} \, \cr
}  \eqno{(\rm \secFORM.7)}$$
where the symmetry of the last line holds for any $Q$ that is invertible
in $\Zed_n$ and those $P_{{\rm m},Q}$ satisfying
$P_{{\rm m},Q} Q^{-1} + \pi z \equiv 0$ mod $\pi$ as well as
$e^{i 2 P_{{\rm m},Q}}$ being an $n$th root of unity.
For very particular values of the momentum $P$ and the chain length $N$ some
projection is necessary in (\secFORM.7).
\endlemma
\mn
The first two lines of (\secFORM.7) follow immediately by looking at
$\Parity H_N^{(n)} \Parity$, keeping in mind that the translation operator
defined in (\secB.10) satisfies $\Parity T_N \Parity = T_N^{-1} = T_N^{+}$.
The derivation of the third line of (\secFORM.7) is more complicated and
relies on a suitable choice of basis. Because of its technicality we have
shifted it to appendix {\appB}.
\mn
\beginremarks
\item{1)}
For the last line of (\secFORM.7) insert the parameterization (\secB.12) and
set $z = {2 \over n}$. Then $P_{{\rm m},Q} = \pi (1 - {2 Q \over n})$ is
a solution to $P_{{\rm m},Q} Q^{-1} + {2 \pi \over n} = 0$ lying in the
interval $\lbrack -\pi,\pi \rbrack$ -- the other solution is shifted
by $\pi$. The solution $P_{{\rm m},Q} = \pi (1 - {2 Q \over n})$
corresponds to the minimum (\secD.4) of the dispersion relation of the
single-particle state in this charge sector.
\item{2)}
The restriction to $Q$ invertible in $\Zed_n$ (those $Q$ whose greatest
common divisor with $n$ is 1) was necessary for technical reasons
but there are indications that lemma {\LemA} is true without this
restriction. Combining this with remark 1) we conclude that
the relation $P_{{\rm m},Q} = \pi (1 - {2 Q \over n})$ seems to
be valid for general $Q$. In fact, close inspection of the argument
in appendix {\appB} shows that the assumptions can be weakened in order
to cover e.g.\ also the case $Q=2$ for $n=4$.
\item{3)}
In the generic case (which is the only case we are really interested in)
one may ignore the projection operators in the equality (\secFORM.7).
\mn
\beginttheorem{\ThB}{Oscillation length \q{\perturb}}
Let the Hamiltonian $H(P,Q)$ restricted to momentum and charge eigenspaces
with eigenvalues $P$ and $Q$ have one of the following symmetries:
$$\Parity H(P_{{\rm m}, Q}+P,Q) \Parity = H(P_{{\rm m}, Q}-P,Q)
  \qquad \hbox{or} \qquad
  \Parity H(P_{{\rm m}, Q}+P,Q) \Parity = \left(H(P_{{\rm m},
Q}-P,Q)\right)^{+}
    \eqno({\rm \secFORM.8})$$
with some $P_{{\rm m}, Q}$ depending on the charge sector $Q$. Assume
furthermore that $P_{\vac} = 0$ and that $\Xi_1 \vac$ has charge $Q$.
Then the oscillation length $L$ of the correlation function $C_{\Xi}(x)$
satisfies
$$L P_{{\rm m}, Q} = 2 \pi \, .
    \eqno({\rm \secFORM.9})$$
\endTheorem
\mn
\beginproof{}
We start from the form factor expansion (\secFORM.1)
which in the present case becomes
$$C_{\Xi}(x)= \sum_{r} \int_0^{2 \pi} {\rm d}P \,
        e^{i P x } \, { \abs{\astate{P,Q;r} \Xi_1 \vac}^2
                 \over \normvac}
    \eqno({\rm \secFORM.10})$$
where we have only written the quantum numbers $P$ and $Q$ explicitly
and comprised the other ones in the label `$r$'. First we observe
that $\Parity \Xi_1 \Parity = \Xi_1$. If the Hamiltonian satisfies
$\Parity H(P_{{\rm m}, Q}+P,Q) \Parity = H(P_{{\rm m}, Q}-P,Q)$, then
eigenstates of momentum $P_{{\rm m}, Q}+P$ are mapped under parity to
eigenstates of momentum $P_{{\rm m}, Q}-P$. This means that
$\astate{(P_{{\rm m}, Q}+P),Q;r} \Xi_1 \vac =
\astate{(P_{{\rm m}, Q}-P),Q;r} \Xi_1 \vac$. If the symmetry involves
the adjoint of the Hamiltonian one finds
$\astate{(P_{{\rm m}, Q}+P),Q;r} \Xi_1 \vac =
\astate{(P_{{\rm m}, Q}-P),Q;r} \Xi_1 \vac^{*}$. Thus, the following
identity is valid in both cases:
$$\abs{ \astate{(P_{{\rm m}, Q}+P),Q;r} \Xi_1 \vac}^2 =
\abs{\astate{(P_{{\rm m}, Q}-P),Q;r} \Xi_1 \vac}^2 \, .
    \eqno({\rm \secFORM.11})$$
Now we turn back to the form factor expansion (\secFORM.10):
$$\eqalign{
C_{\Xi}(x) &= \sum_{r} \left\{
        \int\limits_{P_{{\rm m}, Q}}^{P_{{\rm m}, Q} + \pi} {\rm d}P \,
        e^{i P x } \, { \abs{\astate{P,Q;r} \Xi_1 \vac}^2 \over \normvac}
     +  \int\limits_{P_{{\rm m}, Q} - \pi}^{P_{{\rm m}, Q}} {\rm d}P \,
        e^{i P x } \, { \abs{\astate{P,Q;r} \Xi_1 \vac}^2 \over \normvac}
     \right\} \cr
&= \sum_{r} \int\limits_{0}^{\pi} {\rm d}P \, \left\{
        e^{i (P_{{\rm m}, Q}+P) x } \,
        { \abs{\astate{(P_{{\rm m}, Q}+P),Q;r} \Xi_1 \vac}^2 \over \normvac}
   \right. \cr & \qquad \qquad \qquad \left.
     +  e^{i (P_{{\rm m}, Q}-P) x } \,
        { \abs{\astate{(P_{{\rm m}, Q}-P),Q;r} \Xi_1 \vac}^2 \over \normvac}
     \right\} \cr
&= e^{i P_{{\rm m}, Q} x} \sum_{r}
        \int\limits_{0}^{\pi} {\rm d}P \, 2 \cos(P x) \,
        { \abs{\astate{(P_{{\rm m}, Q}+P),Q;r} \Xi_1 \vac}^2 \over \normvac}
         \cr
}    \eqno({\rm \secFORM.12})$$
where the last equality follows from (\secFORM.11).
This shows that $C_{\Xi}(x)$ is of the form
$$C_{\Xi}(x) = e^{{2 \pi i x \over L}} f(x)
    \eqno({\rm \secFORM.13})$$
with $L$ satisfying (\secFORM.9) and $f(x)$ is given by the remaining
integral in (\secFORM.12) which is clearly real.
\endproof
\mn
\beginremarks
\item{1)}
Note that theorem {\ThB} is true for more general Hamiltonians $H(P,Q)$, but
it covers in particular the case (\secFORM.7) for the $\Zed_n$-chiral Potts
model.
\item{2)}
Note that if the Hamiltonian has several different $P_{{\rm m},Q}$ such
that (\secFORM.8) holds (which applies to (\secFORM.7)) one obtains different
expressions for $C_{\Xi}(x)$ involving different $L$ and $f(x)$. We have
argued before that form factors are maximal for smallest energy. Therefore
one may expect to obtain the representation with $f(x) > 0$ on $x \in [0, y]$
with maximal $y$ choosing the value $P_{{\rm m},Q}$ that corresponds
to the {\it minimum} of the dispersion relation of the fundamental
particle.
\endremark
\mn
{}From lemma {\LemA} and theorem {\ThB} we immediately obtain the following
\mn
\begincorollary
$$C_{\Ga_{}^Q}^{}(x) = e^{2 \pi i x \over L} f_{Q,r}(x)
    \eqno({\rm \secFORM.14a})$$
with
$$\eqalign{
f_{Q,r}(x) \in \Real & \qquad \forall x \, , \cr
L = \infty & \qquad \hbox{for } \phi = r \pi, r \in \Zed \ \hbox{ or } \
                                \vphi \in \Real, {\rm Re}(\phi) = 0 \, , \cr
L = {2n \over n-2Q} \, & \qquad \hbox{for }\vphi\in\Real, {\rm Re}(\phi)=\pi
                      \hbox{ and } 0 < Q < n \, . \cr
}    \eqno({\rm \secFORM.14b})$$
\endcorollary
\vfill
\eject
\rightline{\helvits Il libro della natura \'e scritto in lingua matematica}
\sn
\rightline{\helvits Galileo Galilei}
\bn
\chaptitle{\secZ.\ Extended conformal algebras}
\mn
Symmetries play a fundamental r\^ole in theoretical physics because they
help to simplify a problem considerably or sometimes they even determine
the relevant quantities completely. For example, the symmetry group $SO(4)$
of the quantum mechanical central force problem allows to determine the
spectrum of the hydrogen atom by purely algebraic computations. In particular,
for quantum field theories it is desirable to have a sufficiently large
symmetry because they can be solved directly in only very few cases.
Such a situation does indeed arise in two space-time dimensions where so-called
`extended conformal algebras' (also called $\w$-algebras) provide infinite
symmetries. The study of the representation theory of these algebras gives
rise to `rational conformal field theories' (RCFTs) which are in some sense
finite. Particular RCFTs describe the second order
phase transitions in $\Zed_n$-spin chains as we have already mentioned
earlier. More generally, a classification of RCFTs would also classify the
associated conformally invariant critical points of statistical models.
Therefore, it is interesting to understand the structure of RCFTs. This
also involves a good understanding of $\w$-algebras and the RCFTs associated
to them (their so-called `rational models'). The remainder of this thesis
is devoted to investigations that either yield new results on the structure
of $\w$-algebras or are relevant for spin quantum chains.
\bn
\chapsubtitle{\secwA.\ Quantum $\bw$-algebras}
\mn
In this section we give a brief review of the notion of a quantum
$\w$-algebra -- for details see e.g.\ $\q{\bouwschou,\flrep}$.
\mn
The ingredients of a quantum $\w$-algebra are:
\item{1)}
A quantum $\w$ algebra is a vector space of local chiral quantum fields,
i.e.\ there is a basis of fields $\phi(z)$ depending on a complex
coordinate $z \in \ComplexBar$ (including one point at infinity)
with definite `{\it conformal dimension}' $d(\phi)$ and mode expansion
$$\phi(z)=
\sum_{n-d(\phi)\in\zed} z^{n-d(\phi)}\phi_{n}
        \eqno({\rm \secwA.1})$$
such that all fields $\phi(z)$ are either bosonic or fermionic and that
their dimensions are half-integer $2 d(\phi) \in \Zed$.
\item{2)}
The algebraic structure is encoded in the so-called `operator product
expansion' (OPE) of two fields $\phi(z)$, $\chi(w)$:
$$\OPE{\phi(z)}{\chi(w)} = \sum_{\psi} {A_{\phi \chi}^{\psi}
                      \psi(w) \over (z-w)^{d(\phi)+d(\chi)-d(\psi)}}
        \eqno({\rm \secwA.2})$$
where the sum on the r.h.s.\ of (\secwA.2) runs over all local chiral
fields in the $\w$-algebra. The $A_{\phi \chi}^{\psi}$ are the structure
constants of the $\w$-algebra. The l.h.s.\ of (\secwA.2) is assumed to
be radially ordered. The OPE (\secwA.2) is required to be associative.
Inserting the mode expansion (\secwA.1) into (\secwA.2) and performing Cauchy
integrals, the singular part of the OPE gives rise to a Lie bracket
whereas the regular part defines a normal ordered product $(\phi
\partial^n \chi)$ (the derivative $\partial$ is defined in the
obvious manner). Note that the associativity of the OPE can be used in order
to compute OPEs involving normal ordered products (also known as
`Wick rules').
\item{3)}
The $\w$-algebra contains a distinguished field, the energy-momentum tensor
$L$ with OPE:
$$\OPE{L(z)}{L(w)} = {c \over 2 (z-w)^4} +
              {2 L(w) \over (z-w)^2} + {L'(w) \over (z-w)}
              + reg.\
        \eqno({\rm \secwA.3})$$
The constant $c$ is called the `{\it central charge}'. The Lie algebra obtained
by a mode expansion of the singular part of (\secwA.3) is a representation
of the `Virasoro algebra' where the central element $C$ is represented by
$c \id$.
\item{4)}
The energy-momentum tensor $L$ defines a grading on the
$\w$-algebra:
$$\OPE{L(z)}{\phi(w)} = {d(\phi) \phi(w) \over (z-w)^2}
                 + \sum_{\psi\atop d(\psi) \ne d(\phi)} {A_{L \phi}^{\psi}
                      \psi(w) \over (z-w)^{2+d(\phi)-d(\psi)}}
        \eqno({\rm \secwA.4})$$
where $d(\phi)$ is the conformal dimension of $\phi$.
\item{5)}
Fields transforming as
$$\OPE{L(z)}{\phi(w)} = {d(\phi) \phi(w) \over (z-w)^2} +
                 {\phi'(w) \over (z-w)}
           + \sum_{\psi\atop \abs{d(\psi) - d(\phi)} > 1} {A_{L \phi}^{\psi}
                      \psi(w) \over (z-w)^{2+d(\phi)-d(\psi)}}
        \eqno({\rm \secwA.5})$$
are called `{\it quasi-primary}'. The field $\phi$ is called `{\it primary}'
if the remaining sum runs only over fields $\psi$ with
$d(\psi) > d(\phi) + 1$. Note that the energy-momentum tensor $L$ itself is
quasi-primary with conformal dimension $d(L) = 2$.
\item{6)}
A basis for all fields in the $\w$-algebra can be obtained from a countable
basic set of quasi-primary fields $\{W^{(i)}\}$ using normal ordered products
and derivatives. The $W^{(i)}$ are called `{\it generators}'.
If the set of generators is minimal, the $\w$-algebra is called a
`$\w(d(W^{(1)}), d(W^{(2)}), \ldots)$'.
\item{7)}
There is an irreducible representation of the mode algebra, the so-called
`{\it vacuum representation}', i.e.\ the modes $\phi_n$ of any field
$\phi(z)$ act as endomorphisms in the vacuum representation. The vacuum
representation has a cyclic highest weight vector $\vac$ with the property
$$\phi_n \vac = 0 \, , \qquad \forall n < d(\phi) \, .
        \eqno({\rm \secwA.6})$$
A normal ordered differential polynomial that acts trivially in the vacuum
representation is identically equal to zero. Fields with this property
are called `{\it null fields}'.
Note that it is sufficient to require (\secwA.6) for quasi-primary fields
only, it then automatically extends to all fields including derivatives.
\mn
Property 7) gives a precise meaning to the fields $\phi(z)$.
The definition of a $\w$-algebra can be made precise using the mathematical
language of vertex operator algebras (see $\q{\wowoPhD}$ for a definition).
Then, property 7) is the starting point of the definition and properties
1) - 6) turn out to be reformulations of the axioms of a vertex operator
algebra. In general, one will have to impose additional finiteness conditions
like requiring the graded components of the vacuum module to be finite
dimensional.
\medskip
The simplest $\w$-algebras are `{\it current algebras}'. They are
generated by currents $J_i(z)$ (fields with dimension $d(J_i) = 1$)
with OPEs
$$\OPE{J_i(z)}{J_j(w)} = {g_{i j} \, k \over (z-w)^2} + \sum_l
                                {f_{i j}^l \, J_l(w) \over (z-w)}
                     + reg.
        \eqno({\rm \secwA.7})$$
where $g_{i j}$ is an invariant metric of some Lie algebra $\Lie_n$ and
the $f_{i j}^l$ are the structure constants of this Lie algebra.
The current algebra defined by (\secwA.7) is denoted by `$\hat{\Lie}_n$' and
$k$ is called the `{\it level}'. The singular part of the OPE of the basic
currents $J_i$ defines an infinite dimensional Lie algebra which is a
so-called `Kac-Moody algebra'. Below we will abuse notation and also refer
to the $\w$-algebras defined by (\secwA.7) with the term Kac-Moody algebra.
In particular we will ignore that it is not entirely trivial that a
Kac-Moody algebra becomes a $\w$-algebra once it is endowed with a normal
ordered product.
\medskip
There are several different constructions for $\w$-algebras, most of them
starting from Kac-Moody algebras (\secwA.7). The so-called `{\it coset
construction}' (also known as the GKO construction since for
$\hat{\Lie}_n = \widehat{sl(2)}$ it is the construction for representations
of the Virasoro algebra (\secwA.3) of ref.\ $\q{\goddard}$) is particularly
simple to formulate. There are two closely related types of cosets:
$${W \over \hat{\Lie}_n} \, , \qquad \qquad
  {W \over \Lie_n} \, .
        \eqno({\rm \secwA.8})$$
By definition, the coset $\w$-algebra $W / \hat{\Lie}_n$
is formed by all fields in the $\w$-algebra $W$ that commute (i.e.\ have
regular OPEs) with the Kac-Moody subalgebra $\hat{\Lie}_n \subset W$.
Similarly, the coset $W / \Lie_n$ is defined as the set of all fields in $W$
that are invariant with respect to the action of the Lie algebra $\Lie_n$
implemented by the zero modes of the currents $\hat{\Lie}_n \subset W$.
If the energy-momentum tensor does not coincide with the Sugawara
energy-momentum tensor of $\hat{\Lie}_n$, the coset (\secwA.8) yields
automatically a closed $\w$-subalgebra of $W$ (see e.g.\ $\q{\bouwschou}$
for more details).
\medskip
Another very important construction is the so-called `{\it quantized
Drinfeld-Sokolov}' (DS) reduction (see e.g.\ $\q{\dBT,\laszlorep}$
for recent descriptions). This reduction is a Hamiltonian reduction
that also gives a precise meaning to the `quantized Miura map' of the
Fateev-Luk'yanov construction (see e.g.\ $\q{\lykyanov,\flrep}$). Here we
just recall some ingredients -- for details see e.g.\ $\q{\dBT,\laszlorep}$.
The data specifying the DS reduction is an embedding of $sl(2)$ into $\Lie_n$.
The Cartan generator $H$ of the $sl(2)$ defines a grading on $\Lie_n$.
By definition, the positive step operator $E_{+}$ of this $sl(2)$ has grade
one. Let $T^{j}$ be the basis of $\Lie_n$ with the metric $g_{ij}$ and
structure constants $f_{ij}^l$ used in (\secwA.7). We introduce a
Lie algebra valued current $(J)$ by $(J)=\sum_j J_j T^{j}$.
The grading on $\Lie_n$ leads to the following decomposition of the
Kac-Moody algebra
$$(J) = (J)_{-} \oplus (J)_0 \oplus (J)_{+}
        \eqno({\rm \secwA.9})$$
where $(J)_0$ refers to the currents of grade zero, $(J)_{-}$ to all currents
with {\it arbitrary} negative grade and $(J)_{+}$ to the currents with
arbitrary positive grade. The DS reduction is defined by imposing the
following first class constraints on the currents with positive grade
\footnote{${}^{5})$}{
There are some subtleties for the grade ${1 \over 2}$ part of a
half-integral embedding which we ignore here because it is not relevant for
the aspects we need below.}:
$$(J)_{+} = E_{+} \, .
        \eqno({\rm \secwA.10})$$
This amounts to equating the positive grade currents to zero except for the
one corresponding to $E_{+}$ that is set to one.
{}From a classical point of view the constraints (\secwA.10) generate
gauge transformations and the problem is to find a suitable representative
for each gauge orbit. At the quantum level the constraints (\secwA.10) are
imposed using a nilpotent BRST operator $Q_{{\rm BRST}}$
($Q_{{\rm BRST}}^2 = 0$). In order to impose the constraints with a
BRST operator one has to introduce some auxiliary fields, so-called
ghost-antighost pairs. Then one considers the cohomology of $Q_{{\rm BRST}}$
and the problem is to find a suitable representative for each cohomology
class.
\mn
Suitable representatives are associated to the so-called lowest weight gauge:
The generators $W^{(j)}$ of the DS $\w$-algebra can be chosen to be the lowest
weight components of the embedded $sl(2)$ in the adjoint representation on
the Kac-Moody algebra $\hat{\Lie_n}$. From the lowest weight gauge one
reads off that the conformal dimension of $W^{(j)}$ is given by the spin
$S_j$ of the $sl(2)$ representation, i.e.\ $d(W^{(j)}) = S_j + 1$.
Furthermore, the generators can be made primary using a certain `dressing'.
In the quantum case, this dressing is provided by the `tic-tac-toe' equations
of $\q{\dBT}$.
\mn
For the principal embedding of $sl(2)$ into $\Lie_n$ the DS reduction gives
rise to the so-called Casimir $\w$-algebras which we denote by ${\cal WL}_n$.
In the case of a nonprincipal $sl(2)$ embedding into a simple Lie algebra
$\Lie_n$ we use the notation $\w^{{\cal L}_n}_{\cal S}$ where ${\cal S}$
denotes the embedding.
For example, for $\Lie_n = \A_n$, ${\cal S}$ can be chosen as the $r$-tuple
of the dimensions of the irreducible $sl(2)$ representations which appear
in the defining representation of $\Lie_n$.
The Polyakov-Bershadsky algebra $\w(1,{3 \over 2}, {3 \over 2}, 2)$
$\q{\poly,\bersh}$ which is obtained by DS reduction of the nonprincipal
$sl(2)$ embedding into $sl(3)$ is abbreviated in this notation by
$\w^{sl(3)}_{2,1}$.
\medskip
In the next section we shall be interested in non-trivial outer automorphisms
of $\w$-algebras and their effect on the HWRs. An automorphism $\rho$ of
a $\w$-algebra is a bijective map of the algebra that satisfies the
following compatibility condition with (\secwA.2):
$$\OPE{\rho(\phi(z))}{\rho(\chi(w))} = \sum_{\psi} {A_{\phi \chi}^{\psi}
                      \rho(\psi(w)) \over (z-w)^{d(\phi)+d(\chi)-d(\psi)}}
        \, .
        \eqno({\rm \secwA.11})$$
$\rho$ is called an `outer' automorphism if it is not generated by
the $\w$-algebra itself.
Each such automorphism enables one to impose non-trivial boundary
conditions on the fields $\phi_j$ in the algebra:
$$\phi_j\left(e^{2 \pi i} z\right) = \rho\left(\phi_j\left( z \right)\right)
  \eqno({\rm \secwA.12})$$
where now the argument $z$ lives on a suitable covering of the complex plane.
This type of boundary condition will be called a `twist'.
\mn
Quite often automorphisms $\rho$ have the form $\rho(W_j(z)) = \pm W_j(z)$
where some generators $W_j(z)$ transform with a plus sign and others with
a minus sign. For any field with
$\phi_j\left(e^{2 \pi i} z\right) = -\phi_j(z)$
the Laurent expansion (\secwA.1) has to be modified to
$$\phi_j(z)=
 \sum_{n-d(\phi_j)\in\zed+{1\over 2}} z^{n-d(\phi_j)}\phi_{j,n}\, .
  \eqno({\rm \secwA.13})$$
This leads to the Ramond-sector of a fermionic $\w$-algebra. Bosonic fields
acquire half-integer modes.
\mn
In the presence of an outer automorphism one can also consider the projection
of the $\w$-algebra onto the invariant subspace. This projection is
called an `{\it orbifold}'.
\bn
\chapsubtitle{\secwB.\ Automorphisms of Casimir $\bw$-algebras
                       and boundary conditions}
\mn
The simplest $\w$-algebras are those with one additional generator, so-called
`$\w(2,\delta)$-algebras'. The automorphisms of these algebras were studied
in detail in $\q{\automos}$ and $\q{\diplom}$. Here we will focus on the
so-called `Casimir' algebras ${\cal WL}_n$ where
the dimensions of the simple fields equal the orders of the Casimir
invariants of a simple Lie algebra $\Lie_n$ (see e.g.\ $\q{\bbss,\bai}$).
One common approach to their study is Toda field theory $\q{\baf,\bal,\blg}$.
As we will see below only Casimir algebras ${\cal WL}_n$ based on
{\it simply-laced} $\Lie_n$ have non-trivial outer automorphisms for
generic $c$. The unitary minimal series of these algebras can also be studied
via GKO-constructions $\q{\goddard}$. Q.\ Ho-Kim and H.B.\ Zheng have noticed
that in this approach outer automorphisms of the Lie algebra give rise to
automorphisms of the $\w$-algebra and argued that there no further ones
$\q{\hokimA,\hokimB,\hokimC}$. Owing to their work twists of the unitary
minimal series of Casimir algebras are well understood. Still, we would
like to comment on Casimir algebras from the point of view of extended
conformal algebras, including the non-unitary minimal models into our
discussion.
\mn
One way to study automorphisms of Casimir $\w$-algebras is by inspection of
the structure constants of known examples. Inspecting the structure constants
of the algebras constructed e.g.\ in
$\q{\blm,\kau,\klausWAN,\hornfeck}$ it was
observed in $\q{\automos}$ that ${\cal WL}_n$ with generic central charge
$c$ has as many automorphisms as ${\cal L}_n$ has.
This was also observed before for the {\it unitary minimal series} in
$\q{\hokimA,\hokimB,\hokimC}$.
In fact, it can easily be seen that using quantized Drinfeld Sokolov (DS)
reduction (see e.g.\ $\q{\dBT,\laszlorep}$) each outer automorphism
of the underlying Lie algebra gives rise to an automorphism of the
associated Casimir $\w$-algebra. In the DS framework Casimir $\w$-algebras
arise from the principal $sl(2)$ embedding. The Cartan- resp.\ step-operators
of this principal $sl(2)$ are sums over all the Cartan- resp.\ step-operators
associated to the simple roots of the underlying Lie algebra. Since outer
automorphisms of simple Lie algebras are nothing but certain permutations
of the roots the principal $sl(2)$ is clearly preserved. Therefore, the
automorphism of the underlying Lie (or Kac-Moody) algebra survives the
reduction. We will argue below for ${\cal WA}_n$ that the automorphism
remains non-trivial after the reduction.
In the case of Casimir $\w$-algebras it seems that this type
of automorphism is in fact the only one $\q{\automos}$ although this is
far from being obvious. In fact, also $\w$-algebras for {\it non-principal}
$sl(2)$ embedding admit automorphisms some of which do not seem to have
any relation to the automorphisms of the underlying Lie algebra
(compare e.g.\ the analogue of the Ramond sector of the
Polyakov-Bershadsky-algebra $\w(1,{3 \over 2},{3 \over 2},2)$
$\q{\poly,\bersh}$).
\mn
In the case of ${\cal WA}_n$ the effect of the automorphism
of the underlying Lie algebra after reduction can easily be controlled.
To this end one uses the lowest weight gauge for the DS reduction, i.e.\
one has to find the lowest weights of the principal $sl(2)$ in the
adjoint representation. These lowest weights are given by the tensor
product of a spin ${n \over 2}$ representation with itself minus a spin
zero representations (in order to ensure tracelessness). By inspection
of the Clebsch-Gordan coefficients one sees that the parity under the
automorphism alternates with the spin $S$, i.e.\ the lowest weights transform
with a factor $(-1)^S$. Since the dimension of the generating fields
$W^{(j)}$ is $d(W^{(j)}) = S_j+1$ and the energy-momentum tensor $L = W^{(1)}$
must be invariant, one sees that the generators with even dimension
are invariant and those with odd dimension pick up a minus sign.
This shows that
$$\rho(W^{(j)}) = (-1)^{d(W^{(j)})} W^{(j)}
   \eqno({\rm \secwB.1})$$
for the generators $W^{(j)}$ of ${\cal WA}_n \cong \w(2,\ldots,n+1)$.
\mn
For certain values of the central charge one finds null fields in the
algebras ${\cal WA}_2 \cong \w(2,3)$ and ${\cal WA}_3 \cong \w(2,3,4)$.
These can be exploited to determine
the conformal dimensions of some minimal models along the lines of
$\q{\wirrep,\rva}$ using the special purpose computer algebra system
$\q{\commute}$. Certain normal ordered products in the twisted sector are
not straightforward to handle (see e.g.\ $\q{\supwirrep}$ for a discussion
of problems of this type) only a few models are accessible to these direct
computations.
The results have been presented in $\q{\automos}$. Here,
we list only the subset of results corresponding to $\Zed_n$ parafermions
(see table 5).
\mn
\centerline{\vbox{
\hbox{
\vrule \hskip 1pt
\vbox{ \offinterlineskip
\def\tablespace{height2pt&\omit&&\omit&&\omit&&\omit&&\omit&\cr}
\def\tablerule{ \tablespace
                \noalign{\hrule}
                \tablespace        }
\hrule
\halign{&\vrule#&
  \strut\hskip 10pt\hfil#\hfil\hskip 10pt\cr
height3pt & \multispan3 && \multispan5 &\cr
& \multispan3 \hfil $\w(2,3)$ \qquad $c={4 \over 5}$ \hfil
     && \multispan5 \hfil $\w(2,3,4)$ \qquad $c=1$ \hfil       &\cr
height3pt & \multispan3 && \multispan5 &\cr
\noalign{\hrule}
height2pt & \omit && \omit && \multispan3 && \omit &\cr
& \hfil {\it untwisted} \hfil && \hfil {\it twisted} \hfil
   && \multispan3 \hfil {\it untwisted} \hfil && {\it twisted}  &\cr
height2pt & \omit && \omit && \multispan3 && \omit &\cr
\noalign{\hrule}
\tablespace
&  $0$            && ${1 \over 8}$
    && $0$            && ${3 \over 4}$  && ${1 \over 16}$   &\cr \tablespace
&  ${2 \over 3}$  && ${1 \over 40}$
    && $1$            && ${9 \over 16}$ && ${3 \over 16}$   &\cr \tablespace
&  ${2 \over 5}$  && \omit
    && ${1 \over 3}$  && ${1 \over 16}$ && ${1 \over 48}$   &\cr \tablespace
&  ${1 \over 15}$ && \omit
    && ${1 \over 12}$ && \omit          && ${25 \over 48}$  &\cr \tablespace
}
\hrule}\hskip 1pt \vrule}
\hbox{Table 5: Conformal dimensions of untwisted and twisted}
\hbox{\phantom{Table 5:} $\w(2,3) \cong {\cal WA}_2$
                    and $\w(2,3,4) \cong {\cal WA}_3$ for the $\Zed_3$ resp.}
\hbox{\phantom{Table 5:} $\Zed_4$ parafermionic models
                    $c_{\A_2}(4,5) = {4 \over 5}$, $c_{\A_3}(5,6) = 1$.}}
}
\mn
{}From the above discussion we conclude that there will be no outer
automorphisms
for ${\cal WA}_1$, ${\cal WB}_n$, ${\cal WC}_n$, ${\cal WE}_7$, ${\cal WE}_8$,
${\cal WF}_4$ and ${\cal WG}_2$. The algebras ${\cal WA}_n$ for $n > 1$,
${\cal WD}_n$ for $n > 4$ and ${\cal WE}_6$ should have exactly one
outer automorphism and correspondingly exactly one twisted sector in
addition to the untwisted one.
For these algebras explicit computations $\q{\automos}$ showed how the
formulae for the $h$-values in $\q{\hokimB,\hokimC}$ might generalize:
$$\eqalign{
c_{\Lie_n}(p,q) &= n - 12 \rho^2 {(p-q)^2 \over p q} \ , \cr
h_{p,q;\lambda,\mu}^{\Lie_n^{(1,2,3)}} &=
   {(p \lambda - q \mu)^2 \over 2 p q}+{c_{p,q}^{\Lie_n}-n \over 24}
               + \tilde{h} n_1 \ , \cr
}\eqno({\rm \secwB.2})$$
where
$$\rho = \sum_{i=1}^{n} \tilde{\Omega}_i \  , \qquad
\lambda = \sum_{i=1}^{n_0} r_i \Omega_i \  , \qquad
\mu = \sum_{i=1}^{n_0} s_i \Omega_i \  ,
\eqno({\rm \secwB.3})$$
and $n_0$, $n_1$ are the dimensions of the invariant subalgebra
$\Lie'_{n_0}$ respectively twisted subalgebra of $\Lie_n$;
$\tilde{\Omega}_i$ are the fundamental weights of $\Lie_n$;
$\Omega_i$ the fundamental weights of $\Lie'_{n_0}$; $\tilde{h}$
is a constant (that can be interpreted as the conformal dimension of some
field) and $r_i$, $s_i$ positive integers subject to certain constraints.
For the case of $\A_n$ the invariant subalgebra is
$\AC_{\lbrack {n+1 \over 2} \rbrack}$. In the unitary minimal series of
${\cal WA}_n$, ${\cal WD}_n$ ($n > 4$) and ${\cal WE}_6$
one has $\tilde{h} = {1 \over 16}$. We observe that (\secwB.2) indeed
reproduces the data in table 5 for $\w(2,3,4)$ if we use the weights
of $\AC_2$ (which is the invariant subalgebra) and $\tilde{h} = {1 \over 16}$.
\mn
For $\Lie_n^{(1)}$ (periodic boundary conditions) eqs.\ (\secwB.2)
with (\secwB.3) have been rigorously
proven in $\q{\frenkel}$ (see e.g.\ $\q{\unicos}$ for a brief summary
including the range of the integers $r_i$ and $s_i$).
It should be possible to generalize the result of $\q{\frenkel}$
to twisted Kac-Moody algebras $\hat{\Lie}_n^{(2)}$ and $\hat{\D}_4^{(3)}$
yielding a rigorous derivation of (\secwB.2) with (\secwB.3).
At least one thing is immediately clear if one assumes that the
representations of a (twisted) $\w$-algebra obtained by DS reduction
are induced by admissible representations of the (twisted) Kac-Moody
algebra: The representations of a Kac-Moody algebra are labeled by
the weights of the Lie algebra formed by the zero modes of all currents.
While for an untwisted Kac-Moody algebra $\hat{\Lie}_n^{(1)}$ the algebra
of zero modes is isomorphic to $\Lie_n$, the zero modes of a {\it twisted}
Kac-Moody algebra form a Lie algebra $\Lie'_{n_0}$ that is precisely
the {\it invariant} subalgebra of $\Lie_n$.
\medskip
The exceptional cases ${\cal WE}_6$ and ${\cal WD}_4$ are particularly
interesting. Especially for ${\cal WD}_4 \cong \w(2,4,4,6)$ the group of outer
automorphisms should be ${\cal S}_3$ for generic $c$. In $\w(2,4,4,6)$ one
structure constant remains a free parameter $\q{\hgkpriv}$. It was observed
in $\q{\automos}$ that one can define an operation of $O(2)$ on the two
generators of conformal dimension 4 under which the structure constants
transform covariant. The algebra is {\it invariant} under the natural
embedding of ${\cal S}_3$ into the group $O(2)$
operating on the fields if one chooses the self coupling
constants of these two generators to be equal.
\sn
Denote the primary fields of dimension 4 in $\w(2,4,4,6)$ by $V(z)$ and
$W(z)$. Then the ${\cal S}_3$-symmetry of this algebra translates into the
following type of boundary conditions
for the given choice of coupling constants:
$$\eqalign{
V(e^{2 \pi i} z) &= \cos(\alpha) V(z) - \sin(\alpha) W(z) \cr
W(e^{2 \pi i} z) &= \sin(\alpha) V(z) + \cos(\alpha) W(z) \cr
} \eqno({\rm \secwB.4})$$
or
$$\eqalign{
V(e^{2 \pi i} z) &= \cos(\alpha) W(z) - \sin(\alpha) V(z) \cr
W(e^{2 \pi i} z) &= \sin(\alpha) W(z) + \cos(\alpha) V(z) \cr
} \eqno({\rm \secwB.5})$$
with $\alpha \in \{ 0, {2 \over 3} \pi, {4 \over 3} \pi \}$.
The three different boundary conditions given by (\secwB.4)
correspond to those elements of ${\cal S}_3$ which under the embedding
yield elements of $SO(2)$. The boundary conditions (\secwB.5) correspond to
the three elements of ${\cal S}_3$ that are mapped to elements in $O(2)$ with
determinant $-1$.
\sn
The $h$-values in the unitary minimal series (\secwB.2) of ${\cal WD}_4$
have been calculated in $\q{\hokimB}$ without having to consider the
boundary conditions of the additional simple fields which look quite strange
at first sight. Note that for
$\D_4$ the invariant subalgebra is the exceptional algebra $\G_2$
and here $\tilde{h}$ in (\secwB.2) satisfies $\tilde{h} = {1 \over 18}$.
\sn
In $\q{\bouwschou}$ it has already been stated that the ${\cal S}_3$-symmetry
should lead to modes in ${\zed \over 3}$ and the precise statement was
presented in $\q{\automos}$ which we now would like to recall.
Let us first focus on (\secwB.4). Set $U^{(1)}(z) := V(z) + i W(z)$ and
$U^{(2)}(z) := V(z) - i W(z)$. Then (\secwB.4) turns into
$U^{(1)}(e^{2 \pi i} z) = e^{i \alpha} U^{(1)}(z)$ and
$U^{(2)}(e^{2 \pi i} z) = e^{-i \alpha} U^{(2)}(z)$ which can be satisfied
by choosing modes in $\Zed + {\alpha \over 2 \pi}$ for $U^{(1)}$ and
those for $U^{(2)}$ in $\Zed - {\alpha \over 2 \pi}$.
Consider now (\secwB.5). For this case set
$Y^{(1)}(z) := \cos(\alpha) V(z) + (\sin(\alpha)+1) W(z)$ and
$Y^{(2)}(z) := \cos(\alpha) V(z) + (\sin(\alpha)-1) W(z)$. Now (\secwB.5)
turns into $Y^{(1)}(e^{2 \pi i} z) = Y^{(1)}(z)$ and
$Y^{(2)}(e^{2 \pi i} z) = -Y^{(2)}(z)$. This can be satisfied by
choosing the modes for $Y^{(1)}$ in $\Zed$ and those for $Y^{(2)}$
in $\Zed+{1 \over 2}$.
\medskip
We would like to conclude this discussion by commenting on the relation
of twisted representations to orbifolds (for a detailed discussion of
orbifolding see e.g.\ $\q{\dijkgraaf}$). For $\w$-algebras with a
$\Zed_2$ automorphism $\rho^2 = \id$ one has two partition functions,
one where only the characters $\chi^W = \tr_V q^{L_0-{c \over 24}}$
of the untwisted sector enter, and one where the characters $\chi^W$
{\it and} $\tilde{\chi}^W = \tr_V \rho q^{L_0-{c \over 24}}$ of both
sectors enter. The latter can be identified with the partition function
$Z$ of the orbifold
\footnote{${}^{6})$}{We simplify notation by absorbing multiplicities
of characters into the index set.}:
$$\eqalign{
Z &= \sum_{k: \ {\rm untwisted}} \left[ (\chi_k^W)^{*} \chi_k^W
                          + (\tilde{\chi}_k^W)^{*} \tilde{\chi}_k^W
                                                \right]
   + \sum_{k: \ {\rm twisted}} \left[ (\chi_k^W)^{*} \chi_k^W
                          + (\tilde{\chi}_k^W)^{*} \tilde{\chi}_k^W
                                                \right] \cr
&= 2 \sum_{k: \ {\rm untwisted} \atop {\rm and \phantom{l} twisted}}
   \left[
   \oh (\chi_k^W + \tilde{\chi}_k^W)^{*} \oh (\chi_k^W + \tilde{\chi}_k^W)
+  \oh (\chi_k^W - \tilde{\chi}_k^W)^{*} \oh (\chi_k^W - \tilde{\chi}_k^W)
   \right] \, . \cr
}    \eqno({\rm \secwB.6})$$
This implies that the characters of the orbifold $\w$-algebra are given
by $\oh (\chi_k^W + \tilde{\chi}_k^W)$ and
$\oh (\chi_k^W - \tilde{\chi}_k^W)$. We conclude that the $h$-values
for the HWRs of the orbifold are those of the original $\w$-algebra
in both sectors in addition to some which differ by (half-) integers.
\medskip
It would be interesting to generalize the observations of this section
to the supersymmetric case. Of special interest is $osp(4 \vert 4)$
which is the supersymmetric analogon of $\D_4$.
The corresponding Casimir algebra is a
${\cal SW}({3 \over 2}, 2, 2, {7 \over 2})$. In $\q{\automos}$ it was
argued on the basis of some explicit results in $\q{\supwir}$ that
${\cal SW}({3 \over 2}, 2, 2, {7 \over 2})$ should be invariant under
the natural embedding of ${\cal S}_3$ into the group $O(2)$ operating
on the fields if one chooses the self coupling constants of the two
generators of dimension two to be equal -- precisely like for
${\cal WD}_4 \cong \w(2,4,4,6)$.
\bn
\chapsubtitle{\secwC.\ Consequences for spin quantum chains}
\mn
As already pointed out in $\q{\hokimA}$ there is a close connection
of the partition function of $\w(2,3)$ at $c={4 \over 5}$
including the twisted sector and the three states Potts model. In fact,
the different boundary conditions of $\w(2,3)$ correspond to the different
boundary conditions of the Potts quantum spin chain at critical temperature.
Choosing the spin shift operator at the end of the chain to be equal to the
spin shift operator at the first site yields the field content of the
`untwisted' sector of $\w(2,3)$ (see e.g.\ $\q{\fateev,\zamzam}$ and
$\q{\gehlen}$ for numerical verification). We recall the remarkable
fact $\q{\hokimA}$ that the
twisted sector of $\w(2,3)$ yields additional representations which can be
identified
with fields in the thermodynamic limit of the three states Potts quantum
spin chain if the spin shift operator at the end of the chain is chosen
to equal the adjoint of the one at the first site $\q{\cardyB}$.
\mn
It would be interesting to know if this observation generalizes to
{\it all} $\Zed_n$. For twisted boundary conditions only partial
results are available in the literature (see e.g.\ $\q{\schuetz}$).
We shall therefore present an explicit verification
of this statement in the case of $\Zed_4$. We will
follow the approach of $\q{\gehlen}$ and study the spectrum
Hamiltonian (\secB.1) at $\phi = \vphi = 0$ numerically.
\mn
Let $E_{N,i}$ be the eigenvalues of $H_N^{(n)}$ with periodic boundary
conditions in ascending order and $\tilde{E}_{N,i}$ those with twisted
boundary conditions. Then the relevant scaling functions are given by
$\q{\cardyA,\gehlenA}$:
$$\eqalign{
\xi_{N,i} &:= {N \over 2 \pi} (\tilde{E}_{N,i} - E_{N,0}) \cr
\xi_i &:= \lim_{N \to \infty} \xi_{N,i}. \cr
}  \eqno({\rm \secwC.1})$$
In the case of periodic and cyclic boundary conditions, the
eigenvalues $Q$ of the charge operator (\secB.9) and momentum $P$ are good
quantum numbers. In the case of twisted boundary conditions
neither charge nor momentum are conserved any more and one
does not have any obvious conserved quantities.
At least for even $n$ the charge $Q \mod 2$ is conserved.
\mn
Assume that (\secB.1) with $\phi = \vphi = 0$ exhibits conformal
invariance at $\la=1$ and
denote the dimensions of the fields in the left chiral
part by $h$ and of those in the right chiral part by
$\bar{h}$. For periodic and twisted boundary conditions
the field theory is diagonal, i.e.\
the fields $\phi(z, \bar{z})$ with dimension $h+\bar{h}$
satisfy $h=\bar{h}$ and thus have vanishing spin $h-\bar{h}$.
Therefore, the modes of the fields $\phi(z, \bar{z})$
yield levels in the spectrum with $\xi = h+\bar{h}+r$ where
$r \in \Zed_{+}$ for periodic boundary conditions and
$r \in {\zed_{+} \over 2}$ for twisted boundary conditions.
\mn
In order to test this method we shall first study the well-known three states
Potts model. For $\Zed_3$ we have studied $3$ to $9$ sites. Thus, it was
necessary to partially diagonalize matrices of dimension
$3^9 = 19683$. The limits $N \to \infty$ of the lowest gaps
$\xi_i$ are given in table 6.
\mn
\centerline{\vbox{
\hbox{
\vrule \hskip 1pt
\vbox{ \offinterlineskip
\def\tablespace{ height2pt&\omit&&\omit&&\omit&\cr }
\def\tablerule{ \tablespace
                \noalign{\hrule}
                \tablespace        }
\hrule
\halign{&\vrule#&
  \strut\hskip 4pt\hfil#\hfil\hskip 4pt\cr
height4pt& \multispan{5} & \cr
& \multispan{5} \hfil $\Zed_3$ \hfil &\cr
height4pt& \multispan{5} & \cr
\noalign{\hrule}
\tablespace
& $i$   &&  $\xi_i$      &&  $h+\bar{h}+r$    &\cr
\tablespace
\noalign{\hrule}
\tablespace
& $0$   && $0.050000(2)$ && ${1\over 40}+{1\over 40}$  &\cr\tablespace
& $1$   && $0.250005(5)$ && ${1\over 8}+{1\over 8}$    &\cr\tablespace
& $2$   && $0.5500(5)$   && ${1\over 40}+{1\over 40}+{1\over 2}$&\cr\tablespace
& $3$   && $1.05(4)$     && ${1\over 40}+{1\over 40}+1$  &\cr\tablespace
}
\hrule}\hskip 1pt \vrule}
\hbox{Table 6: The low-lying spectrum of the}
\hbox{\phantom{Table 6:} twisted $\Zed_3$-chain at $\la = 1$}}
}
\mn
The numbers in brackets indicate the estimated error in the
last given digit. For details on the extrapolation procedures
and error estimation see e.g.\ $\q{\henkel}$.
We do not give more than four levels because
the errors of the next levels make an accurate identification
impossible.
Note that we can nicely identify the dimensions
${1 \over 40}$ and ${1 \over 8}$ of the chiral fields -- as expected.
\medskip
Let us now turn to $\Zed_4$. The spectrum
of the $\Zed_4$-chain was already discussed in $\q{\baakeB}$
also for twisted boundary condition. This was done applying
Kac-Moody algebra and numerical techniques. Since the former considerations
are analogous to the discussion of the previous section and the numerical
results are unpublished, we will
present results of a direct calculation here because we would like
to demonstrate the correspondence between boundary conditions
in statistical mechanics and conformal field theory.
Note that the $\Zed_4$-version of (\secB.1) at $\phi = \vphi = 0$
is a special case of the Ashkin-Teller quantum chain which was
introduced in $\q{\kohmoto}$ setting the parameter $h={1 \over 3}$
(in the notations of $\q{\baakeA}$).
\sn
For $\Zed_4$ we have at least a splitting of the spectrum into two
sectors of $Q \mod 2$. We have studied $4$ to $8$ sites, implying
the partial diagonalization of matrices in dimensions up to
$4^8 / 2 = 32768$.
\mn
\centerline{\vbox{
\hbox{
\vrule \hskip 1pt
\vbox{ \offinterlineskip
\def\tablespace{
height2pt&\omit&&\omit&&\omit&\hskip 1pt \vrule&\omit&&\omit&&\omit&\cr }
\def\tablerule{ \tablespace
                \noalign{\hrule}
                \tablespace        }
\hrule
\halign{&\vrule#&
  \strut\hskip 4pt\hfil#\hfil\hskip 4pt\cr
height4pt& \multispan{11} & \cr
& \multispan{11} \hfil $\Zed_4$ \hfil &\cr
height4pt& \multispan{11} & \cr
\noalign{\hrule}
height3pt & \multispan5 &\hskip 1pt \vrule& \multispan5 &\cr
& \multispan5 \hfil $Q \mod 2 = 0$ \hfil  &\hskip 1pt \vrule&
          \multispan5 \hfil $Q \mod 2 = 1$ \hfil       &\cr
height3pt & \multispan5 &\hskip 1pt \vrule& \multispan5 &\cr
\noalign{\hrule}
\tablespace
& $i$   &&  $\xi_i$     &&  $h+\bar{h}+r$  &\hskip 1pt \vrule&
    $i$   &&  $\xi_i$    &&  $h+\bar{h}+r$  &\cr
\tablespace
\noalign{\hrule}
\tablespace
& $0$   && $0.04167(2)$ && ${1\over 48}+{1\over 48}$ &\hskip 1pt \vrule&
     $0$  &&  $0.1254(1)$ && ${1\over 16}+{1\over 16}$ &\cr\tablespace
& $1$   && $0.375(2)$   && ${3\over 16}+{3\over 16}$ &\hskip 1pt \vrule&
     $1$  &&  $0.6231(3)$ && ${1\over 16}+{1\over 16}+{1\over 2}$
              &\cr\tablespace
& $2$   && $1.040(2)$   && ${25\over 48}+{25\over 48}$ &\hskip 1pt \vrule&
     $2$  &&  $0.623(7)$  && ${1\over 16}+{1\over 16}+{1\over 2}$
              &\cr\tablespace
& \omit && \omit        && \omit &\hskip 1pt \vrule&
     $3$  &&  $1.12(1)$   && ${1\over 16}+{1\over 16}+1$ &\cr\tablespace
}
\hrule}\hskip 1pt \vrule}
\hbox{Table 7: The low-lying spectrum of the twisted $\Zed_4$-chain
               at $\la = 1$}}
}
\mn
The dimensions ${1 \over 48}$, ${1 \over 16}$, ${3 \over 16}$ and
${25 \over 48}$ of the chiral field theory can be nicely seen in
the explicit results of table 7.
\medskip
The field content of the $\Zed_n$-spin quantum chains (\secB.1) at their
second order phase transition $\phi = \vphi = 0$, $\la = 1$ is given by
the first unitary representation of ${\cal WA}_{n-1}$, i.e.\ by
(\secwB.2) with $\hat{\Lie}_{n_0-1} = \Lie_{n-1} = \A_{n-1}$ and $p=n+1$,
$q=n+2$. In particular, the central charge for a $\Zed_n$-model
equals $c = {2 (n-1) \over n+2}$.
In this section we have explicitly verified for
$n=3$ and $4$ that the representations of the twisted sector
of ${\cal WA}_{n-1}$ correspond to the spectrum of the
$\Zed_{n}$-model with twisted boundary conditions. In fact, this is
also true for the $\Zed_5$-version $\q{\gehlenunpub}$.
These explicit results are in agreement with the statement that
the field content of the spin quantum chain (\secB.1) at $\phi = \vphi = 0$,
$\la=1$ with twisted boundary conditions $\Ga_{N+1} = \Ga_1^{+}$
can be described by a representation of a twisted ${\cal WA}_{n-1}$
for all $n$. Thus, it is possible to calculate
the spectrum of the twisted $\Zed_n$-quantum chain by
(\secwB.2) using $\Lie_{n-1} = \A_{n-1}$,
$\hat{\Lie}_{n_0} = \AC_{\lbrack {n \over 2} \rbrack}$
and $p=n+1$, $q=n+2$.
\mn
For cyclic boundary conditions ($\Ga_{N+1} = \om^{-R} \Ga_1$,
$0 < R < n$) the diagonal symmetry of the statistical mechanics
model is known to be broken such that the spin $h-\bar{h}$
takes on rational values. The dimensions of the chiral fields,
however, are unaffected by this change of boundary conditions.
This has been verified in $\q{\cardyB}$ and $\q{\gehlen}$ for
the case $n=3$, in $\q{\baakeB}$ for $n=4$ and more abstractly
for general $n$ in $\q{\gepner}$.
In $\q{\cardyB}$ a similar result has been obtained
for $\Ga_{N+1} = \om^{-R} \Ga_1^{+}$, $0 < R < 3$, $n=3$
and the only effect of a factor $\om^{-R}$
for all $n$ should be to combine the left- and right-chiral parts
in a non-diagonal way.
\mn
We would like to conclude by mentioning that the case of {\it free}
boundary conditions should be described by the representations of the
orbifold of ${\cal WA}_{n-1}$ at its first unitary minimal model.
This can be read off from the one-to-one correspondence of boundary
conditions of $\w$-algebras and spin chains which we have just
observed. After mapping the cylinder on which the CFT lives to a strip
in the plane with open ends, only those fields can survive which are
{\it invariant} with respect to transformations at the boundary which
implies that precisely the orbifold is expected to survive in the case
of free boundary conditions.
\bn
\chapsubtitle{\secwCl.\ Classical $\bw$-algebras and Virasoro structure}
\mn
In this section we first give a brief review of the notion of classical
$\w$-algebras -- for details see e.g.\ $\q{\FORT,\laszlorep}$.
Afterwards, we make some general statements following from Virasoro
covariance which are only partially known in the literature.
\mn
For a characterizations of a classical $\w$-algebra we closely follow
the quantum case -- compare section {\secwA}:
\item{1')}
A classical $\w$ algebra is a vector space of local chiral classical fields,
i.e.\ there is a basis of fields $\phi(z)$ depending on a complex coordinate
$z \in \ComplexBar$ with definite `{\it conformal dimension}' $d(\phi)$ and
mode expansion
$$\phi(z)=
\sum_{n-d(\phi)\in\zed} z^{n-d(\phi)}\phi_{n}
        \eqno({\rm \secwCl.1})$$
such that all fields $\phi(z)$ are either bosonic or fermionic and that
their dimensions are half-integer $2 d(\phi) \in \Zed$.
\item{2')}
These fields form a Poisson bracket algebra which we write
in OPE form. The `{\it classical OPE}' of two fields $\phi(z)$, $\chi(w)$
is of the form:
$$\clOPE{\phi(z)}{\chi(w)} = \sum_{\psi \atop d(\phi)+d(\chi) > d(\psi)}
                           {A_{\phi \chi}^{\psi}
                      \psi(w) \over (z-w)^{d(\phi)+d(\chi)-d(\psi)}} +reg.
        \eqno({\rm \secwCl.2})$$
The $A_{\phi \chi}^{\psi}$ are the structure constants of the
classical $\w$-algebra. Instead of the Wick rules of the the
quantum case, the classical OPE has the derivation property of
Poisson brackets. For example, $\clOPE{\phi(z)}{\chi(w) \psi(w)} =
\chi(w) \clOPE{\phi(z)}{\psi(w)} + \clOPE{\phi(z)}{\chi(w)} \psi(w)$.
Inserting the mode expansion (\secwCl.1) into (\secwCl.2) and performing
Cauchy integrals, the classical OPE gives rise to a Lie bracket
-- like on the quantum level.
\item{3')}
The classical $\w$-algebra contains a distinguished field, the energy-momentum
tensor $L$ with classical OPE:
$$\clOPE{L(z)}{L(w)} = {c \over 2 (z-w)^4} +
              {2 L(w) \over (z-w)^2} + {L'(w) \over (z-w)} +reg.
        \eqno({\rm \secwCl.3})$$
\item{4')}
The energy-momentum tensor $L$ defines a grading on the classical $\w$-algebra:
$$\clOPE{L(z)}{\phi(w)} = {d(\phi) \phi(w) \over (z-w)^2}
                 + \sum_{\psi\atop {d(\psi) \ne d(\phi) \atop
                                2 + d(\phi)  > d(\psi)}} {A_{L \phi}^{\psi}
                      \psi(w) \over (z-w)^{2+d(\phi)-d(\psi)}} +reg.
        \eqno({\rm \secwCl.4})$$
where $d(\phi)$ is the conformal dimension of the classical field $\phi$.
\item{5')}
Classical fields transforming as
$$\clOPE{L(z)}{\phi(w)} = {d(\phi) \phi(w) \over (z-w)^2} +
                 {\phi'(w) \over (z-w)}
           + \sum_{\psi\atop d(\phi) > d(\psi) + 1} {A_{L \phi}^{\psi}
                      \psi(w) \over (z-w)^{2+d(\phi)-d(\psi)}}+reg.
        \eqno({\rm \secwCl.5})$$
are called `{\it quasi-primary}'. The field $\phi$ is called `{\it primary}'
if the remaining sum is absent.
\item{6')}
All classical fields in the classical $\w$-algebra are differential
polynomials (with finite order) in a countable basic set of quasi-primary
fields $\{W^{(i)}\}$ -- the so-called `{\it generators}' $W^{(i)}$. In
general, this differential polynomial ring will not be freely generated
but may have relations, i.e.\ differential polynomials that vanish
identically. Note that in contrast to the normal ordered product of the
quantum level, the product of two classical fields is (anti-)commutative.
\item{7')}
The `{\it phase space}' of the classical $\w$-algebra is the vector space
of meromorphic functions from the Riemannian sphere $\ComplexBar$ into the
complex numbers $\Complex$. The classical fields are defined as integral
kernels of functionals on this phase space. In particular, all classical OPEs
(\secwCl.2) have to be understood after Cauchy integration with some
meromorphic test function.
\mn
Note that usually a notation different from the above is used in the
literature. Usually one uses the compact manifold $S^1$ instead of the
Riemannian sphere $\ComplexBar$. Then the poles in the OPE are replaced
by $\delta$-functions on the r.h.s.\ of the Poisson brackets. Test functions
are smooth square-integrable functions on $S^1$, not meromorphic functions.
However, this is only a minor difference in the point of view since
after all we are interested in the algebraic structure of a classical
$\w$-algebra, and here both formulations lead to the same results.
\mn
The description in section {\secwA} of Kac-Moody algebras, cosets,
orbifolds and DS reduction apply at the classical level as well with
some obvious minor modifications. Therefore we do not explicitly repeat
the corresponding discussion of section {\secwA}.
\medskip
In contrast to the quantum case one does not have a bilinear form
on a classical $\w$-algebra which one can use to prove basic theorem
e.g.\ about the $su(1,1)$ structure of the space of fields. Nevertheless,
explicit constructions are known on the quantum level (see e.g.\
$\q{\blm}$) which can be transferred to classical $\w$-algebras.
Then one can use these explicit constructions in order to prove e.g.\
that the generators of a classical $\w$-algebra can be chosen primary
in quite general situations. In the remainder of this section we wish
to indicate how this should be done.
\mn
A first result is a formula for a quasi-primary projection of two
classical fields including some derivatives:
\mn
\beginllemma{\LemCla}{Classical quasi-primary product}
For any two quasi-primary fields $\phi$, $\chi$ let
$$\Qp^{(r)}(\phi,\chi) :=
  \sum_{n+m = r} (-1)^n {2 d(\phi) + r - 1 \choose m}
        {2 d(\chi) + r - 1 \choose n} (\partial^n \phi) (\partial^m \chi)
          \, .
        \eqno({\rm \secwCl.6})$$
Then $\Qp^{(r)}(\phi,\chi)$ is a quasi-primary field of conformal
dimension $d(\phi) + d(\chi) + r$.
\endlemma
\mn
In order to prove this statement one has to apply (\secwCl.4) to the r.h.s.\
of (\secwCl.6) and check that it has the correct transformation properties.
Of course, an equivalent expression can be obtained by applying a suitable
classical limit $\q{\unicos}$ to the formula for the quantum case $\q{\blm}$.
Covariance problems of this type have been treated in the mathematical
literature already many years ago. For example,
finding the projection of the $r$th derivative of the product of
two modular forms onto a modular form (see e.g.\ $\q{\don}$) is equivalent
to determining the quasi-primary projection. In fact, we have taken
Eq.\ (\secwCl.6) from the formula (1) of $\q{\don}$ which is called the
`$r$th Rankin-Cohen bracket'.
\mn
Note that it is possible that the classical quasi-primary product (\secwCl.6)
vanishes identically.
\medskip
Now we turn to the question of Virasoro covariance of the space of fields
of a classical $\w$-algebra. To this end we need a weak property of the
generating set. Then we can show by construction that for any $\w$-algebra
with this property the space of fields decomposes into quasi-primary fields
and derivatives thereof. We will also say that a $\w$-algebra has a
quasi-primary basis if it admits such a decomposition.
\mn
\begindef{\DefClA}
A classical $\w$-algebra is called `regularly generated' if for each
fixed dimension $\Delta$ there are only finitely many generators with
conformal dimension less than $\Delta$.
\enddef
\mn
\beginclaim{\ClaimA}
Any classical $\w$-algebra that is regularly generated admits a basis
of quasi-primary fields and derivatives thereof.
\endclaim
\mn
This statement should be proven for homogeneous polynomials of a fixed
degree in the generators. For the linear polynomials the claim is
true by definition. Also for the second order polynomials it is
immediately clear that the formula (\secwCl.6) yields precisely as many
linearly independent quasi-primary fields as are needed in order to
complement the derivatives to a basis. More precisely, all linear relations
among the second order quasi-primary fields are given by
$\Qp^{(r)}(\phi,\chi) = \pm \Qp^{(r)}(\chi,\phi)$ and
$\Qp^{(r)}(\phi,\phi) = 0$ for $\phi$ bosonic and $r$ odd or
$\phi$ fermionic and $r$ even.
For polynomials of higher degree we forget at the moment about the
relations possibly present in the space of fields (they can be imposed
at the very end). Then one has to gain control about the action of the
derivative. This can e.g.\ be done by using partitions into different
colours (compare $\q{\flohrdipl}$ for a presentation in the quantum case).
Now a basis of the quasi-primary fields can be labeled by these partitions.
It remains to be checked that those quasi-primary polynomials associated
to the partitions are indeed linearly independent. We leave this
inspection for future investigations.
\mn
Note that this claim should in particular be valid for $\w$-algebras
that are generated by finitely many fields whose dimensions are strictly
positive. However, the above argument should apply to regularly generated
classical $\w$-algebras without additional effort. Claim {\ClaimA} might
be true in even more general situations but this will certainly give rise
to extra complications in the proof.
\mn
If a classical $\w$-algebra admits a basis of quasi-primary fields and
derivatives thereof one can derive a specialization of the classical OPE
(\secwCl.2) when applied to quasi-primary fields:
\mn
\beginllemma{\LemClA}{\q{\fehort}}
For any classical $\w$-algebra with a quasi-primary basis the Poisson
brackets of two classical quasi-primary fields $\phi(z)$, $\chi(w)$ have
the form
$$\eqalign{
\clOPE{\phi(z)}{\chi(w)} = & {d_{\phi,\chi} \delta_{d(\phi),d(\chi)}
                           \over (z-w)^{d(\phi)+d(\chi)} }+
              \sum_{\psi \hbox{ \fiverm quasi-primary}
                                  \atop d(\phi)+d(\chi) > d(\psi)}
                           \!\!\! C_{\phi \chi}^{\psi} \!\!\!
                  \sum_{r=0}^{d(\phi)+d(\chi)-d(\psi)-1} \cr
                & \qquad { {d(\phi) - d(\chi) + d(\psi) + r - 1 \choose r}
                    \partial^r \psi(w) \over r! {2 d(\psi) + r - 1 \choose r}
                    (z-w)^{d(\phi)+d(\chi)-d(\psi)-r}}  +reg. \cr
}        \eqno({\rm \secwCl.7})$$
\endTheorem
\mn
This statement is also true at the quantum level (see e.g.\ $\q{\blm}$),
and in fact was first discovered there. It can be proven along different
lines. One can either use the realization of $L_{-1}$ as the differential
operator $L_{-1} \phi(z) = z (z \partial + 2 d(\phi)) \phi(z)$ acting
on quasi-primary fields $\phi(z)$. Then one uses covariance of (\secwCl.2)
with respect to this differential operator $\q{\fehort}$. Equivalently,
lemma {\LemClA} can be derived from the Jacobi identity for the Poisson
brackets $\{L(y), \{\phi(z), \chi(w) \} \} + {\rm cyclic} = 0$ using
(\secwCl.5).
\mn
In view of lemma {\LemClA} we can introduce the following shorthand notation
for the Poisson brackets of two classical quasi-primary fields
$\phi$, $\chi$ writing only the quasi-primary fields on the r.h.s.:
$$\phi \star \chi = \sum_{\psi \hbox{ \fiverm quasi-primary}
                                  \atop d(\phi)+d(\chi) > d(\psi)}
                           C_{\phi \chi}^{\psi} \psi  \, .
        \eqno({\rm \secwCl.8})$$
The notation (\secwCl.8) has to be understood in the sense that the arguments
and poles have to be re-instituted in the form of (\secwCl.7). Similarly,
the coupling constants involving derivatives can be recovered from (\secwCl.7)
since they are uniquely given by those involving only quasi-primary fields.
In this shorthand notation the classical Virasoro algebra (\secwCl.3) becomes
$L \star L = {c \over 2} + 2 L$.
\mn
This shorthand notation is useful for arguing that one can choose the
generators of a classical $\w$-algebra primary in a quite general situation.
\mn
\beginclaim{\ClaimB}
All generators of a classical $\w$-algebra except for the Virasoro field $L$
can be chosen to be primary if the $\w$-algebra is regularly generated,
the Virasoro centre is non-zero $c \ne 0$ and the conformal dimensions
of the generators are strictly positive.
\endclaim
\mn
A sketch of the proof of this statement is the following: If claim {\ClaimA}
is true there is a quasi-primary basis according to the assumptions and
we can use the shorthand notation (\secwCl.8). First we note that the
generators $W^{(j)}$ of conformal dimension $d(W^{(j)}) < 2$ must already
be primary because of the assumptions. Now consider any
generator $W^{(j)}$ of conformal dimension $d(W^{(j)}) = 2$.
Since $W^{(j)}$ is quasi-primary, the OPE with $L$ must read
$L \star W^{(j)} = \alpha_j c + 2 W^{(j)}$. Then
$\hat{W}^{(j)} = W^{(j)} - 2 \alpha_j L$ is primary. For all other dimensions
we first observe that
$$\eqalign{
L \star \Qp^{(r)}(L,\phi) = & {d(\phi)+r-1 \choose r}
                            {c (r+4)! \over 12} \phi
                            + (d(\phi)+r+2) \Qp^{(r)}(L,\phi) \cr
                 & + \sum_{\psi,s \atop d(\psi)+s+2 \le d(\phi)+r}
                   \alpha^s_{\phi,\psi} \Qp^{(s)}(L,\psi) \cr
}        \eqno({\rm \secwCl.9})$$
The $\alpha^s_{\phi,\psi}$ are certain structure constants.
For the generators $W^{(j)}$ of conformal dimension not equal to two, the OPE
with $L$ must have the form
$$L \star W^{(j)} = d(W^{(j)}) W^{(j)} + \sum_{k \le d(W^{(j)})-2}
                           \alpha_k \phi^{(k)}
        \eqno({\rm \secwCl.10})$$
with one quasi-primary field $\phi^{(k)}$ per conformal dimension $k$.
Now we can use (\secwCl.9) to add fields $\Qp^{(r)}(L,\phi^{(k)})$ with
suitable parameters. Since there may be further fields $\psi$ of the
same dimensions on the r.h.s.\ of (\secwCl.9) this procedure may have
to be iterated adding further fields with suitable parameters. According
to the assumptions there are only finitely many fields below a fixed
conformal dimension. Therefore, after adding finitely many counterterms
one is left with a linear system of equations that must be solved.
Once this system is solved, the generator becomes primary and claim
{\ClaimB} is verified.
{}From (\secwCl.9) one obviously needs $c \ne 0$ in order to have an invertible
system. For low conformal dimensions (e.g.\ $d(W^{(j)}) < 4$) one can argue
that
this system is indeed invertible. The general case may become quite complicated
and we will therefore leave the proof of invertibility to future
investigations.
\medskip
We would like to conclude this section by remarking that from here on it
is in principle straightforward to construct classical $\w$-algebras by
making some assumptions e.g.\ on the conformal dimensions and then checking
Jacobi identities. For example, one can check that there is a unique
classical $\w(2,3)$ (up to isomorphism) if the central terms are chosen
to be non-zero.
\bn
\chapsubtitle{\secwCO.\ Classical cosets and orbifolds}
\mn
In this section we discuss orbifolds and cosets of classical $\w$-algebras.
During the attempts to understand a mysterious algebra of type
$\w(2,4,6)$ $\q{\howcl}$ using some classical limit evidence emerged that
a classical analogon would need more generators than this quantum
$\w$-algebra and that the same statement should be true for the orbifold of
of the $N=1$ quantum super Virasoro algebra (for more details see
$\q{\unicos}$). It was then argued in $\q{\ajl}$ that generic classical
cosets and orbifolds are --in contrast to the quantum level-- infinitely
generated and satisfy infinitely many constraints. Let us now briefly
summarize the results of $\q{\ajl}$.
\mn
Classical orbifolds are defined in the same manner as they are defined at
the quantum level (see the end of section {\secwA}). Below we restrict
to $\Zed_2$ automorphisms $\rho$ that act on the finitely many generators
$\{W_a \mid a \in \I \cup \K \}$ as follows:
$$\eqalign{
\rho(W_a) =& W_a \phantom{-} \qquad \forall a \in \K \ , \cr
\rho(W_b) =& -W_b \qquad \forall b \in \I \ . \cr
}\eqno(\secwCO.1)$$
We can divide the index set $\I$ into two subsets: A set
$\I_1$ referring to {\it bosonic} fields and a set $\I_2$
referring to {\it fermionic} fields transforming
nontrivially under the automorphism $\rho$.
It is easy to determine a generating
set {\it classically}. Note that the nontrivial $\rho$-invariant
differential polynomials are even order in the $\{W_b \mid b \in \I \}$.
Plainly, every even order polynomial can be regarded as a polynomial
in quadratic expressions. Therefore the quadratic expressions formed
out of the $\{W_b \mid b \in \I \}$ generate the orbifold together
with the invariant fields $\{W_a \mid a \in \K \}$. A redundant
set of quadratic generators is given by:
$$X_{b,c}^{i,j} := W_b^{(i)} W_c^{(j)}
\qquad b,c \in \I, \ 0 \le i,j \in \Zed
\eqno(\secwCO.2)$$
where $W_b^{(i)} := \partial^i W_b$. The derivative acts on the generators
(\secwCO.2) as follows:
$$\partial X_{b,c}^{i,j} = X_{b,c}^{i+1,j} + X_{b,c}^{i,j+1} .
\eqno(\secwCO.3)$$
Using the action of the derivative (\secwCO.3) and paying
attention to the Pauli principle for the fermionic generators,
i.e.\ that fermions have odd Grassmann parity, one can choose
the following minimal set of generators for the orbifold:
$$\eqalign{
W_a , & \ a \in \K \qquad \qquad \qquad \qquad \qquad \qquad \qquad \qquad
\hbox{(invariant generators)} \, , \cr
X_{b,c}^{0,j} &:= W_b \partial^j W_c , \quad b<c , \ b,c \in
         \I, \ 0 \le j\in\Zed \, , \cr
X_{d,d}^{0,j} &:= W_d \partial^j W_d , \quad d \in \I_1, \ 0 \le j \in 2 \Zed
          \qquad \qquad \hbox{(square of bosons)} \, , \cr
X_{e,e}^{0,j} &:= W_e \partial^j W_e , \quad e \in \I_2, \ 0 < j \in 2 \Zed + 1
          \qquad \ \hbox{(square of fermions)} \cr
}\eqno(\secwCO.4)$$
where `$b<c$' denotes some ordering of the original generators. Eq.\
(\secwCO.4)
shows that $\Zed_2$ orbifolds are always {\it infinitely} generated at
the classical level (if there is one field transforming non-trivially under
$\rho$).
\mn
In order to find the complete set of relations we first regard all
$W_a^{(i)}$ as independent. The complete set of relations satisfied
by the redundant set of generators (\secwCO.2) is generated by
$$\eqalignno{
X_{b,c}^{i,j} - \epsilon_{b,c} X_{c,b}^{j,i} &= 0, &({\rm \secwCO.5a}) \cr
X_{b,c}^{i,j} X_{d,e}^{k,l} -
\epsilon_{c,d} X_{b,d}^{i,k} X_{c,e}^{j,l} &= 0, &({\rm \secwCO.5b}) \cr
}$$
where $\epsilon_{b,c} = -1$ if {\it both} $W_b$ and $W_c$ are
fermions, and $\epsilon_{b,c} = 1$ otherwise. (Clearly,
choosing certain indices in (\secwCO.5) equal leads to trivial relations).
The proof that (\secwCO.5) indeed generate all relations is a simple
sorting argument. Plainly, a basis for invariant polynomials is given
by those differential monomials that have even degree in the generators
transforming non-trivially under $\rho$ if the generators are arranged
in a suitable order. Obviously, any given monomial in the fields
(\secwCO.2) can be re-sorted into this order using the relations (\secwCO.5)
and (anti-)commutativity of any two fields.
\mn
It is straightforward to derive the relations satisfied by the
nonredundant set of generators (\secwCO.4) from (\secwCO.5). One simply has to
recursively apply (\secwCO.3) (which encodes the action of the derivative) in
order to express the relations (\secwCO.5) in terms of the generators
(\secwCO.4).
\medskip
Next, we further elaborate some of these relations for two examples.
One of the simplest examples of orbifolds is the bosonic projection
of the $N=1$ super Virasoro algebra. The $N=1$ super Virasoro algebra
is the extension of the Virasoro algebra $L$ by a primary dimension
${3 \over 2}$ fermion $G$. According to (\secwCO.4)
a nonredundant set of generators for the classical orbifold is
$$L, \qquad \Phi^n := G \partial^n G \qquad \hbox{for all odd $n$}.
\eqno(\secwCO.6)$$
In particular, this orbifold has one generator at each positive
{\it even} scale dimension. Using the notation of (\secwCO.4) we have
the identification $\Phi^n = X^{0,n}$ if we omit the irrelevant
lower indices. From (\secwCO.5b) one reads off
$0 = X^{0,j} X^{0,l} + X^{0,0} X^{j,l} = X^{0,j} X^{0,l}$ because
$X^{0,0} = 0$. In terms of the generators (\secwCO.6) these
infinitely many relations read
$$\Phi^n \Phi^m = 0 \qquad \hbox{for any} \quad 0 < n,m \in 2 \Zed + 1.
\eqno(\secwCO.7)$$
In this case the particular subset (\secwCO.7) of relations for
the nonredundant set of generators can also immediately be
inferred from the Pauli-principle:
$\Phi^n \Phi^m = (G \partial^n G) (G \partial^m G)
    = -G^2 (\partial^n G) (\partial^m G) = 0$.
However, (\secwCO.5) encodes more relations. For example
(\secwCO.5) and (\secwCO.3) imply $X^{0,1} \partial^2 X^{0,1} =
X^{0,1} (X^{2,1} + 2 X^{1,2} + X^{0,3} )
= X^{0,1} X^{1,2} = -X^{0,1} X^{1,2}$. For the nonredundant
set of generators this implies the following relation at
scale dimension 10:
$$\Phi^1 \partial^2 \Phi^1 = 0.
\eqno(\secwCO.8)$$
\indent
As a simple second example we consider two commuting
copies of the Virasoro algebra ($L_1$ and $L_2$) with
equal central charges. Then $W := L_1 - L_2$ is primary with
respect to $L := L_1 + L_2$. Furthermore, $\rho(L) = L$
and $\rho(W) = -W$ is an automorphism of this $\w(2,2)$.
According to (\secwCO.4) the subspace invariant under $\rho$
is generated by the following fields:
$$L, \qquad \tilde{\Phi}^n := W \partial^n W \qquad \hbox{for all even $n$}.
\eqno(\secwCO.9)$$
Again, we obtain one generator at each positive even scale
dimension. In this case rewriting the relations (\secwCO.5) for
the redundant set (\secwCO.3) in terms of the nonredundant set
(\secwCO.4) is slightly more complicated. Using the notation
$X^{i,j} = W^{(i)} W^{(j)}$ one checks that from (\secwCO.3)
and (\secwCO.5a)
$\partial^2 (X^{0,0} X^{0,0}) = 8 X^{0,1} X^{0,1}
            + 2 X^{0,0} \partial^2 X^{0,0}$
and $\partial^2 X^{0,0} = 2 X^{1,1} + 2 X^{0,2}$. Using (\secwCO.5b)
it is straightforward to check that
$\partial^2 (X^{0,0} X^{0,0}) + 6 X^{0,0} \partial^2 X^{0,0}
            + 8 X^{0,0} X^{0,2} = 0$.
This relation arises at scale dimension 10, which is the lowest scale
dimension admitting a relation. In terms of the generators (\secwCO.9) it reads
$$\partial^2 (\tilde{\Phi}^0 \tilde{\Phi}^0)
- 6 \tilde{\Phi}^0 \partial^2 \tilde{\Phi}^0
+ 8 \tilde{\Phi}^0 \tilde{\Phi}^2 = 0.
\eqno(\secwCO.10)$$
\indent
Now we turn to a discussion of cosets. First, we briefly discuss cosets
of the type $W / \G$ with $\G$ as simple Lie algebra. By definition,
the generating set of the coset is a generating set for those differential
polynomials in $W$ that are invariant with respect to the action of $\G$.
Problems of this type are known in the mathematical literature under the name
of
`{\it invariant theory}' and one can use their results (see e.g.\ $\q{\weyl}$)
in order to obtain a generating set. Observing that $\hat{\G}_k \subset W$
and that there is always an invariant bilinear form on $\G$ it is
immediately clear that the coset will have infinitely many generators
because any bilinear expression in the currents of $\hat{\G}_k$ with
an arbitrary number of derivatives is an invariant field.
Concerning the relations one also sees that there will be infinitely
many of them because non-trivial invariant polynomials in {\it finitely}
many variables always satisfy some relations. Now one can put infinitely
many derivatives (like for the generators) and it is clear that there
will be infinitely many relations.
\mn
Cosets of the type $W / \hat{\G}_k$ are a bit more complicated.
The following result was obtained in $\q{\ajl}$:
Assume that the restriction of the central term of $W$ to
$\hat{\G}_k$ is nondegenerate (we do note assume $\G$ to be simple)
and that it is possible to partition the generating fields of $W$
into the generating fields $J_a$ of $\hat{\G}_k$ and a complementary
set of generating fields $J_i^{\perp}$ that form primary multiplets
with respect to the current algebra $\hat{\G}_k$ (to avoid confusion
note that the $J_i^{\perp}$ need not have conformal dimension one).
Under these conditions the adjoint action of $\hat{\G}_k$ on its
complement $W$ is a representation of $\G$:
$$\clOPE{J_a(z)}{J_i^{\perp}(w)} = \sum_j {T(a)^j_i J_j^{\perp}(w)
                                   \over (z-w)}
  \eqno(\secwCO.11)$$
in terms of matrices $T(a)$. Now one introduces a covariant
derivative $\D$ acting on the fields $J_i^{\perp}$
$$\D J_i^{\perp} := \partial J_i^{\perp} - {1 \over k}
                   \sum_{a,b,j} J_b g^{b a} T(a)^j_i J_j^{\perp}
  \eqno(\secwCO.12)$$
where $g^{b a}$ is the inverse of the metric on $\G$.
Note that the covariant derivative $\D$ is indeed a derivation.
The reason simply is that if a field $\phi$ transforms under a representation
$T$ and $\chi$ under a representation $\hat{T}$ then $\phi \partial^n \chi$
transforms under the tensor product of the representations $T$ and
$\hat{T}$.
\mn
Now the coset $W / \hat{\G}_k$ is the space of those
$\G$-invariant differential polynomials in the fields $J_i^{\perp}$
where as derivative the covariant derivative (\secwCO.12) has to be used.
{}From this characterization it is immediately clear that $W / \hat{\G}_k$
must have infinitely many generators satisfying infinitely many
constraints precisely for the same reasons as for $W / \G$.
\mn
A simple consequence of this general result is that for classical cosets of
type $\hat{{\cal G}}_k \oplus \hat{{\cal H}}_m / \hat{{\cal H}}_{i k +m}$
(where $i$ denotes the Dynkin index of the embedding ${\cal H} \subset \G$)
one has
$$\lim_{m \to \infty} {\hat{\G}_k \oplus \hat{{\cal H}}_m \over
   \hat{{\cal H}}_{i k +m}} = {\hat{\G}_k \over {\cal H}}
  \eqno(\secwCO.13)$$
and that the coset on the l.h.s.\ for generic $m$ is a deformation of the
coset on the r.h.s. Eq.\ (\secwCO.13) is also true for cosets of
quantum $\w$-algebras  (see e.g.\ $\q{\bogo}$).
\medskip
Let us now conclude this general discussion of classical cosets and turn
to some more concrete statements. First we observe that
if one wants to perform quasi-primary projections in coset $\w$-algebras
one may use the formula (\secwCl.6) if one replaces the ordinary
derivative $\partial$ by the covariant derivative $\D$. The reason is mainly
that $\D - \partial$ commutes with the coset energy-momentum tensor $L$.
\mn
For cosets of Kac-Moody algebras the coset energy-momentum tensor $L$ is
always composite which implies $c=0$. Then $L$ is obviously primary.
Under these conditions the following fields are primary as well if $\phi$
and $\chi$ are primary fields of dimension $d(\phi)$ and $d(\chi)$
respectively:
$$\Qp^{(0)}(\phi,\chi) = \phi \chi \, , \qquad
  \oh \Qp^{(1)}(\phi,\chi) = d(\phi) \phi  \chi' - d(\chi) \phi' \chi \, .
  \eqno(\secwCO.14)$$
As a curiosity aside we note that the covariance properties of the second
expression have already been noted several decades ago in a different
context (see e.g.\ eq.\ (24) on page 23 of $\q{\wilczynski}$). It should
be clear to the reader that the assumptions of claim {\ClaimB} are not
satisfied for composite $L$ and thus there need not be a primary
generating set.
\mn
Let us now illustrate these basic remarks with a few statements on some
classical cosets. The absence of a primary generating set can be seen
very nicely in the classical coset $\widehat{sl(2)}/\widehat{U(1)}$
(compare also $\q{\toppan}$). The fields
$$W_{n+2} := \Qp^{(n)}(J^{+}, J^{-}) \, , \qquad 0 \le n \in \Zed
  \eqno(\secwCO.15)$$
form a quasi-primary generating set of the classical coset
$\widehat{sl(2)}/\widehat{U(1)}$ (for the relations see $\q{\ajl}$).
The fields (\secwCO.15) cannot have a primary projection unless they are
already primary. It can be checked explicitly that $W_2 = k L$ and $W_3$
are primary whereas $W_4$, $W_5$ and $W_6$ are {\it not} primary. This
demonstrates that there is no primary basis in the classical coset
$\widehat{sl(2)}/\widehat{U(1)}$
\footnote{${}^{7})$}{
Note, however, that due to (\secwCO.14) there are more primary fields in this
coset than claimed in $\q{\toppan}$.}.
\mn
Also the classical coset $SVIR(N=2)/\widehat{U(1)}$ can be controlled quite
well. One finds precisely one generator for each integer scale dimension
greater or equal to two. For this coset the energy-momentum tensor
$L$ has a non-vanishing central term because it is not composite.
Thus, according to claim {\ClaimB} it should be possible to choose
the generators with dimension greater than two such that they are primary.
This can in fact be checked explicitly for small conformal dimensions
(we have performed checks up to conformal dimension five). For more
details on this coset see $\q{\ajl}$.
\mn
Finally, let us briefly consider the classical coset $\widehat{sl(2)}_k/sl(2)$.
{}From the discussion of the $\beta-\gamma$-system in $\q{\ajl}$ one
expects that the classical coset $\widehat{sl(2)}_k/sl(2)$ should have
a specific Poisson subalgebra at $k=- {1 \over 2}$. In order to
check this one can perform explicit computations of Poisson
brackets in $\widehat{sl(2)}_k/sl(2)$ at low scale dimension. We use
normalization conventions analogous to $\q{\unicos}$. According to
$\q{\ajl}$ we have to consider the following generators:
$$\eqalign{
S_{m,n} :&=
 (\partial^m J^{+}) \partial^n J^{-}
+ {\textstyle {1 \over 2}} (\partial^m J^{0}) \partial^n J^{0}
+ (\partial^m J^{-}) \partial^n J^{+} \, , \cr
S_{m,n,k} :&=
 (\partial^m J^{-}) (\partial^n J^{0}) \partial^k J^{+}
- (\partial^m J^{-}) (\partial^n J^{+}) \partial^k J^{0}
- (\partial^m J^{0}) (\partial^n J^{-}) \partial^k J^{+} \cr
&+ (\partial^m J^{0}) (\partial^n J^{+}) \partial^k J^{-}
+ (\partial^m J^{+}) (\partial^n J^{-}) \partial^k J^{0}
- (\partial^m J^{+}) (\partial^n J^{0}) \partial^k J^{-}
\, . \cr
}  \eqno(\secwCO.16)$$
Eliminating the redundant generators from (\secwCO.16) we see that
this classical coset is of type $\w(2,4,6,6,8,8,9,10,10,\ldots)$.
We can choose the following non-redundant quasi-primary generators
up to scale dimension 6:
$$\eqalign{
L &= {\textstyle {1 \over 2 k}} S_{0,0} \, , \cr
W_4 &= S_{0,2} + {\textstyle {3 \over 2}} S_{1,1} \, , \cr
W_{\rm 6a} &= S_{0,4} - 10 S_{1,3} + 10 S_{2,2} \, , \cr
W_{\rm 6b} &= S_{0,1,2} . \cr
}  \eqno(\secwCO.17)$$
One can check that $L$ and $W_{\rm 6b}$ are primary whereas $W_4$
and $W_{\rm 6a}$ are only quasi-primary which once again demonstrates
the absence of a primary generating set. Looking for a Poisson subring
one finds the following useful linear combination of generators:
$$\hat{W}_6 := {k \over 7} W_{\rm 6a} - 2 W_{\rm 6b} \, .
  \eqno(\secwCO.18)$$
We have checked up to scale dimension 6 that $L$, $W_4$ and $\hat{W}_6$
close among themselves under Poisson brackets for all $k$
(i.e.\ we have not looked at any fields beyond scale dimension 6 on the
r.h.s.\ of the classical OPEs -- still a quite non-trivial fact).
{}From this observation and the study of the $\beta-\gamma$-system in
$\q{\ajl}$ it seems sensible to expect that the classical coset
$\widehat{sl(2)}_k/sl(2)$ has a Poisson subalgebra generated by specific even
dimensional fields for all values of the level $k$.
\bigskip
Concerning the quantum case of orbifolds and cosets in general
there is the following hypothesis:
Upon normal ordering of the classical relations correction terms emerge
that contain classical generators such that one can eliminate precisely one
generator per generator of the classical relations. This `{\it cancellation}'
has been checked so far rigorously only for a very simple coset $\q{\ajl}$
and for a few dimensions in some other cosets and orbifolds $\q{\ajl,\unicos}$.
Note that the character argument which has been used in the argumentation
supporting the results of the next section relies heavily on the assumption
that this cancellation always happens.
\bn
\chapsubtitle{\secwSQ.\ The structure of quantum $\bw$-algebras}
\mn
One of the main aims of the explicit constructions of quantum $\w$-algebras
e.g.\ in refs.\ $\q{\blm,\kau,\hornfeck,\howcl}$ was to gain some insight into
the structure of quantum $\w$-algebras and the associated rational conformal
field theories (RCFTs). The rational models of those
$\w$-algebras existing for isolated $c$ only can be either interpreted as
extensions or truncations of $\w$-algebras obtained by DS reduction, or the
effective central charge is integer (see $\q{\proceedings}$ for a summary).
After identifying the Casimir $\w$-algebras and the orbifold of the $N=1$
super Virasoro algebra among the $\w$-algebras existing for generic $c$ two
unexplained solutions remained: An algebra of type $\w(2,4,6)$ $\q{\kau}$ and
an algebra of type $\w(2,3,4,5)$ $\q{\hornfeck}$ which both give rise to
generic null fields. In particular the origin of this $\w(2,4,6)$ was
unexplained for some time $\q{\howcl}$. In the meantime both of these
algebras have been identified $\q{\ajl,\unilet,\unicos}$ and it was
realized that they fit in a general new structure: They play the r\^ole
of `{\it unifying $\w$-algebras}' for Casimir $\w$-algebras $\q{\unilet,
\unicos}$. The purpose of this section is to explain this new
unifying structure.
\mn
One of the basic observations is that truncations of Casimir $\w$-algebras
are quite general, i.e.\ that for certain values of the central charge $c$
some of the generators turn out to be null fields. This can be investigated
systematically using the Kac determinant $\q{\bouwschou}$ for the vacuum
representation. This has been carried out in $\q{\unilet}$ leading
to the truncations in table 8.
\mn
\centerline{\vbox{
\hbox{
\vrule \hskip 1pt
\vbox{ \offinterlineskip
\def\tablespace{ height2pt&\omit&&\omit&&\omit&\cr }
\def\tablerule{ \tablespace\noalign{\hrule}\tablespace }
\hrule\halign{&\vrule#&\strut\quad\hfil#\hfil\quad\cr
\tablespace
\tablespace
& {\it Casimir algebra}  && $c$ && {\it truncated algebra} &\cr
\tablerule
& ${\cal WA}_n$  && $c_{{\cal A}_n}(n+r,n+s)$
&& $\w(2,\dots,rs-1)$ &\cr
\tablerule
&${\cal WB}_n$   && $c_{{\cal B}_n}(2 n -2+r,2 n -1+s) $
&& $\w(2,4,\dots,rs-2)$ for $r\Cdot s$ even   &\cr
\tablerule
&${\cal WC}_n$   && $c_{{\cal C}_n}(n+r,2 n -1+s) $
&& $\w(2,4,\dots,rs-2)$  for $r\Cdot s$ even  &\cr
\tablerule
& $\orb{{\cal WD}_n}$   && $c_{{\cal D}_n}(2 n - 3+r,2 n - 3+s) $
&& $\w(2,4,\dots,rs-2)$   for $r\Cdot s$ even &\cr
\tablespace}
\hrule}\hskip 1pt \vrule}
\hbox{\quad Table 8: Truncations of Casimir $\w$-algebras}}
}
\mn
The parameterization of the central charge in table 8 is given by
the following generalization of (\secwB.2):
$$c_{\Lie_n}(p,q) = n - 12 {(q \rho - p \rho^{\vee})^2 \over p q}
       \eqno({\rm \secwSQ.1})$$
where $\rho^{\vee}$ is the dual Weyl vector (for simply laced $\Lie_n$
one has $\rho^{\vee} = \rho$).
\mn
The result of table 8 gives rise to the structure indicated for
the Casimir $\w$-algebras ${\cal WA}_n$ in Fig.\ 4. The horizontal
lines correspond to the algebras ${\cal WA}_n$
that exist for generic central charge $c$ (or equivalently, generic
`level' $k$). For the $k$th unitary minimal model all generators
of ${\cal WA}_n$ with dimension greater than $k^2+3k+1$ become null
fields leading to a truncation of ${\cal WA}_n$ to an algebra of
type $\w(2,\ldots,k^2+3k+1)$ for all sufficiently large $n$.
It was argued in $\q{\unilet}$ that one can interpolate these
truncations at fixed $k$ to non-integer $n$ giving rise to a new
$\w$-algebra existing for generic $c$ -- a `{\it unifying $\w$-algebra}'.
\mn
In general we will call a $\w$-algebra a unifying $\w$-algebra if it is
finitely generated, exists for generic $c$ and has the property that a
discrete family of $\w$-algebras exists (e.g.\ Casimir $\w$-algebras
based on a Lie algebra of fixed type) such that infinitely many members
of this family at certain distinct values of the central charge truncate
to this unifying algebra. Since unifying $\w$-algebras exist for generic
$c$ they can e.g.\ be thought of as continuations of Casimir $\w$-algebras
${\cal WL}_n$ to real values of the rank $n$ for certain values of the
central charge. Usually, unifying $\w$-algebras give rise to null fields,
i.e.\ they are {\it non-freely} generated.
\mn
The part of Fig.\ 4 corresponding to unitary minimal models ($k$ a positive
integer) can be understood in terms of level-rank duality
$\q{\altschuler,\bogo}$ using the coset realization of ${\cal WA}_{n-1}$:
$${\cal WA}_{n-1} \cong
   {\widehat{sl(n)}_k\oplus \widehat{sl(n)}_1
       \over {\widehat{sl(n)}_{k+1}}} \cong
   {\widehat{sl(k+1)}_n \over
     {\widehat{sl(k)}_n\oplus \widehat{U(1)}}} = {\cal CP}(k) \, .
       \eqno({\rm \secwSQ.2})$$
\mn
$$\verline{0}{0}{100}
  \leftputbox{99}{-0.1}{$\rightarrow$}
  \putbox{-5}{103}{$n$}
  \putbox{10}{-5}{$1$}
  \putbox{20}{-5}{$2$}
  \putbox{30}{-5}{$3$}
  \putbox{40}{-5}{$4$}
  \putbox{50}{-5}{$5$}
  \putbox{60}{-5}{$6$}
  \putbox{70}{-5}{$7$}
  \putbox{80}{-5}{$8$}
  \putbox{90}{-5}{$9$}
  \Verline{10}{-1.5}{3}
  \Verline{20}{-1.5}{3}
  \Verline{30}{-1.5}{3}
  \Verline{40}{-1.5}{3}
  \Verline{50}{-1.5}{3}
  \Verline{60}{-1.5}{3}
  \Verline{70}{-1.5}{3}
  \Verline{80}{-1.5}{3}
  \Verline{90}{-1.5}{3}
  \horline{0}{0}{100}
  \topputbox{0.1}{99}{$\uparrow$}
  \putbox{103}{-5}{$k$}
  \putbox{-5}{10}{$1$}
  \putbox{-5}{20}{$2$}
  \putbox{-5}{30}{$3$}
  \putbox{-5}{40}{$4$}
  \putbox{-5}{50}{$5$}
  \putbox{-5}{60}{$6$}
  \putbox{-5}{70}{$7$}
  \putbox{-5}{80}{$8$}
  \putbox{-5}{90}{$9$}
  \Horline{-1.5}{10}{3}
  \Horline{-1.5}{20}{3}
  \Horline{-1.5}{30}{3}
  \Horline{-1.5}{40}{3}
  \Horline{-1.5}{50}{3}
  \Horline{-1.5}{60}{3}
  \Horline{-1.5}{70}{3}
  \Horline{-1.5}{80}{3}
  \Horline{-1.5}{30}{3}
  \horline{0}{10}{100}
  \horline{0}{20}{100}
  \horline{0}{30}{100}
  \horline{0}{40}{100}
  \horline{0}{50}{100}
  \horline{0}{60}{100}
  \horline{0}{70}{100}
  \horline{0}{80}{100}
  \horline{0}{90}{100}
  \leftputbox{105}{10}{${\cal WA}_1 \cong {\rm Vir}$}
  \leftputbox{105}{20}{${\cal WA}_2 \cong \w(2,3)$}
  \putbox{120}{110}{${\cal WA}_n \cong \w(2,\ldots,n+1)$}
  \putbox{120}{105}{($c$ generic)}
  \putbox{120}{95}{$\downarrow$}
  \putbox{120}{85}{$\w(2,\ldots,k^2+3k+1)$}
  \putbox{120}{80}{(at level $k$)}
  \topputbox{10}{113}{$\w(2,3,4,5)$}
  \topputbox{10}{105}{$\displaystyle \cong {\widehat{sl(2)}_n \over
                       \widehat{U(1)}}$}
  \verline{10}{0}{100}
  \verline{20}{0}{100}
  \verline{30}{0}{100}
  \verline{40}{0}{100}
  \verline{50}{0}{100}
  \verline{60}{0}{100}
  \verline{70}{0}{100}
  \verline{80}{0}{100}
  \verline{90}{0}{100}
  \cross{10}{10.2}
  \cross{10}{20.2}
  \cross{10}{30.2}
  \cross{10}{40.2}
  \cross{10}{50.2}
  \cross{10}{60.2}
  \cross{10}{70.2}
  \cross{10}{80.2}
  \cross{10}{90.2}
  \leftputbox{11.5}{7.3}{$\Zed_2$}
  \leftputbox{11.5}{17.3}{$\Zed_3$}
  \leftputbox{11.5}{27.3}{$\Zed_4$}
  \leftputbox{11.5}{37.3}{$\Zed_5$}
  \leftputbox{11.5}{47.3}{$\Zed_6$}
  \leftputbox{11.5}{57.3}{$\Zed_7$}
  \leftputbox{11.5}{67.3}{$\Zed_8$}
  \leftputbox{11.5}{77.3}{$\Zed_9$}
  \leftputbox{11.5}{87.3}{$\Zed_{10}$}
  \verline{66}{-5}{2}
  \verline{66}{-2.5}{2}
  \verline{66}{0}{2}
  \verline{66}{2.5}{2}
  \verline{66}{5}{2}
  \verline{66}{7.5}{2}
  \verline{66}{10}{2}
  \verline{66}{12.5}{2}
  \verline{66}{15}{2}
  \verline{66}{17.5}{2}
  \verline{66}{20}{2}
  \verline{66}{22.5}{2}
  \verline{66}{25}{2}
  \verline{66}{27.5}{2}
  \verline{66}{30}{2}
  \verline{66}{32.5}{2}
  \verline{66}{35}{2}
  \verline{66}{37.5}{2}
  \verline{66}{40}{2}
  \verline{66}{42.5}{2}
  \verline{66}{45}{2}
  \verline{66}{47.5}{2}
  \verline{66}{50}{2}
  \verline{66}{52.5}{2}
  \verline{66}{55}{2}
  \verline{66}{57.5}{2}
  \verline{66}{60}{2}
  \verline{66}{62.5}{2}
  \verline{66}{65}{2}
  \verline{66}{67.5}{2}
  \verline{66}{70}{2}
  \verline{66}{72.5}{2}
  \verline{66}{75}{2}
  \verline{66}{77.5}{2}
  \verline{66}{80}{2}
  \verline{66}{82.5}{2}
  \verline{66}{85}{2}
  \verline{66}{87.5}{2}
  \verline{66}{90}{2}
  \verline{66}{92.5}{2}
  \verline{66}{95}{2}
  \verline{66}{97.5}{2}
  \verline{66}{100}{2}
  \verline{66}{102.5}{2}
  \verline{66}{105}{2}
  \verline{66}{107.5}{2}
  \putbox{66}{113}{$k \in \Rational$}
  \hskip 130 mm \hfill
$$
\sn
{\par\noindent\figindents
{\bf Fig.\ 4:}
The structure of the Casimir $\w$-algebras ${\cal WA}_n$ and their
unifying $\w$-algebras.
\par\noindent}
\mn
More precisely, it was shown $\q{\altschuler,\bogo}$ that the cosets
in (\secwSQ.2) give rise to equivalent energy-momentum tensors. In
$\q{\unicos}$ evidence was collected that (\secwSQ.2) is in fact true on the
level of $\w$-algebras. Thus, the symmetry algebra of the ${\cal CP}(k)$
model is a unifying $\w$-algebra for the $k$th unitary minimal
model of ${\cal WA}_{n-1}$. Note that the l.h.s.\ of (\secwSQ.2) is defined for
integer $n$ and arbitrary $k$, whereas the r.h.s.\ is defined for integer
$k$ and general $n$. The isomorphism in (\secwSQ.2) is valid iff $k$ and $n$
are both positive integers. Note that ${\cal CP}(k)$ does indeed exist
for generic $c$ or $k$, thus giving rise to a coset realization of the
unifying $\w$-algebra predicted by inspection of the Kac-determinant.
\mn
It was argued in $\q{\unicos}$ that the field content of ${\cal CP}(k)$
is given by
$${\cal CP}(k) \cong \w(2,3,\ldots,k^2+3 k +1)
       \eqno({\rm \secwSQ.3})$$
with two generic null fields at dimension $k^2+3 k + 4$ which is consistent
with the prediction from the Kac-determinant.
\mn
Now we would like to discuss the first unitary minimal models of the
algebras ${\cal WA}_{n-1}$ in some more detail. On the one hand, the
case $k=1$ corresponds to the $\Zed_n$ parafermions which are relevant
to the $\Zed_n$-quantum spin chains that we have discussed earlier.
On the other hand, the coset $\widehat{sl(2)}_n / \widehat{U(1)}$
arising at $k=1$ obviously has two effective degrees of freedom, and
using the character argument of P.\ Bouwknegt (see e.g.\ $\q{\bouwschou}$)
it is obvious that it should be finitely generated. Nevertheless, there
is some confusion about this question of the generating set in the literature
(see e.g.\ $\q{\bakri,\narganes,\yuwu}$). Therefore we would like to
explicitly recall what we have been able to show with some effort in
$\q{\unicos}$:
\mn
\beginpprop{\PropSQ}{\q{\unicos}}
The quantum coset $\widehat{sl(2)}_n / \widehat{U(1)}$ has a closed
$\w(2,3,4,5)$ subalgebra with two generic null fields at conformal dimension 8.
Furthermore, there are no additional generators of conformal dimension 6, 7
or 8 in the coset $\widehat{sl(2)}_n / \widehat{U(1)}$.
\endprop
\mn
\beginremarks{}
\item{1)}
More strongly, a character argument $\q{\unicos}$ indicates that the coset
$\widehat{sl(2)}_n / \widehat{U(1)}$ is isomorphic to the $\w(2,3,4,5)$.
Unfortunately, we have not been able to prove this stronger statement.
\item{2)}
This $\w(2,3,4,5)$ is the previously unexplained algebra found in
$\q{\hornfeck}$.
\endremark
\mn
We have already mentioned that this $\w(2,3,4,5)$ unifies the first
unitary minimal models of ${\cal WA}_{n-1}$ that are related to critical
points in the $\Zed_n$-spin quantum chains of earlier sections. In
view of sections {\secwB} and {\secwC} we would be interested in the
twisted sector of this algebra. Indeed, an inner automorphism of
$\widehat{sl(2)}_n$ gives rise to an outer $\Zed_2$ automorphism of the
coset $\widehat{sl(2)}_n / \widehat{U(1)}$ $\q{\unicos}$. Unfortunately,
at the moment we do not know how to apply the standard machinery of
coset representations in the presence of an inner automorphism which means
that in contrast to the periodic sector the twisted sector cannot be easily
obtained from the coset construction. Nevertheless, one can argue that the
orbifold is of type $\w(2,4,6,8,10)$ with a generic null field at scale
dimension $14$ for generic $n$. According to section {\secwC} this orbifold
should be relevant for the $\Zed_n$-spin chain with {\it free}
boundary conditions. We conclude this discussion by recalling the simple
observation that due to the coset realization the irreducible modules of
$\Zed_n$ parafermionic model can grow at most as fast as a free module
of two chiral bosonic fields. It would be interesting to see if the
presence of a unifying object at generic real $n$ can be used for
obtaining more insights into $\Zed_n$ parafermionic conformal field theories.
\medskip
The investigation of the Kac determinant is not restricted to integer $k$
but also applies to $k \in \Rational$ -- including in particular the
non-unitary minimal models. Using the parameterization (\secwSQ.1)
the field content can be read off from table 8. As before, one can
interpolate these further unifying $\w$-algebras to generic $n$ (or $c$).
These unifying $\w$-algebras can also be realized in terms of cosets.
Now, one needs Drinfeld-Sokolov reductions based on $sl(r)$ for a
{\it non-principal} embedding such that they have a
$\widehat{sl(k)} \oplus \widehat{U(1)}$ Kac-Moody subalgebra:
$$\w^{sl(r)}_{r-k,1^{k}} = \w(1^{k^2},2,3,\ldots,r-k,
\left({\textstyle {r-k+1 \over 2}}\right)^{2 k}) .
\eqno(\secwSQ.4)$$
The $k^2$ currents form a $\widehat{sl(k)} \oplus \widehat{U(1)}$
Kac-Moody algebra, the fields of dimension 2, $\ldots$, $r-k$ are singlets
with respect to this Kac-Moody and the $2 k$ fields of dimension
${r-k+1 \over 2}$ are a Kac-Moody multiplet. More precisely, these $2 k$
fields transform as two $U(1)$-charge conjugate defining representations of
$sl(k)$. In $\q{\unicos}$ it was argued that
$${\cal WA}_{n-1} \cong
{\w^{sl(r)}_{r-k,1^{k}} \over \widehat{sl(k)} \oplus \widehat{U(1)}}
\cong \w(2,3,\ldots,(k + 1) r + k)
 \qquad \hbox{at} \qquad c_{{\cal A}_{n-1}}(n+k,n+r)
\eqno(\secwSQ.5)$$
with two generic null fields at dimension $(k+1) r + k + 3$.
Identifying $\w^{sl(r+1)}_{1^{r+1}}$ with the unconstrained
$\widehat{sl(r+1)}$ Kac-Moody algebra, the case $r = k+1$ is identical
to the ${\cal CP}(k)$ cosets which we have already discussed.
One can also apply (\secwSQ.5) to the case $k=0$. In this case, the $sl(2)$
embedding is the principal one and there is no Kac-Moody subalgebra.
Thus, the coset in (\secwSQ.5) is just ${\cal WA}_{r-1}$ and we recover
eq.\ (2.4) of ref.\ $\q{\unilet}$ (note that this equality was originally
observed in $\q{\nakan}$).
\medskip
The Casimir algebras ${\cal WA}_n$ in Fig.\ 4 obviously exist also for
negative $k$ and from the preceding discussion we conclude that one can
make sense of the unifying $\w$-algebras also for negative $n$. However,
it is also possible to directly make sense of ${\cal WA}_n$ at negative
$n$ $\q{\kneu}$ (compare also the comments on ${\cal WD}_{-m}$ below)
or even for complex $n \in \Complex$ $\q{\enriquez}$.
Among those, the ${\cal WA}_n$ at negative $n$ are defined less formally
since they can be given a coset realization
${\cal WA}_{-n-1} \cong (\widehat{sl(n)}_k\oplus \widehat{sl(n)}_{-1})
/ {\widehat{sl(n)}_{k-1}}$ and have a definite finite generating
set $\q{\kneu}$.
\mn
Using ${\cal WA}_n$ for real $n$ or the fact that Fig.\ 4 is densely covered
by unifying $\w$-algebras one can define $\q{\kneu}$ a `{\it universal}'
$\w$-algebra for all Casimir algebra based on $\A_n$ that depends on
{\it two} parameters, e.g.\ $k$ and $n$. For generic irrational
$k \not\in \Rational$ this universal $\w$-algebra will have
infinitely many generators. Doing so, one can nicely fit the $\w_\infty$
and $\w_{1+\infty}$ algebras (see e.g.\ $\q{\bakinf}$) in the picture
of Fig.\ 4: They can be recovered from the topmost border $n = \infty$
respective right border $k = \infty$. Then, e.g.\ the truncations of
$\w_{1+\infty}$ at $c=n$ $\q{\fkacr}$ arise naturally from Fig.\ 4
(for more details see $\q{\unicos}$).
\medskip
To summarize, the picture of the space of all Casimir algebras ${\cal WA}_n$
and their unifying $\w$-algebras as presented in Fig.\ 4 is mainly a
conjecture. However, this conjecture has been thoroughly checked in
$\q{\unilet,\unicos}$: Beyond the inspection of the Kac determinant and
cross-checks with limiting cases which we mentioned above one can
perform character computations for the cosets and compare structure constants
or minimal models (where known). Even more, the simplest of the relevant
cosets can be constructed explicitly $\q{\unicos}$ and e.g.\ the statement
about the generating set can be quite rigorously verified.
\medskip
The structure of the Casimir algebras ${\cal WD}_n$, ${\cal WC}_n$ and
${\cal WB}_n$ is similar to Fig.\ 4 which we have discussed in some detail.
The truncations inferred from the Kac determinant are presented in table
8. Whereas in the case of ${\cal WA}_n$ we were able to provide a coset
realization for all unifying $\w$-algebras, coset realizations of unifying
$\w$-algebras for the other Casimir algebras are only partially known.
The known coset realizations for unifying $\w$-algebras of Casimir algebras
are summarized in table 9 (for a detailed discussion see $\q{\unicos}$).
Sometimes an additional $\Zed_2$ orbifold is needed which is denoted
by `${\rm Orb}$' in table 9. If `${\rm Orb}$' is put in brackets
only part of the algebras give rise to non-trivial outer automorphisms --
for the other ones we do not have to take an orbifold.
\mn
\centerline{\vbox{\hbox{
\vrule \hskip 1pt
\vbox{ \offinterlineskip
\def\tablespace{ height2pt&\omit&&\omit&&\omit&&\omit&&\omit&\cr }
\def\tablerule{ \tablespace\noalign{\hrule}\tablespace }
\hrule\halign{&\vrule#&\strut\hskip1mm\hfil#\hfil\hskip1mm\cr
\tablespace\tablespace
& {\it Casimir}  && {\it central charge} && {\it coset realization} &&
  {\it dimensions of} && {\it dimension of} & \cr
\tablespace
& {\it algebra}  && $c$ && {\it of unifying algebra} &&
  {\it simple fields} && {\it first null field} & \cr
\tablerule
& ${\cal WA}_{n-1}$  && $c_{{\cal A}_{n-1}}(n \pl k,n \pl r)$ &&
  ${\w^{sl(r)}_{r-k,1^{k}} \over \widehat{sl(k)} \oplus \widehat{U(1)}}$ &&
    $2,3,\ldots,kr+r+k$ && $kr+r+k+3$ & \cr
\tablerule
&${\cal WB}_n$   && ${\hbox{$c_{{\cal B}_n}(2n \pl k \hbox{$-$} 1, 2n\pl 1)$}
                    \atop \hbox{$c_{{\cal B}_n}(2n , 2n \pl k)$}}$ &&
 ${\rm (Orb)}\left({\widehat{so(k)}_\kappa \oplus \widehat{so(k)}_1 \over
    \widehat{so(k)}_{\kappa+1}}\right)$ &&
    $2,4,\ldots,2 k$ && $2k + 4$ & \cr
\tablerule
&${\cal WC}_n$   && $c_{{\cal C}_n}(n \pl k \pl 1,2 n \pl 2 k \pl 1)$ &&
 ${\widehat{sp(2k)}_n \oplus \widehat{sp(2k)}_{-{1 \over 2}} \over
    \widehat{sp(2k)}_{n-{1 \over 2}} }$ &&
    $2,4,\ldots,2 k^2 + 4  k$ && $2 k^2 + 4 k +5$ & \cr
\tablerule
& $\orb{{\cal WD}_n}$ && $c_{{\cal D}_n}(2n \pl k \mi 2,2n \pl k \mi 1)$ &&
   $\orb{ {\widehat{so(k+1)}_{2 n} \over \widehat{so(k)}_{2n} } }$ &&
    $2,4,\ldots,k^2+3 k$ && $k^2+3k+4$ & \cr
\tablespace}
\hrule}\hskip 1pt \vrule}
\hbox{\quad Table 9: Coset realizations of unifying $\w$-algebras}
}}
\mn
Among the unifying $\w$-algebras for these three families of Casimir
algebras those for ${\cal WC}_n$ are particularly interesting because
they can in some sense be regarded as algebras of type `${\cal WD}_{-k}$' --
continuations of the ${\cal WD}_m$ series of Casimir algebras to negative
$m$. Historically, the mysterious $\w(2,4,6)$ was the starting point for
the discovery of unifying $\w$-algebras in $\q{\unilet,\unicos}$. It was
proposed in $\q{\klausREP}$ that this $\w(2,4,6)$ can in some sense be
regarded as belonging to the series of ${\cal WD}_n$ Casimir algebras:
$\w(2,4,6) \cong {\cal WD}_{-1}$. Furthermore, all known minimal models
of this $\w(2,4,6)$ $\q{\howcl}$ were identical with minimal models of some
${\cal WC}_n$ Casimir algebra. Only after generalizing these observations
we became aware that the algebras ${\cal WD}_{-k}$ can be given a group
theoretical meaning.
\mn
Negative dimensional groups are known in the literature (see e.g.\
$\q{\dunne}$). More precisely, some aspects of the representation
theory of the classical groups can be understood naturally after continuation
to negative dimensions. For example, the dimension formulas for $SO(2n)$
and $Sp(2m)$ can be related by transposition of the Young tableaux for
$m = -n$. Similar relations can also be established for the Casimir operators
upon interchanging symmetrization and antisymmetrization. These observations
can be written in a compact form e.g.\ $SO(-2n) \cong \overline{Sp(2n})$
where the overbar means interchange of symmetrization and antisymmetrization.
For the Kac-Moody algebras one can now establish a relation for the levels
by equating the central charges of the Sugawara energy-momentum tensors.
For the case we are interested in one finds $\q{\unicos}$:
$$\widehat{so(-2k)_\kappa} = \widehat{sp(2k)}_{-{\kappa \over 2}} \, .
     \eqno(\secwSQ.6)$$
Thus, from the coset realization of ${\cal WD}_m$ one finds that
$${\cal WD}_{-k} \cong
{\widehat{sp(2k)}_{\kappa} \otimes \widehat{sp(2k)}_{-{1 \over 2}} \over
 \widehat{sp(2k)}_{\kappa-{1 \over 2}}}
 \cong \w(2,4,\ldots,2k(k+2))
     \eqno(\secwSQ.7)$$
with a first generic null field at conformal dimension $2k^2+4k+5$.
This fields content can e.g.\ be inferred from a character argument
$\q{\unicos}$. The algebras ${\cal WD}_{-k}$ are also supposed to be
unifying $\w$-algebras for the ${\cal WC}_n$ Casimir algebras
$\q{\unilet}$:
$${\cal WD}_{-k} \cong {\cal WC}_n
 \qquad \hbox{at} \qquad c_{{\cal C}_n}(n+k+1,2n+2k+1) \, .
     \eqno(\secwSQ.8)$$
The field content of (\secwSQ.7) is consistent with the
truncations of table 8 predicted for the minimal models of
(\secwSQ.8).
\medskip
To summarize, the space of all $\w$-algebras gives rise to complicated
structures including a `{\it unifying structure}' that we have discussed
in this section. However, there are indications $\q{\unicos}$ that the
only rational models of unifying $\w$-algebras are located at intersection
points with the Casimir $\w$-algebras. This implies that unifying $\w$-algebras
do probably not give rise to new RCFTs, and that the classification problem
of RCFTs (which is far from being solved) might in fact be simpler than a
classification of $\w$-algebras. One might hope that (super-symmetric)
quantized DS exhausts all possible RCFTs (with exceptions at (half-)integer
effective central charge). Thus, it would be desirable to have at least a good
understanding of representations of {\it non-principal} DS $\w$-algebras
where very little is known so far. Nevertheless, the unifying $\w$-algebras
could still be of some use because on the one hand they might lead to
conceptual simplifications and on the other hand not all physical phenomena
are necessarily described by RCFTs.
\vfill
\eject
{\helvits
{\par\noindent
\leftskip=1cm
\baselineskip=11pt
... In these colleges, the professors contrive new rules and methods of
agriculture and building, and new instruments and tools for all trades
and manufactures, whereby, as they undertake, one man shall do the work of ten;
a palace may be built in a week, of materials so durable as to last
forever without repairing. All the fruits of the earth shall come to
maturity at whatever season we think fit to choose, and increase an
hundred-fold more than they do at present; with innumerable other happy
proposals. The only inconvenience is, that none of these projects are
yet brought to perfection; ...
\par\noindent}
\sn
\rightline{Jonathan Swift in ``A Voyage to Balnibarbi''}
\par\noindent}
\bn
\chaptitle{\secJ.\ Conclusion}
\mn
In this thesis we have discussed several aspects of quantum spin models and
extended conformal algebras.
\medskip
We have investigated an algebraic condition that guarantees integrability of
quantum spin models -- the so-called `Dolan-Grady-condition' (also called
superintegrability). In one dimension we achieved a classification of
superintegrable nearest neighbour interaction quantum spin models based
on $sl(2)$. For two dimensions the only result we obtained so far is a
non-existence statement of certain superintegrable spin-$1/2$ models.
Nevertheless, this approach seems to be feasible also in higher dimensions
and deserves further attention.
\medskip
In a large part of this thesis we have studied the $\Zed_n$-chiral Potts
model without necessarily demanding integrability.
We have presented an argument using perturbation theory that shows that the
massive high-temperature phases of all $\Zed_n$-spin quantum chains exhibit
quasiparticle spectra with $n-1$ fundamental particles. Since the
argument relies on perturbation theory it applies rigorously only to
very high temperatures. Due to the perturbative nature of the details
we were not able to give it any predictive power for those case where some
of the fundamental particles cross with scattering states. For these cases
one needs different methods, e.g.\ Bethe ansatz techniques
$\q{\dkcoy,\kedem}$ or numerical methods $\q{\lett}$.
Nevertheless, the basic idea of approximating a multi-particle state
by single-particle states sitting on subparts of the chain might be
applicable in the entire massive high-temperature phase.
\sn
Our derivation of the quasiparticle picture involving $n-1$ fundamental
particles is valid also for the scaling region near the conformal point
$\la = 1$, $\phi = \vphi = 0$, the only non-rigorous part of the proof being
the radius of convergence.
\mn
Having derived such a quasiparticle picture the main open problem
to obtain the scattering matrix of the corresponding massive field theory.
\mn
Furthermore, we discussed a general duality property stating
equality of spectra in the low- and high-temperature phase.
Thus, the quasiparticle interpretation for the high-temperature
phase of the general $\Zed_n$-chiral Potts quantum chain
can be pulled over to the low-temperature phase. However, charge
$Q$ and boundary conditions $R$ are interchanged by the duality
transformation. In particular, for periodic boundary conditions one sees
only energy levels above the ground state that correspond to composite particle
states.
\medskip
We also studied correlation functions using a perturbation expansion
for the ground state of the $\Zed_3$-model and numerical techniques.
Although these approaches are limited to short ranges, we were not
only able to estimate correlation lengths in the massive high-temperature
phase but it also turned out that the correlation functions have
oscillatory contributions. For very high temperatures the oscillation
length is proportional to the inverse of one of the chiral
angles $L \sim \phi^{-1}$. We further observed that the oscillation
length is closely related to the minimum of the dispersion relations
for general values of the parameters. The relation $L P_{\min}
= 2 \pi$ is valid on the lattice with a much better accuracy than
the well-known relation $\xi \sim m^{-1}$. For special values of the
parameters we were able to derive the relation $L P_{\min} = 2 \pi$
from a form factor decomposition but one should certainly try to gain more
insight into the correlation functions, in particular into the
long-distance behaviour of the amplitude.
\medskip
We have then turned to extended conformal algebras. First, we have shown
how automorphisms of $\w$-algebras can be used to enlarge minimal models
by imposing twisted boundary conditions on the chiral fields. Some of these
rational conformal fields theories are relevant for the $\Zed_n$-spin
quantum chains. Furthermore, there is a striking one-to-one correspondence
of boundary conditions for the chiral fields of a $\w$-algebra and the
quantum spin chain.
\mn
Classical $\w$-algebras are an ideal testing ground for the study of
construction principles. Surprisingly, certain basic properties can
not as easily be derived for them as for quantum $\w$-algebras.
We have argued that the space of fields of a classical $\w$-algebra should
under very general conditions decompose into $su(1,1)$ highest weight
representations, i.e.\ it should admit a basis of quasi-primary fields
and derivatives thereof. We have also explained why we expect that it
should be possible to choose the generators primary for a classical
$\w$-algebra if, in addition, the conformal dimensions of all generators
are strictly positive and the Virasoro centre $c$ is non-zero.
\mn
Classical cosets and orbifolds can be treated systematically. For orbifolds
with respect to a $\Zed_2$ automorphism the generators and relations are
completely under control. Concerning classical cosets one can in general
reduce the problem of finding the generators and relations to a problem
in the invariant theory of finite-dimensional Lie algebra. Where the
solution to the problem in invariant theory is known (e.g.\ for $sl(2)$
-- see $\q{\weyl}$) the space of fields is also completely under control.
It follows from general considerations that the space of fields of
classical cosets and orbifolds is always an infinitely generated differential
polynomial ring subject to infinitely many relations.
\mn
The situation is contrary for quantum $\w$-algebras. It seems that on the
quantum level cosets and orbifolds are always {\it finitely} generated.
This difference is due to a cancellation mechanism between classical
generators and relations. Note however that this mechanism has been
rigorously checked only for a few cases $\q{\ajl,\unicos}$. Nevertheless,
there had been some puzzles concerning the space of $\w$-algebras that
could be explained once this difference was recognized.
\mn
A further difference between classical and quantum $\w$-algebras is that
quantum $\w$-algebras can truncate at certain values of the central charge
because some of the generators become null fields. One can argue by
inspection of the Kac determinant that such truncations are in fact a
very general phenomenon for quantum Casimir $\w$-algebras. These truncations
give rise to new `unifying' quantum $\w$-algebras that interpolate these
truncations of Casimir $\w$-algebras. The notion of `unifying $\w$-algebras'
is a generalization and more precise version of level-rank-duality.
Unifying $\w$-algebras can frequently be realized in terms of cosets, not only
but also if this is expected from level-rank-duality. In the coset
realization of unifying $\w$-algebras non-principal DS reductions and
also negative dimensional groups play a certain r\^ole. In particular,
algebras of type ${\cal WD}_{-k}$ arise as unifying $\w$-algebras for
the Casimir algebras ${\cal WC}_n$.
\vfill
\chapsubtitle{Acknowledgements}
\mn
I would like to thank everybody who has in some way or other contributed to
the successful completion of this thesis.
\sn
First of all, I would like to thank my teachers Prof.\ G.\ von Gehlen and
Prof.\ W.\ Nahm for all their support, encouragement and further help.
\sn
I am grateful to W.\ Eholzer and R.\ H\"ubel for sharing most of the
experience from the very beginning of our studies until now.
\sn
Thanks go also to my other collaborators including R.\ Blumenhagen, J.\ de
Boer, M.\ Flohr, N.S.\ Han, K.\ Hornfeck and R.\ Varnhagen. In particular,
I would like to thank L.\ Feh\'er who taught me much of what I know e.g.\
about classical $\w$-algebras.
\sn
The warm atmosphere at the Physikalisches Institut in Bonn has been of
great help for my work. Without being able to mention everybody
I would like to thank F.\ E{\ss}ler, H.\ Hinrichsen, J.\ Kellendonk,
A.\ Recknagel, M.\ R\"osgen, R.\ Schimmrigk, M.\ Terhoeven, A.\ Wi{\ss}kirchen
and K.\ Yildirim for many useful discussions.
\sn
I am indebted to the Max-Planck-Institut f\"ur Mathematik in Bonn-Beuel for
providing the computer time necessary for performing the more complicated
algebraic and numerical calculations reported in this work.
\sn
Last but not least I would like to thank my wife Gabriele not only for her
tolerance with respect to the long hours spent at the institute but also
for her financial support during the initial stages of this thesis.
\eject
\chapsubtitle{Appendix {\appA}: Perturbation expansions for two-particle
states}
\mn
In this appendix we explicitly diagonalize the first order of the perturbation
expansion for two identical particles.
\sn
Let us first consider $Q_1 \ne Q_2$. In this case we specialize to $n=3$
such that we can choose $Q_1 = 1$, $Q_2 = 2$.
Now, the potential $V$ acts in the space (\secD.9) as:
$$\eqalign{
q r(V) \pstate{t^{1,2}_1}_P =&
-{2 \over \sqrt{3}} \left(
e^{-i(\phit - P)} \pstate{t^{1,2}_{N-1}}_P
+ 2 \cPh e^{-i(\phit + \Ph)} \pstate{t^{1,2}_2}_P \right)
\cr
q r(V) \pstate{t^{1,2}_j}_P =&
-{2 \over \sqrt{3}} \left(
2 \cPh e^{i(\phit + \Ph)} \pstate{t^{1,2}_{j-1}}_P
+ 2 \cPh e^{-i(\phit + \Ph)} \pstate{t^{1,2}_{j+1}}_P \right) \cr
& \hskip 5cm
\qquad 1 < j < N-1 \cr
q r(V) \pstate{t^{1,2}_{N-1}}_P =&
-{2 \over \sqrt{3}} \left(
2 \cPh e^{i(\phit + \Ph)} \pstate{t^{1,2}_{N-2}}_P
+ e^{i(\phit - P)} \pstate{t^{1,2}_{1}}_P \right)
\cr
} \eqno({\rm \appA.1})$$
where $q$ is the projector onto the space (\secD.9).
Although it is not difficult to diagonalize (\appA.1) numerically
for comparably long chains (e.g.\ $N=100$), we did not succeed in
obtaining a closed expression for the eigenvalues or eigenvectors.
\mn
In the second case, i.e.\ $Q := Q_1 = Q_2$ introduce the abbreviation $W$ by:
$$-{2 \over \bsin{{\pi Q \over n}}}  \bcos{\Ph - \PMQ{Q}} W
  \pstate{t^{Q,Q}_j}_P :=
  q r(V) \pstate{t^{Q,Q}_j}_P.
    \eqno({\rm \appA.2})$$
In the case of two identical excitations we will also have to
distinguish between even and odd momenta in terms of lattice
sites which is conveniently encoded in the abbreviation $\delta_P^N$
defined by (\secD.10).
The action of the potential $V$ now is
$$\eqalign{
W \pstate{t^{Q,Q}_1}_P =& \left(
e^{-i \Ph} \pstate{t^{Q,Q}_2}_P \right)
\cr
W \pstate{t^{Q,Q}_j}_P =& \left(
e^{i \Ph} \pstate{t^{Q,Q}_{j-1}}_P
+ e^{-i \Ph} \pstate{t^{Q,Q}_{j+1}}_P \right)
\qquad 1 < j < \left\lb{\textstyle {N \over 2}}\right\rb-1 \cr
W \pstate{t^{Q,Q}_{\lb{N \over 2}\rb-1}}_P =&
\cases{
\left(e^{i \Ph} \pstate{t^{Q,Q}_{{N-5 \over 2}}}_P
+ e^{-i \Ph} \pstate{t^{Q,Q}_{{N-1 \over 2}}}_P \right),
& if $N$ odd; \cr
\left(e^{i \Ph} \pstate{t^{Q,Q}_{{N \over 2}-2}}_P
+ \delta_P^N \sqrt{2} e^{-i \Ph} \pstate{t^{Q,Q}_{{N \over 2}}}_P \right),
& if $N$ even \cr
} \cr
W \pstate{t^{Q,Q}_{\lb{N \over 2}\rb}}_P =&
\cases{
\left(e^{i \Ph} \pstate{t^{Q,Q}_{{N-3 \over 2}}}_P
- (-1)^{\delta_P^N} \pstate{t^{Q,Q}_{{N-1 \over 2}}}_P \right),
& if $N$ odd; \cr
\delta_P^N \sqrt{2} e^{i \Ph} \pstate{t^{Q,Q}_{{N \over 2}-1}}_P,
& if $N$ even. \cr
} \cr
} \eqno({\rm \appA.3})$$
At first sight (\appA.3) looks much more complicated than (\appA.1).
This is however misleading and the matrix $W$ can be diagonalized explicitly.
In order to do so, we exploit a connection to graph theory
(see e.g.\ $\q{\jones}$). Before performing an explicit diagonalization
we first present a graphical representation of (\appA.1) and (\appA.3).
Each vector $\pstate{t_j^{Q_1,Q_2}}_P$
will be symbolized as a `$\bullet$' with the index written above.
The action of the potential $V$ is symbolized by lines, with the
square of the matrix elements (up to an isomorphism to be presented
below) attached to them. Assume first that
we could distinguish the two flips we make. Then the graphical
representation for the action of the potential $V$ (or $W$)
would be
$$\bu{1}\mskp\Hr{1}\mskp\bu{2}\cdots\cdots\mkern-12mu
\bu{N - 2}\mkern-12mu
\mskp\Hr{1}\mskp\mkern-12mu
\bu{N - 1} = \left({\cal A}_{N-1}\right).
  \eqno({\rm \appA.4})$$
Here `$\left({\cal L}_k\right)$' denotes the incidence matrix
derived from the Cartan matrix of a Lie algebra ${\cal L}_k$.
However, the states $\pstate{t^{Q,Q}_j}_P$ and $\pstate{t^{Q,Q}_{N-j}}_P$
are proportional to each other and must therefore be identified.
Furthermore,
it turns out that for $N$ even and ${N P \over 2 \pi}$ odd
$\pstate{t^{Q,Q}_{N \over 2}}_P  = 0$ vanishes identically. This
already splits the graph (\appA.4) into two disjoint parts.
Therefore, a graphical representation of (\appA.3) is given by:
$$\eqalign{
W \cong \bu{1}\mskp\Hr{1}\mskp\bu{2}\cdots\cdots\mkern-12mu
\bu{{N\over 2} - 2}\mkern-12mu
\mskp\Hr{1}\mskp\mkern-12mu
\bu{{N\over 2} - 1} & = \left({\cal A}_{{N\over 2} -1}\right)
\qquad {\rm for} \ N \ {\rm even} , \ {N P \over 2 \pi} \ {\rm odd} \cr
W \cong \bu{1}\mskp\Hr{1}\mskp\bu{2}\cdots\cdots\mkern-12mu
\bu{{N-3\over 2}}\mkern-12mu
\mskp\Hr{1}\mskp\mkern-12mu
\bu{{N-1\over 2}}\mkern-21mu\bigcirc \ {\scriptstyle 1}&
 = \left({\cal T}_{{N-1\over 2}}\right)
\qquad \ {\rm for} \ N \ {\rm odd} \cr
W \cong \bu{1}\mskp\Hr{1}\mskp\bu{2}\cdots\cdots\mkern-12mu
\bu{{N\over 2} - 2}\mkern-12mu\mskp\Hr{1}\mkern-12mu
\mskp\bu{{N\over 2} - 1}\mkern-14mu\mskp\Hrr{2}
\mkern-4mu\mskp\bu{{N\over 2}} &
 = \left({\cal B}_{{N\over 2}}\right)
\qquad \ \ \ {\rm for} \ N \ {\rm even} ,
 \ {N P \over 2 \pi} \ {\rm even}. \cr
}  \eqno({\rm \appA.5})$$
Fortunately, all the graphs (\appA.5) have norm less or equal to $2$
\footnote{${}^{8})$}{$\left({\cal T}_k\right)$ is the Tadpole graph.}.
Because the eigenvalues of such graphs are classified
$\q{\jones}$ we can derive the first order explicitly.
\medskip
In the case of (\appA.1) the situation is different. In order to
simplify the discussion consider the case $P = \phi = 0$. Then one
can represent (\appA.1) as
$$V \approx
\cdots\cdots\mkern-12mu\bu{N-1}\mkern-12mu\mskp\Hr{1}
\mskp\bu{1}\mskp\Hrf{4}\mskp\bu{2}\cdots\cdots\mkern-12mu
\bu{N - 2}\mkern-12mu
\mskp\Hrf{4}\mskp\mkern-12mu
\bu{N - 1}\mkern-12mu
\mskp\Hr{1}\cdots\cdots
\ .  \eqno({\rm \appA.6})$$
Note that instead of drawing a closed diagram we have represented
part of it twice. It is easy to see that the norm of (\appA.6) is
larger than 3 (it tends to 4 for $N \to \infty$). The absence
of explicit expressions for the eigenvalues of such graphs prevented
us from deriving an explicit expression for the first order
of two-particle states in the $Q=0$ sector even for the $\Zed_3$-chain.
\medskip
Let us now explicitly diagonalize the matrix (\appA.3).
Although this can be done for all four different cases simultaneously
it is more illustrative to treat first one case separately.
The simplest case is actually $N$ even and
${N P \over 2 \pi}$ odd. First, we perform the following
diagonal change of bases
$$B \pstate{t^{Q,Q}_j}_P := e^{-i \Ph (j-1)} \pstate{t^{Q,Q}_j}_P
. \eqno({\rm \appA.7})$$
If we also identify the $\pstate{t^{Q,Q}_j}_P$ with the standard
basis of $\Complex^{{N \over 2} -1}$
the ${N \over 2} -1$ times ${N \over 2} -1$ matrix $W$ satisfies
$$W = B \left(
\matrix{0 & 1 & 0 & \cdots & 0 \cr
        1 & 0 & 1 & \ddots & \vdots \cr
        0 & 1 & \ddots & \ddots & 0 \cr
   \vdots & \ddots & \ddots & 0 & 1 \cr
        0 & \cdots & 0 & 1 & 0 \cr
}\right) B^{-1}
. \eqno({\rm \appA.8})$$
The eigenvalues and eigenvectors of the matrix on the r.h.s.\ of
(\appA.8) are well-known in the literature (see e.g.\ $\q{\jones}$,
example 1.2.5).
We can use them to write down immediately the eigenvalues and eigenvectors
of the matrix $W$:
$$\eqalign{
\pstate{\tau^{Q,Q}_k}_P :&= B {2 \over \sqrt{N}}
\left( \matrix{\bsin{{2 k \pi \over N}} \cr
           \bsin{{2 k 2 \pi \over N}} \cr
           \vdots \cr
           \bsin{{2 k ({N \over 2} - 1) \pi \over N}} \cr
}\right), \cr
W \pstate{\tau^{Q,Q}_k}_P &= 2 \bcos{{2 k \pi \over N}}
           \pstate{\tau^{Q,Q}_k}_P
\ , \qquad 1 \le k \le {N \over 2} - 1. \cr
}  \eqno({\rm \appA.9})$$
Putting things together one obtains the first order expansions
(\secD.11) and (\secD.12) for $4 \le N$ even, ${N P \over 2 \pi}$ odd.
\mn
The same argument can be applied to the remaining cases. The only
additional consideration is that we need a second change of bases $M$ in
$\Complex^{N-1}$. For $N$ odd it is defined by
$$
M := {1\over \sqrt{2}} \left(
\matrix{1 & 0 & \cdots & \cdots & 0 & 1 \cr
        0 & \ddots & \ddots & \Ddots & \Ddots & 0 \cr
   \vdots & \ddots &  1 & 1 & \Ddots & \vdots \cr
   \vdots & \Ddots & -1 & 1 & \ddots & \vdots \cr
        0 & \Ddots & \Ddots & \ddots & \ddots & 0 \cr
        -1 & 0 & \cdots & \cdots & 0 & 1 \cr
}\right) \, \qquad N-1 \times N-1, N \ {\rm odd}
  \eqno({\rm \appA.10a})$$
whereas for $N$ even it is defined as follows
$$
M := {1\over \sqrt{2}} \left(
\matrix{1 & 0 & \cdots & \cdots & \cdots & 0 & 1 \cr
        0 & \ddots & \ddots & & \Ddots & \Ddots & 0 \cr
   \vdots & \ddots &  1 & 0 & 1 & \Ddots & \vdots \cr
   \vdots &   & 0 & \sqrt{2} & 0 & & \vdots \cr
   \vdots & \Ddots & -1 & 0 & 1 & \ddots & \vdots \cr
        0 & \Ddots & \Ddots &  & \ddots & \ddots & 0 \cr
        -1 & 0 & \cdots & \cdots & \cdots & 0 & 1 \cr
}\right) \, \qquad N-1 \times N-1, N \ {\rm even}.
  \eqno({\rm \appA.10b})$$
With the definition (\appA.7) of the $N-1 \times N-1$
matrix $B$ we can write $W$ as
$$\pmatrix{W^{\rm even} & 0 \cr
           0 & W^{\rm odd} \cr} = B M \left(
\matrix{0 & 1 & 0 & \cdots & 0 \cr
        1 & 0 & 1 & \ddots & \vdots \cr
        0 & 1 & \ddots & \ddots & 0 \cr
   \vdots & \ddots & \ddots & 0 & 1 \cr
        0 & \cdots & 0 & 1 & 0 \cr
}\right) M^{-1} B^{-1}
 \eqno({\rm \appA.11})$$
where $W^{\rm even}$ is the $\lb {N \over 2} \rb \times
\lb {N \over 2} \rb$ matrix $W$ for even lattice momenta
($\delta_P^N = 1$) and $W^{\rm odd}$ is the
$\lb {N-1 \over 2} \rb \times \lb {N-1 \over 2} \rb$ matrix
$W$ for odd lattice momenta (with reversed order of basis vectors).
\mn
Again, we can use the well-known results $\q{\jones}$ to write
down the eigenvalues and eigenvectors of the matrix $W$:
$$\eqalign{
\pstate{\tau^{Q,Q}_k}_P :&= B {2 \over \sqrt{N}}
\left( \matrix{\bsin{{(2 k - \delta_P^N) \pi \over N}} \cr
         \bsin{{(2 k - \delta_P^N) 2 \pi \over N}} \cr
         \vdots_{} \cr
         \bsin{{(2 k - \delta_P^N) (\lb{N \over 2}\rb - 1) \pi \over N}} \cr
         \bsin{{(2 k - \delta_P^N) \lb{N \over 2}\rb \pi \over N}}
\cases{{1 \over \sqrt{2}} & for $N$ even \cr
                        1 & for $N$ odd \cr
} \cr
}\right), \cr
W \pstate{\tau^{Q,Q}_k}_P &=
2 \bcos{{(2 k - \delta_P^N) \pi \over N}} \pstate{\tau^{Q,Q}_k}_P
\ , \qquad 1 \le k \le \left\lb{N \over 2}\right\rb. \cr
}  \eqno({\rm \appA.12})$$
Inserting definitions one immediately obtains the final results
(\secD.11) and (\secD.12) from (\appA.12).
\medskip
Let us now turn to the second order. For simplicity we restrict
to the $Q=2$-sector for $n=3$ and odd $N$. We abbreviate the resolvent by $g$
and its values by:
$$R_{3,1} := - 4 \sqrt{3} \bcos{{\vphi \over 3}} \, , \qquad
R_{1,2} := - 4 \sqrt{3} \bcos{{\pi - \vphi \over 3}} \, , \qquad
R_{0,1} := 4 \sqrt{3} \bcos{{\pi + \vphi \over 3}} \, .
  \eqno({\rm \appA.13})$$
Then we obtain for the matrix elements between the states
$\pstate{t^{1,1}_k}_P$ of the second order expression:
$$\eqalign{
B^{-1} \, & q \, r(V) \, g \, r(V) \, q \, B =
{8 \over 3}\Biggl\{\left\lb
{\bcos{P-{2 \phi \over 3}} + 1 \over R_{0,1}}
 -{1 \over R_{1,2}} + {1 \over R_{3,1}}\right\rb
\left(
\matrix{1 & 0 & \cdots & 0 \cr
        0 & 0 & \cdots & 0 \cr
   \vdots & \vdots & \ddots & \vdots \cr
        0 & 0 & \cdots & 0 \cr
}\right)\cr
&+{\bcos{{P \over 2} + {2 \phi \over 3}} \over R_{1,2}}
\left(
\matrix{0 & 1 & 0 & \cdots & 0 \cr
        1 & 0 & 1 & \ddots & \vdots \cr
        0 & 1 & \ddots & \ddots & 0 \cr
   \vdots & \ddots & \ddots & 0 & 1 \cr
        0 & \cdots & 0 & 1 & 0 \cr
}\right)
+{\bcos{P - {2 \phi \over 3}} \over R_{3,1}}
\left(
\matrix{0 & 0 & 1 & \cdots & 0 \cr
        0 & 0 & 0 & \ddots & \vdots \cr
        1 & 0 & \ddots & \ddots & 1 \cr
   \vdots & \ddots & \ddots & 0 & 1 \cr
        0 & \cdots & 1 & 1 & 0 \cr
}\right) \cr
&\qquad
+\left\lb{2 \over R_{1,2}} + {N-4 \over R_{3,1}} \right\rb \id
\Biggr\} . \cr
}  \eqno({\rm \appA.14})$$
For a chain of length $N$ ($N$ odd), the matrix (\appA.14) is of
size $\lb{N \over 2}\rb \times \lb{N \over 2}\rb$. It is straightforward
to take the matrix elements of this matrix between the first order
eigenstates (\appA.12). We omit the explicit form of this second
order expression because it would be almost as long as (\appA.14).
\vfill
\eject
\chapsubtitle{Appendix {\appB}: Proof of the symmetries of the Hamiltonian}
\mn
This appendix contains a proof of the last line of (\secFORM.7).
\sn
In order to show the third line of (\secFORM.7) we look at the action of
the Hamiltonian (\secB.1) on the states (\secB.11):
$$\eqalign{
H_N^{(n)} & \pstate{i_1 \ldots i_N}_P = -
  \left( \sum_{j=1}^N \sum_{k=1}^{n-1} \ab_k \om^{k \, i_j} \right)
   \pstate{i_1 \ldots i_N}_P \cr
&- \la \sum_{\{a_1,\ldots,a_N\}} \sum_{k=1}^{n-1}
        \a_k V^{\{a_1,\ldots,a_N\}}_{k,\{i_1,\ldots,i_N\}}
        e^{i P_{{\rm m},Q} f^{\{a_1,\ldots,a_N\}}_{k,\{i_1,\ldots,a_N\}} }
        e^{i (P-P_{{\rm m},Q}) g^{\{a_1,\ldots,a_N\}}_{k,\{i_1,\ldots,a_N\}} }
        \pstate{a_1 \ldots a_N}_P  \cr
}  \eqno{(\rm \appB.1)}$$
where the $\pstate{a_1 \ldots a_N}_P$ form a suitable basis of the space
with momentum $P$ and charge $Q$.
The $f^{\{a_1,\ldots,a_N\}}_{k,\{i_1,\ldots,a_N\}}$ and
$g^{\{a_1,\ldots,a_N\}}_{k,\{i_1,\ldots,a_N\}}$ are certain integers
depending on the choice of basis. The
$V^{\{a_1,\ldots,a_N\}}_{k,\{i_1,\ldots,i_N\}}$ are certain {\it real}
constants. The crucial step in the derivation of
the above identity is to show that the basis can be chosen such that
$$f^{\{a_1,\ldots,a_N\}}_{k,\{i_1,\ldots,a_N\}} = k Q^{-1} \,
  \eqno{(\rm \appB.2)}$$
if $Q^{-1} \in \Zed_n$.
The existence of a basis $\pstate{a_1 \ldots a_N}_P$ satisfying
(\appB.1) and (\appB.2) can be seen as follows. Starting from an arbitrary
state with charge $Q$ and momentum $P$, e.g.\ the state $\pstate{s^Q}_P$
defined in (\secD.2), one can iteratively {\it define} the phase factors
of the new basis vectors $\pstate{a_1 \ldots a_N}_P$ appearing in
(\appB.1) such that (\appB.2) holds. It remains to check that this definition
is consistent which is equivalent to showing that a basis vector which
is already introduced turns up with the proper phase factor in (\appB.1).
Since all states $\pstate{a_1 \ldots a_N}_P$ and $\pstate{b_1 \ldots b_N}_P$
with $0 = \sum_{j=1}^N a_j - b_j$ have an equivalent behaviour under the
action (\appB.1) of the Hamiltonian, one can restrict to the states
$\pstate{s^Q}_P$ when checking consistency.
\mn
Let us now consider the states $\pstate{s^Q}_P$ more closely. First we
observe that one application of the potential $V$ has contributions
also to $\pstate{s^Q}_P$. The two relevant terms in $V \pstate{s^Q}_P$
are $\a_Q e^{i P} \pstate{s^Q}_P$ and $\a_{n-Q} e^{-i P} \pstate{s^Q}_P$.
These phase factors are compatible with (\appB.1) and (\appB.2) because
$e^{i 2 P_{{\rm m},Q}}$ is an $n$th root of unity. Now it remains to
check that iterations of (\appB.2) that do not lead to states $\pstate{s^Q}_P$
in intermediate steps yield the correct phase factors. It is easy to see that
the only non-trivial conditions come from iterations of type
$$\pstate{s^Q}_P \, \mathop{\mapsto}\limits_{\a_{k_1}}
                         \, \pstate{(Q+k_1) \, (-k_1) \, 0 \ldots}_P
                 \, \mathop{\mapsto}\limits_{\a_{k_2}}
                          \, \pstate{(Q+k_1+k_2) \, (-k_1-k_2) \, 0 \ldots}_P
\, \ldots \, \mathop{\mapsto}\limits_{\a_{k_r}} \, \pstate{s^Q}_P
  \eqno{(\rm \appB.3)}$$
{\it without} a final shift in the state $\pstate{s^Q}_P$.
The choice (\appB.2) for $Q^{-1}$ an integer is obviously consistent
with eq.\ (\appB.1) since $\sum_{s=1}^r k_s$ has to be a multiple of $n$
if $\pstate{s^Q}_P$ is required to reappear without a shift in (\appB.3).
This concludes the proof that one can indeed choose a basis such that
(\appB.2) is valid.
\mn
The isomorphism stated in (\secFORM.7) can now be read off from the
following actions of $H_N^{(n)}$ and $(H_N^{(n)})^{+}$:
$$\eqalign{
H_N^{(n)} \pstate{i_1 \ldots i_N}_{P_{{\rm m},Q}+P} =
  - & \left( \sum_{j=1}^N \sum_{k=1}^{n-1} \ab_k \om^{k \, i_j} \right)
   \pstate{i_1 \ldots i_N}_{P_{{\rm m},Q}+P} \cr
- \la \! \! \! \sum_{\{a_1,\ldots,a_N\}} \sum_{k=1}^{n-1}
       \a_k & V^{\{a_1,\ldots,a_N\}}_{k,\{i_1,\ldots,i_N\}}
       e^{i P_{{\rm m},Q} Q^{-1} k}
       e^{i P g^{\{a_1,\ldots,a_N\}}_{k,\{i_1,\ldots,a_N\}} }
         \pstate{a_1 \ldots a_N}_{P_{{\rm m},Q}+P}  \, , \cr
(H_N^{(n)})^{+} \pstate{i_1 \ldots i_N}_{P_{{\rm m},Q}-P} =
  - & \left( \sum_{j=1}^N \sum_{k=1}^{n-1} \ab_{n-k} \om^{-k \, i_j} \right)
   \pstate{i_1 \ldots i_N}_{P_{{\rm m},Q}-P} \cr
- \la \! \! \! \sum_{\{a_1,\ldots,a_N\}} \sum_{k=1}^{n-1}
       \a_k V^{\{a_1,\ldots,a_N\}}_{k,\{i_1,\ldots,i_N\}} &
       e^{-i 2 \pi z k} e^{-i P_{{\rm m},Q} Q^{-1} k}
       e^{i P g^{\{a_1,\ldots,a_N\}}_{k,\{i_1,\ldots,a_N\}} }
       \pstate{a_1 \ldots a_N}_{P_{{\rm m},Q}-P} \, .\cr
}  \eqno{(\rm \appB.4)}$$
In (\appB.4) we have used $\ab_k^{*} = \ab_{n-k}$ and $\a_k^{*} =
e^{-{2 \pi i z k}} \a_k$.
We would like to conclude the proof by noting that for very particular
values of the momentum $P$ and the chain length $N$ some of the
eigenstates $\pstate{a_1 \ldots a_N}_P$ in (\appB.4) might turn out to be
identically zero. In this case, the last line of (\secFORM.7) is strictly
true only after inserting projection operators.
\vfill
\eject
\chapsubtitle{References}
\skip103=\parindent
\def\pno{}
\mn
\bibitem{\ahn} C.\ Ahn, K.\ Shigemoto, {\it Onsager Algebra and
              Integrable Lattice Models},
              Mod.\ Phys.\ Lett.\ {\bf A6} (1991) \pno 3509
\bibitem{\albCONF} G.\ Albertini, S.\ Dasmahapatra, B.M.\ McCoy, {\it Spectrum
              Doubling and the Extended Brillouin Zone in the Excitations
              of the Three States Potts Spin Chain}, Phys.\ Lett.\
              {\bf A170} (1992) \pno 397
\bibitem{\albcoy} G.\ Albertini, B.M.\ McCoy,
              {\it Correlation Functions of the Chiral Potts Chain
              from Conformal Field Theory and Finite-Size Corrections},
              Nucl.\ Phys.\ {\bf B350} (1990) \pno 745
\bibitem{\mccoyadv} G.\ Albertini, B.M.\ McCoy, J.H.H.\ Perk,
              {\it Eigenvalue Spectrum of the
              Superintegrable Chiral Potts Model},
              Adv.\ Studies in Pure Math.\ {\bf 19} (1989) \pno 1
\bibitem{\albertiniA} G.\ Albertini, B.M.\ McCoy, J.H.H.\ Perk,
              {\it Commensurate-Incommensurate Transition in the Ground
              State of the Superintegrable Chiral Potts Model},
              Phys.\ Lett.\ {\bf 135A} (1989) \pno 159
\bibitem{\albertiniB} G.\ Albertini, B.M.\ McCoy, J.H.H.\ Perk,
              {\it Level Crossing Transitions and the Massless
              Phases of the Superintegrable Chiral Potts Chain},
              Phys.\ Lett.\ {\bf 139A} (1989) \pno 204
\bibitem{\tang} G.\ Albertini, B.M.\ McCoy, J.H.H.\ Perk, S.\ Tang,
              {\it Excitation Spectrum and Order Parameter for the
              Integrable $N$-State Chiral Potts Model},
              Nucl.\ Phys.\ {\bf B314} (1989) \pno 741
\bibitem{\alcaraz} F.C.\ Alcaraz, A.L.\ Santos, {\it Conservation Laws for
              $\Zed(N)$ Symmetric Quantum
              Spin Models and Their Exact Ground State Energies},
              Nucl.\ Phys.\ {\bf B275} (1986) \pno 436
\bibitem{\altschuler} D.\ Altschuler, {\it Quantum Equivalence of Coset
              Space Models}, Nucl.\ Phys.\ {\bf B313} (1989) \pno 293
\bibitem{\baxterA} H.\ Au-Yang, R.J.\ Baxter, J.H.H.\ Perk,
              {\it New Solutions of the Star-Triangle Relations for
              the Chiral Potts Model},
              Phys.\ Lett.\ {\bf 128A} (1988) \pno 138
\bibitem{\yang} H.\ Au-Yang, B.M.\ McCoy, J.H.H.\ Perk, Sh.\ Tang, M.L.\ Yan,
              {\it Commuting Transfer Matrices in the Chiral Potts Models:
              Solutions of Star-Triangle Equations with Genus $>1$},
              Phys.\ Lett.\ {\bf 123A} (1987) \pno 219
\bibitem{\perkadv} H.\ Au-Yang, J.H.H.\ Perk, {\it Onsager's Star-Triangle
              Equation: Master Key to Integrability},
              Adv.\ Studies in Pure Math.\ {\bf 19} (1989) \pno 57
\bibitem{\baakeA} M.\ Baake, G.\ von Gehlen, V.\ Rittenberg, {\it Operator
              Content of the Ashkin-Teller Quantum Chain -- Superconformal
              and Zamolodchikov-Fateev Invariance: I.\ Free Boundary
              Conditions}, J.\ Phys.\ A: Math.\ Gen.\ {\bf 20} (1987) \pno L479
\bibitem{\baakeB} M.\ Baake, G.\ von Gehlen, V.\ Rittenberg, {\it Operator
              Content of the Ashkin-Teller Quantum Chain -- Superconformal
              and Zamolodchikov-Fateev Invariance: II.\ Boundary Conditions
              Compatible with the Torus}, J.\ Phys.\ A: Math.\ Gen.\ {\bf 20}
              (1987) \pno L487
\bibitem{\bbss} F.A.\ Bais, P.\ Bouwknegt, K.\ Schoutens, M.\ Surridge, {\it
              Extensions of the Virasoro Algebra Constructed from Kac-Moody
              Algebras Using Higher Order Casimir Invariants}, Nucl.\ Phys.\
              {\bf B304} (1988) \pno 348
\bibitem{\bai} F.A.\ Bais, P.\ Bouwknegt, K.\ Schoutens, M.\ Surridge, {\it
              Coset Construction for Extended Virasoro Algebras}, Nucl.\
              Phys.\ {\bf B304} (1988) \pno 371
\bibitem{\bakinf} I.\ Bakas, {\it The Large-$N$ Limit of Extended
              Conformal Symmetries}, Phys.~Lett.~{\bf B228} (1989) \pno 57
\bibitem{\bakri} I.\ Bakas, E.\ Kiritsis, {\it Beyond the Large $N$ Limit:
              Non-Linear $\w_\infty$ as Symmetry of the $SL(2,\Real)/U(1)$
              Coset Model},
              Int.~Jour.~of Mod.~Phys.~{\bf A7}, Suppl.~1A (1992) \pno 55
\bibitem{\baf} J.\ Balog, L.\ Feh\'er, P.\ Forg\'acs, L.\ O'Raifeartaigh, A.\
              Wipf, {\it Toda Theory and $\w$-Algebra from a Gauged WZNW
              Point of View}, Ann.\ Phys.\ {\bf 203} (1990) \pno 76
\bibitem{\bal} J.\ Balog, L.\ Feh\'er, P.\ Forg\'acs, L.\ O'Raifeartaigh, A.\
              Wipf, {\it Kac-Moody Realization of $\w$-Algebras}, Phys.\
              Lett.\ {\bf B244} (1990) \pno 435
\bibitem{\baxterB} R.J.\ Baxter, {\it The Superintegrable Chiral Potts Model},
              Phys.\ Lett.\ {\bf 133A} (1988) \pno 185
\bibitem{\baxterC} R.J.\ Baxter, {\it Superintegrable Chiral Potts Model:
              Thermodynamic Properties, an ``Inverse'' Model, and a Simple
              Associated Hamiltonian}, J.\ Stat.\ Phys.\ {\bf 57} (1989) \pno
              1
\bibitem{\baym} G.\ Baym, {\it Lectures on Quantum Mechanics},
              Benjamin/Cummings Publishing (1969), 3rd printing (1974),
              chapter 11
\bibitem{\bpz} A.A.\ Belavin, A.M.\ Polyakov, A.B.\ Zamolodchikov,
              {\it Infinite Conformal Symmetry in Two-Dimensional Quantum
              Field Theory}, Nucl.\ Phys.\ {\bf B241} (1984) \pno 333
\bibitem{\bersh} M.\ Bershadsky, {\it Conformal Field Theories via Hamiltonian
              Reduction}, Commun.\ Math.\ Phys.\ {\bf 139} (1991) \pno 71
\bibitem{\blg} A.\ Bilal, J.L.\ Gervais, {\it Systematic Construction of
              Conformal Theories with Higher-Spin Virasoro Symmetries}, Nucl.\
              Phys.\ {\bf B318} (1989) \pno 579
\bibitem{\unilet} R.\ Blumenhagen, W.\ Eholzer, A.\ Honecker, K.\ Hornfeck,
              R.\ H\"ubel, {\it Unifying $\w$-Algebras},
              Phys.\ Lett.\ {\bf B332} (1994) \pno 51
\bibitem{\unicos} R.\ Blumenhagen, W.\ Eholzer, A.\ Honecker,
              K.\ Hornfeck, R.\ H\"ubel,
              {\it Coset Realization of Unifying $\w$-Algebras},
              preprint BONN-TH-94-11, DFTT-25/94, hep-th/9406203,
              to appear in Int.\ Jour.\ of Mod.\ Phys.\ {\bf A}
\bibitem{\supwir} R.\ Blumenhagen, W.\ Eholzer, A.\ Honecker, R.\ H\"ubel, {\it
              New $N=1$ Extended Superconformal Algebras with Two and Three
              Generators}, Int.\ Jour.\ of Mod.\ Phys.\ {\bf A7} (1992) \pno
              7841
\bibitem{\blm} R.\ Blumenhagen, M.\ Flohr, A.\ Kliem, W.\ Nahm, A.\ Recknagel,
              R.\ Varnhagen, {\it $\w$-Algebras with Two and Three
              Generators}, Nucl.\ Phys.\ {\bf B361} (1991) \pno 255
\bibitem{\ajl} J.\ de Boer, A.\ Honecker, L.\ Feh\'er,
              {\it A Class of $\w$-Algebras with Infinitely Generated
              Classical Limit},
              Nucl.~Phys.~{\bf B420} (1994) \pno 409
\bibitem{\dBT} J.\ de Boer, T.\ Tjin, {\it The Relation between Quantum
              $\w$ Algebras and Lie Algebras},
              Commun.\ Math.\ Phys.\ {\bf 160} (1994) \pno 317
\bibitem{\bouwschou} P.\ Bouwknegt, K.\ Schoutens, {\it $\w$-Symmetry in
              Conformal Field Theory}, Phys.\ Rep.\ {\bf 223} (1993) \pno 183
\bibitem{\bogo} P.\ Bowcock, P.\ Goddard, {\it Coset Constructions and
              Extended Conformal Algebras},
              Nucl.\ Phys.\ {\bf B305} (1988) \pno 685
\bibitem{\camp} W.J.\ Camp, {\it Decay of Order in Classical Many-Body
Systems.\
              II.\ Ising Model at High Temperatures}, Phys.\ Rev.\ {\bf B6}
              (1972) \pno 960
\bibitem{\cardyA} J.L.\ Cardy, {\it Conformal Invariance and Surface Critical
              Behaviour}, Nucl.\ Phys.\ {\bf B240} (1984) \pno 514
\bibitem{\cardyB} J.L.\ Cardy, {\it Effect of Boundary Conditions on
              the Operator Content of Two-Dimen\-sional
              Conformally Invariant Theories},
              Nucl.\ Phys.\ {\bf B275} (1986) \pno 200
\bibitem{\cardy} J.L.\ Cardy, {\it Critical Exponents of the Chiral
              Potts Model from Conformal Field Theory},
              Nucl.\ Phys.\ {\bf B389} (1993) \pno 577
\bibitem{\rietalB} P.\ Centen, M.\ Marcu, V.\ Rittenberg,
              {\it Non-Universality in $\Zed_3$ Symmetric Spin Systems},
              Nucl.\ Phys.\ {\bf B205} (1982) \pno 585
\bibitem{\chrihen} Ph.\ Christe, M.\ Henkel, {\it Introduction to Conformal
              Invariance and its Applications to Critical Phenomena},
              Lecture Notes in Physics m16, Springer-Verlag (1993)
\bibitem{\dkcoy} S.\ Dasmahapatra, R.\ Kedem, B.M.\ McCoy,
              {\it Spectrum and Completeness of the 3 State
              Superintegrable Chiral Potts Model},
              Nucl.\ Phys.\ {\bf B396} (1993) \pno 506
\bibitem{\daviesA} B.\ Davies, {\it Onsager's Algebra and Superintegrability},
              J.\ Phys.\ A: Math.\ Gen.\ {\bf 23} (1990) \pno 2245
\bibitem{\daviesB} B.\ Davies, {\it Onsager's Algebra and the Dolan-Grady
              Condition in the non-Self-Dual Case},
              J.\ Math.\ Phys\ {\bf 32} (1991) \pno 2945
\bibitem{\dijkgraaf} R.\ Dijkgraaf, C.\ Vafa, E.\ Verlinde, H.\ Verlinde,
              {\it The Operator Algebra of Orbifold Models},
              Commun.\ Math.\ Phys.\ 123 (1989) \pno 485
\bibitem{\dogra} L.\ Dolan, M.\ Grady, {\it Conserved Charges from
              Self-Duality},
              Phys.\ Rev.\ {\bf D25} (1982) \pno 1587
\bibitem{\dunne} G.V.\ Dunne, {\it Negative-Dimensional Groups in
              Quantum Physics}, J.\ Phys.\ A: Math.\ Gen.\ {\bf 22}
              (1989) \pno 1719
\bibitem{\wowoPhD} W.\ Eholzer, {\it Fusion Algebras and Characters of
              Rational Conformal Field Theories}, Ph.D.\ thesis
              BONN-IR-95-10, hep-th/9502160 (1995)
\bibitem{\wirrep} W.\ Eholzer, M.\ Flohr, A.\ Honecker,
              R.\ H{\"u}bel, W.\ Nahm, R.\ Varnhagen,
              {\it Representations of $\w$-Algebras with Two Generators
              and New Rational Models},
              Nucl.\ Phys.\ {\bf B383} (1992) \pno 249
\bibitem{\proceedings} W.\ Eholzer, M.\ Flohr, A.\ Honecker,
              R.\ H{\"u}bel, R.\ Varnhagen,
              {\it $\w$-Algebras in Conformal Field Theory},
              to appear in the proceedings of the workshop
              `Superstrings and Related Topics' Trieste, July 1993
\bibitem{\supwirrep} W.\ Eholzer, A.\ Honecker, R.\ H{\"u}bel,
              {\it Representations of N=1 Extended Superconformal
              Algebras with Two Generators},
              Mod.\ Phys.\ Lett.\ {\bf A8} (1993) \pno 725
\bibitem{\howcl} W.~Eholzer, A.~Honecker, R.~H\"ubel,
              {\it How Complete is the Classification
              of $\w$-Sym\-me\-tries ?}, Phys.~Lett.~{\bf B308}
              (1993) \pno 42
\bibitem{\elitzur} S.\ Elitzur, R.B.\ Pearson, J.\ Shigemitsu,
              {\it Phase Structure of Discrete Abelian Spin and Gauge
              Systems}, Phys.\ Rev.\ {\bf D19} (1979) \pno 3698
\bibitem{\enriquez} B.\ Enriquez, {\it Complex Parametrized $\w$-Algebras:
              The $gl$ Case}, Lett.\ Math.\ Phys.\ {\bf 31} (1994) \pno 15
\bibitem{\lykyanov} V.A.\ Fateev, S.L.\ Luk'yanov,
              {\it The Models of Two-Dimensional Conformal Quantum
              Field Theory with $\Zed_n$ Symmetry},
              Int.\ Jour.\ of Mod.\ Phys.\ {\bf A3} (1988) \pno 507
\bibitem{\flrep} V.A.\ Fateev, S.L.\ Luk'yanov, {\it Additional Symmetries
              and Exactly-Soluble Models in Two-Dimensional Conformal Field
              Theory}, Sov.\ Sci.\ Rev.\ A.\ Phys.\ {\bf 15/2} (1990)
\bibitem{\fatzamA} V.A.\ Fateev, A.B.\ Zamolodchikov,
              {\it Nonlocal (Parafermion) Currents in Two-Dimen\-sional
              Conformal Quantum Field Theory and Self-Dual Critical Points
              in $\Zed_N$-Sym\-metric Statistical Systems},
              Sov.\ Phys.\ JETP {\bf 62} (1985) \pno 215
\bibitem{\fateev} V.A.\ Fateev, A.B.\ Zamolodchikov,
              {\it Conformal Quantum Field Theory Models in Two
              Dimensions Having $\Zed_3$ Symmetry},
              Nucl.\ Phys.\ {\bf B280} (1987) \pno 644
\bibitem{\fatzamB} V.A.\ Fateev, A.B.\ Zamolodchikov,
              {\it Integrable Perturbations of $\Zed_N$ Parafermion
              Models and the $O(3)$ Sigma Model},
              Phys.\ Lett.\ {\bf B271} (1991) \pno 91
\bibitem{\FORT} L.\ Feh\'er, L.\ O'Raifeartaigh, P.\ Ruelle, I.\ Tsutsui,
              {\it On the Completeness of the Set of Classical $\w$-Algebras
              Obtained from DS Reductions}, Commun.\ Math.\ Phys.\ {\bf 162}
              (1994) \pno 399
\bibitem{\laszlorep} L.\ Feh\'er,  L.\ O'Raifeartaigh, P.\ Ruelle,
              I.\ Tsutsui, A.\ Wipf, {\it On Hamiltonian Reductions of the
              Wess-Zumino-Novikov-Witten Theories},
              Phys.\ Rep.\ {\bf 222} (1992) \pno 1
\bibitem{\fehort} L.\ Feh\'{e}r, L.\ O'Raifeartaigh, I.\ Tsutsui,
              {\it The Vacuum Preserving Lie Algebra of a Classical
              $\w$-Algebra}, Phys.\ Lett.\ {\bf B316} (1993) \pno 275
\bibitem{\flohrdipl} M.\ Flohr, {\it $\w$-Algebren, Quasiprim\"are Felder \&
              Nicht-Minimale Modelle}, Diplomarbeit BONN-IR-91-30 (1991)
\bibitem{\frenkel} E.V.\ Frenkel, V.G.\ Kac, M.\ Wakimoto, {\it Characters and
              Fusion Rules for $\w$-Algebras via Quantized Drinfeld-Sokolov
              Reduction}, Commun.\ Math.\ Phys.\ {\bf 147} (1992) \pno 295
\bibitem{\fkacr} E.V.\ Frenkel, V.G.\ Kac, A.\ Radul, W.\ Wang,
              {\it $\w_{1+\infty}$ and $\w(gl_N)$ with Central Charge $N$},
              preprint hep-th/9405121
\bibitem{\gehlenunpub} G.\ von Gehlen, unpublished results from 1986
\bibitem{\gehlenph} G.\ von Gehlen, {\it Two-Particle Structure
              of the $\Zed_3$-Chiral Potts Model}, Proceedings of
              {\it International Symposium on Advanced Topics of
              Quantum Physics}, ed.\ J.Q.\ Liang, M.L.\ Wang, S.N.\ Qiao,
              D.C.\ Su, Science Press Beijing (1992) \pno 248
\bibitem{\hoeger} G.\ von Gehlen, C.\ Hoeger, V.\ Rittenberg,
              {\it Finite-Size Scaling for Quantum Chains with an
              Oscillatory Energy Gap},
              J.\ Phys.\ A: Math.\ Gen.\ {\bf 18} (1985) \pno 1813
\bibitem{\weA} G.\ von Gehlen, A.\ Honecker, {\it Multi-Particle Structure
              in the $\Zed_n$-Chiral Potts Models},
              J.\ Phys.\ A: Math.\ Gen.\ {\bf 26} (1993) \pno 1275
\bibitem{\lett} G.\ von Gehlen, A.\ Honecker, {\it Excitation Spectrum and
              Correlation Functions of the $\Zed_3$-Chiral Potts Quantum
              Spin Chain}, Nucl.\ Phys.\ {\bf B435} (1995) \pno 505
\bibitem{\gehri} G.\ von Gehlen, V.\ Rittenberg, {\it $\Zed_n$-Symmetric
              Quantum Chains with an Infinite Set
              of Conserved Charges and $\Zed_n$ Zero Modes},
              Nucl.\ Phys.\ {\bf B257} (1985) \pno 351
\bibitem{\gehlen} G.\ von Gehlen, V.\ Rittenberg, {\it Operator Content of the
              Three-States Potts Quantum Chain}, J.\ Phys.\ A: Math.\ Gen.\
              {\bf 19} (1986) \pno L625
\bibitem{\gehlenA} G.\ von Gehlen, V.\ Rittenberg, {\it The Spectra of Quantum
              Chains with Free Boundary Conditions and Virasoro Algebras}, J.\
              Phys.\ A: Math.\ Gen.\ {\bf 19} (1986) \pno L631
\bibitem{\gehritell} G.\ von Gehlen, V.\ Rittenberg, {\it The Ashkin-Teller
              Quantum Chain and Conformal Invariance},
              J.\ Phys.\ A: Math.\ Gen.\ {\bf 20} (1987) \pno 227
\bibitem{\schuetz} G.\ von Gehlen, V.\ Rittenberg, G.\ Sch\"utz,
              {\it Operator Content of $n$-State Quantum Chains in the
              $c=1$ Region},
              J.\ Phys.\ A: Math.\ Gen.\ {\bf 21} (1988) \pno 2805
\bibitem{\stringA} D.\ Gepner, {\it Exactly Solvable String Compactifications
              on Manifolds of $SU(N)$ Holonomy}, Phys.\ Lett.\ {\bf B199}
              (1987) \pno 380
\bibitem{\stringB} D.\ Gepner, {\it Space-Time Supersymmetry in Compactified
              String Theory and Superconformal Models}, Nucl.\ Phys.\ {\bf
              B296} (1988) \pno 757
\bibitem{\gepner} D.\ Gepner, Z.\ Qiu, {\it Modular Invariant Partition
              Functions for Parafermionic Field Theories}, Nucl.\ Phys.\
              {\bf B285} (1987) \pno 423
\bibitem{\goddard} P.\ Goddard, A.\ Kent, D.\ Olive, {\it Unitary
              Representations of the Virasoro and Super-Virasoro Algebras},
              Commun.\ Math.\ Phys.\ {\bf 103} (1986) \pno 105
\bibitem{\lacaze} O.\ Golinelli, T.\ Jolicoeur, R.\ Lacaze, {\it Finite-Lattice
              Extrapolations for a Haldane Gap Antiferromagnet}, preprint
              cond-mat/9402043, T94/007
\bibitem{\jones} F.M.\ Goodman, P.\ de la Harpe, V.F.R.\ Jones,
              {\it Coxeter Graphs and Towers of Algebras},
              Springer-Verlag (1989)
\bibitem{\grabo} M.P.\ Grabowski, P.\ Mathieu, {\it Integrability Test for Spin
              Chains}, preprint hep-th/ 9412039, LAVAL-PHY-22/94
\bibitem{\han} N.S.\ Han, A.\ Honecker,
              {\it Low-Temperature Expansions and Correlation
              Functions of the $\Zed_3$-Chiral Potts Model},
              J.\ Phys.\ A: Math.\ Gen.\ {\bf 27} (1994) \pno 9
\bibitem{\hela} M.\ Henkel, J.\ Lacki, {\it Integrable Chiral $\Zed_n$
              Quantum Chains and a New Class of Trigonometric Sums},
              Phys.\ Lett.\ {\bf 138A} (1989) \pno 105
\bibitem{\henkel} M.\ Henkel, G.\ Sch\"utz, {\it Finite-Lattice Extrapolation
              Algorithms}, J.\ Phys.\ A: Math.\ Gen.\ {\bf 21} (1988) \pno
              2617
\bibitem{\hokimA} Q.\ Ho-Kim, H.B.\ Zheng, {\it Twisted Conformal Field
Theories
              with $\Zed_3$ Invariance}, Phys.\ Lett.\ {\bf B212} (1988) \pno
              71
\bibitem{\hokimB} Q.\ Ho-Kim, H.B.\ Zheng, {\it Twisted Structures of Extended
              Virasoro Algebras}, Phys.\ Lett.\ {\bf B223} (1989) \pno 57
\bibitem{\hokimC} Q.\ Ho-Kim, H.B.\ Zheng, {\it Twisted Characters and
Partition
              Functions in Extended Virasoro Algebras}, Mod.\ Phys.\ Lett.\
              {\bf A5} (1990) \pno 1181
\bibitem{\diplom} A.\ Honecker, {\it Darstellungstheorie von $\w$-Algebren und
              Rationale Modelle in der Konformen Feldtheorie},
              Diplomarbeit BONN-IR-92-09 (1992)
\bibitem{\commute} A.\ Honecker, {\it A Note on the Algebraic
              Evaluation of Correlators in Local Chiral Conformal Field
Theory},
              preprint BONN-HE-92-25 (1992), hep-th/9209029
\bibitem{\automos} A.\ Honecker, {\it Automorphisms of ${\cal W}$-Algebras
              and Extended Rational Conformal Field Theories},
              Nucl.\ Phys.\ {\bf B400} (1993) \pno 574
\bibitem{\perturb} A.\ Honecker, {\it A Perturbative Approach to the
              Chiral Potts Model}, preprint BONN-TH-94-21,
              hep-th/9409122
\bibitem{\horn} D.\ Horn, M.\ Weinstein, S.\ Yankielowicz,
              {\it Hamiltonian Approach to $\Zed(N)$ Lattice Gauge Theories},
              Phys.\ Rev.\ {\bf D19} (1979) \pno 3715
\bibitem{\klausWAN} K.~Hornfeck, {\it The Minimal
              Supersymmetric Extension of ${\cal WA}_{n-1}$},
              Phys.~Lett.~{\bf B275} (1992) \pno 355
\bibitem{\hornfeck} K.\ Hornfeck, {\it $\w$-Algebras with Set of Primary Fields
              of Dimensions (3,4,5) and (3,4,5,6)}, Nucl.\ Phys.\ {\bf B407}
              (1993) \pno 237
\bibitem{\klausREP} K.~Hornfeck, {\it Classification of
              Structure Constants for $\w$-Algebras from Highest Weights},
              Nucl.~Phys.~{\bf B411} (1994) \pno 307
\bibitem{\kneu} K.\ Hornfeck, {\it $\w$-Algebras of Negative Rank},
              Phys.\ Lett.\ {\bf B343} (1995) \pno 94
\bibitem{\hkn} S.\ Howes, L.P.\ Kadanoff, M.\ den Nijs,
              {\it Quantum Model for Commensurate-In\-com\-men\-su\-rate
              Transitions},
              Nucl.\ Phys.\ {\bf B215} (1983) \pno 169
\bibitem{\kohmoto} L.P.\ Kadanoff, M.\ Kohmoto, M.\ den Nijs, {\it Hamiltonian
              Studies of the $d=2$ Ashkin-Teller Model}, Phys.\ Rev.\ {\bf
              B24} (1981) \pno 5229
\bibitem{\kato} T.\ Kato, {\it On the Convergence of the Perturbation
              Method. I.}, Progr.\ Theor.\ Phys.\ {\bf 4} (1949) \pno 514
\bibitem{\katobook} T.\ Kato, {\it Perturbation Theory for Linear Operators},
              Springer-Verlag (1976), second edition
\bibitem{\hgkpriv} H.G.\ Kausch, private communication
\bibitem{\kau} H.G.\ Kausch, G.M.T.\ Watts, {\it A Study of $\w$-Algebras
              Using Jacobi Identities}, Nucl.\ Phys.\ {\bf B354} (1991) \pno
              740
\bibitem{\kedem} R.\ Kedem, B.M.\ McCoy, {\it Quasi-Particles in the
              Chiral Potts Model},
              Int.\ Jour.\ of Mod.\ Phys.\ {\bf B8} (1994) \pno 3601
\bibitem{\kogut} J.B.\ Kogut, {\it An Introduction to Lattice Gauge
              Theory and Spin Systems},
              Reviews of Mod.\ Phys.\ {\bf 51} (1979) \pno 659
\bibitem{\krallm} T.W.\ Krallmann, {\it Phasendiagramm des Chiralen
              Pottsmodells}, Diplomarbeit BONN-IR-91-11 (1991)
\bibitem{\nakan} A.\ Kuniba, T.\ Nakanishi, J.\ Suzuki, {\it Ferro and
              Antiferro-Magnetizations in RSOS Models}, Nucl.\ Phys.\
              {\bf B356} (1991) \pno 750
\bibitem{\lehmann} H.\ Lehmann, {\it \"Uber Eigenschaften von
              Ausbreitungsfunktionen und Renormierungs\-kon\-stanten
              Quantisierter Felder}, Il Nuovo Cimento
              {\bf 11} (1954) \pno 342
\bibitem{\rietalA} M.\ Marcu, A.\ Regev, V.\ Rittenberg,
              {\it The Global Symmetries of Spin Systems Defined on
              Abelian Groups. I},
              J.\ Math.\ Phys.\ {\bf 22} (1981) \pno 2740
\bibitem{\scm} B.M.\ McCoy, {\it The Chiral Potts Model: from Physics to
              Mathematics and back},
              Special functions ICM 90, Satellite Conf.\ Proc., ed.\
              M.\ Kashiwara and T.\ Miwa, Springer (1991) \pno 245
\bibitem{\mcroan} B.M.\ McCoy, S.-S.\ Roan, {\it Excitation Spectrum and Phase
              Structure of the Chiral Potts Model}, Phys.\ Lett.\ {\bf A150}
              (1990) \pno 347
\bibitem{\mussardo} G.\ Mussardo, {\it Off-Critical Statistical Models:
              Factorized Scattering Theories and Bootstrap Program},
              Phys.\ Rep.\ {\bf 218} (1992) \pno 215
\bibitem{\narganes} F.J.\ Narganes-Quijano, {\it On the Parafermionic $\w_N$
              Algebra}, Int.\ Jour.\ of Mod.\ Phys.\ {\bf A6} (1991) \pno 2611
\bibitem{\onsager} L.\ Onsager, {\it Crystal Statistics.\ I.\ A
              Two-Dimensional Model with an Order-Disorder Transition},
              Physical Review {\bf 65} (1944) \pno 117
\bibitem{\ostlund} S.\ Ostlund, {\it Incommensurate and Commensurate Phases
              in Asymmetric Clock Models},
              Phys.\ Rev.\ {\bf B24} (1981) \pno 398
\bibitem{\perk} J.H.H.\ Perk, {\it Star-Triangle Equations, Quantum Lax Pairs,
              and Higher Genus Curves}, Proc.\ Symp.\ Pure Math.\ {\bf 49}
              (1989) \pno 341
\bibitem{\poly} A.M.\ Polyakov, {\it Gauge Transformations and
              Diffeomorphisms},
              Int.\ Jour.\ of Mod.\ Phys.\ {\bf A5} (1990) \pno 833
\bibitem{\reedsimon}  M.\ Reed, B.\ Simon, {\it Methods of Modern Mathematical
              Physics IV: Analysis of Operators},
              Academic Press (1978), chapter XII
\bibitem{\rellich} F.\ Rellich,
              {\it St\"orungstheorie der Spektralzerlegung.\ IV.},
              Math.\ Ann.\ {\bf 117} (1940) \pno 356
\bibitem{\roanA} S.-S.\ Roan, {\it A Characterization of ``Rapidity''
              Curve in the Chiral Potts Model},
              Commun.\ Math.\ Phys.\ {\bf 145} (1992) \pno 605
\bibitem{\roanB} S.-S.\ Roan, {\it Onsager's Algebra, Loop Algebra and
              Chiral Potts Model},
              preprint Max-Planck-Institut f\"ur Mathematik
              Bonn MPI/91-70
\bibitem{\tarasov} V.O.\ Tarasov, {\it Transfer Matrix of the Superintegrable
              Chiral Potts Model.\ Bethe Ansatz Spectrum}, Phys.\ Lett.\ {\bf
              A147} (1990) \pno 487
\bibitem{\toppan} F.\ Toppan, {\it Generalized NLS Hierarchies from
              Rational $\w$-Algebras}, Phys.\ Lett.\ {\bf B327} (1994) \pno 249
\bibitem{\rva} R.\ Varnhagen, {\it Characters and Representations of
              New Fermionic $\w$-Algebras},
              Phys.\ Lett.\ {\bf B275} (1992) \pno 87
\bibitem{\weyl} H.\ Weyl, {\it The Classical Groups, Their Invariants
              and Representations}, Princeton, New Jersey, Princeton
              University Press (1946)
\bibitem{\wilczynski} E.J.\ Wilczynski, {\it Projective Differential Geometry
              of Curves and Ruled Surfaces}, Teubner Leipzig (1906)
\bibitem{\yildirim} K.\ Yildirim, {\it Parit\"atssymmetrie im Chiralen
              $\Zed_n$-Symmetrischen Integrablen Potts-Modell},
              Diplomarbeit BONN-IB-95-02 (1995)
\bibitem{\yuwu} F.\ Yu, Y.-S.\ Wu, {\it On the Kadomtsev-Petviashvili
              Hierarchy, $\hat W_\infty$ Algebra, and Conformal
              $SL(2,\Real)/U(1)$ Model, I.\ The Classical Case,
              II.\ The Quantum Case}, J.\ Math.\ Phys.\ {\bf 34} (1993)
              \pno 5851; \pno 5872
\bibitem{\don} D.\ Zagier, {\it Modular Forms and Differential Operators},
              preprint Max-Planck-Institut f\"ur Mathematik MPI/94-13
\bibitem{\zam} A.B.\ Zamolodchikov,
              {\it Infinite Additional Symmetries in Two-Dimensional
              Conformal Quantum Field Theory},
              Theor.\ Math.\ Phys.\ {\bf 65} (1986) \pno 1205
\bibitem{\zamPA} A.B.\ Zamolodchikov,
              {\it Higher-Order Integrals of Motion in Two-Dimensional
              Models of the Field Theory with Broken Conformal Symmetry},
              JETP Lett.\ {\bf 46} (1987) \pno 160
\bibitem{\zamPB} A.B.\ Zamolodchikov, {\it Integrals of Motion in Scaling
              3-State Potts Model Field Theory},
              Int.\ Jour.\ of Mod.\ Phys.\ {\bf A3} (1988) \pno 743
\bibitem{\zamPC} A.B.\ Zamolodchikov,
              {\it Integrable Field Theory from Conformal Field Theory},
              Adv.\ Studies in Pure Math.\ {\bf 19} (1989) \pno 641
\bibitem{\zamzam} A.B.\ Zamolodchikov, Al.B.\ Zamolodchikov, {\it Conformal
              Field Theory and Critical Phenomena in Two-Dimensional Systems},
              Physics Rev.\ Vol.\ {\bf 10,4} (1989) \pno 269
\vfill
\eject
\chapsubtitle{Translations of citations}
\parindent=\skip103
\helvit
\mn
Aristotle, ``The Physics'':
\item{}
In all sciences that are concerned with principles or causes or elements,
it is acquaintance with these that constitutes knowledge or understanding.
For we conceive ourselves to know about a thing when we are acquainted with
its ultimate causes and first principles, and have got down to its elements.
Obviously, then, in the study of Nature too, our first object must be to
establish principles.
\bigskip\noindent
Immanuel Kant, Preface to the second edition of ``The Critique of Pure
Reason'':
\item{}
{\it Mathematics} and {\it physics} are the two theoretical sciences
which have to determine their objects {\it a priori}. The former is
purely {\it a priori}, the latter is partially so, but is also dependent
on other sources of cognition.
\bigskip\noindent
Werner Heisenberg:
\item{}
So far we do not know yet in which language we can talk about events in the
atom. We have a mathematical language, i.e.\ a mathematical
scheme, which allows us to compute the stationary states of the atom or
the probabilities of a transition from one state to another. But we do
not know yet --at least not in general-- how this language is connected
with the usual language.
\bigskip\noindent
Niels Bohr:
\item{}
On the other hand we encounter difficulties which lie so deep that we do
not have any idea of the way of their solution; it is my personal opinion
that these difficulties are of such a nature that they hardly allow us to
hope that we shall be able, inside the world of the atoms, to carry through a
description in space and time of the kind that corresponds to our ordinary
images.
\bigskip\noindent
H.A.\ Lorentz:
\item{}
All this has great beauty and extreme importance, but unfortunately we do
not really understand it. We do not understand Planck's hypothesis concerning
oscillators nor the exclusion of the nonstationary orbits, and we do not
understand, how, after all, according to Bohr's theory, light is produced.
There can be no doubt, a mechanics of quanta, a mechanics of discontinuities,
has still to be made.
\vfill
\end